\documentclass[a4paper,11pt]{amsart}
\usepackage{graphicx} 
\usepackage{xcolor}
\usepackage[utf8]{inputenc}
\usepackage{amsmath}
\usepackage{amsfonts}
\usepackage{amsthm}
\usepackage{float}
\usepackage{cite}
\newcommand{\R}{\mathbb{R}}
\newcommand{\N}{\mathbb{N}}

\newcommand*{\dd}{\mathrm{d}}

\title{On visible effects in the double Schwarzschild solution}
\author{Eddy Brandon de Le\'on}
\email{eddy.de.leon@tum.de}
\address{Department of Mathematics, Boltzmannstrasse 3, TU München, D-85748 Garching bei
München, Germany}
\author{Jörg Frauendiener}
\email{joerg.frauendiener@otago.ac.nz}
\address{Department of Mathematics and Statistics, University of Otago, PO Box 56, Dunedin
9054, New Zealand}
\author{Christian Klein}
\email{christian.klein@u-bourgogne.fr}
\address{Institut de Mathématiques de Bourgogne, Institut 
Universitaire de France, Université Bourgogne Europe, 9 avenue Alain Savary, 21078 Dijon Cedex, France}

\thanks{E.B. de Le\'on is funded by the Deutsche Forschungsgemeinschaft (DFG, German Research Foundation) – TRR 352 – Project-ID 470903074.}

\begin{document}
\begin{abstract}
  Physical aspects of a static solution to the Einstein equations 
  with two black holes are studied via ray tracing. The exact 
  solution for this double Schwarzschild solution is known in explicit form.  The black holes are separated by a singularity called \emph{Weyl strut}. The effect of this strut on null geodesics is shown to be defocusing in contrast to the focusing effect of the black holes. It is shown that black holes with a large separation essentially lead to similar behavior of the null geodesics as a single black hole, whereas nearby holes display a widely changed behavior due to the Weyl strut.
\end{abstract}
\maketitle

\section{Introduction}
\label{sec:introduction}

One of the main objectives of Physics is to provide models which allow us to understand the phenomena that we experience in Nature. Models are usually obtained as (approximate) solutions to the equations which, we hope, provide an accurate enough encapsulation of the laws of Nature. Studying the properties of such solutions allows us to draw conclusions about observable phenomena that are possible within the considered part of Physics.

Here, we are concerned with Einstein's theory of gravity which describes the global structure of space-times, which themselves are obtained as solutions to Einstein's field equations. There exist many ways to obtain at least approximate solutions: most prominently are all the different scenarios that have been developed to find perturbative solutions near a fixed background space-time, such as linearized gravitational waves on Minkowski or black-hole space-times, or the immense field of cosmological perturbations, based on a background from the class of Friedmann-Lema\^{i}tre-Robertson-Walker (FLRW) space-times. Other methods are based on the initial-value formulation of the Einstein equations which can be used to evolve a space-time from suitable initial conditions using analytical or numerical methods.

Since the early days of general relativity, exact solutions have played a significant role. As mentioned above, they are used as background space-times for perturbative approaches. The most used exact solutions for this purpose are the ones mentioned above: the Kerr-Newman family (including the Maxwell, Schwarzschild, Reissner-Nordström solutions) and the cosmological space-times from the FLRW family. This is, in fact, not surprising since these are the most explored space-times, in the sense that many of their properties have been determined over time. This is quite in contrast to the many other exact solutions that exist for the Einstein equations about which not much more is known than a coordinate expression for the metric. This is even more the case for exact solutions for various flavors of alternative theories of gravity.

It is therefore an important task to examine a given space-time and explore its properties. This can be done in several ways. In ``simple enough'' space-times many properties can be obtained by analytic/algebraic methods. However, this becomes increasingly difficult, when the metric is given in terms of complicated algebraic constructions.

An increasingly popular method to explore unknown space-times is the 
use of visualisation based on ray-tracing. Such techniques have been 
applied to simulate visualisations around black
holes from \cite{luminet,mueller} to professional
renderings in the movie “Interstellar” 
\cite{interstellar,interstellar2} to the first 
actual picture \cite{EHT}.  In black hole space-times it is possible 
that no light from
certain regions of the space-time reaches an observer which leads to a 
\emph{shadow}, see for instance \cite{CHR,CEHL,NCS,SD,YNCS}. 
In~\cite{DeLeon:2024}, we presented a Matlab code for a ray-tracing 
algorithm making use of the highly accurate barycentric 
interpolation, see~\cite{barycentric} for a description. In the 
present work, we turn our attention to one of the simplest 
multi-black hole solutions, the double Schwarzschild 
solution~\cite{ernst,exac,buch}. It describes a static, axisymmetric 
space-time in which two non-rotating black holes generate an 
asymptotically flat gravitational field. In order for this 
configuration to remain static, it is necessary that the black holes 
are kept at fixed locations. Due to the axisymmetry, this results in 
a so called Weyl strut, a region along the symmetry axis between the 
black holes, where the space-time shows singular behavior usually 
referred to as a deficit/excess angle or conical singularity. The 
presence of the Weyl strut and of two horizons leads to interesting 
interactions between the influences of these geometric features on 
light rays traversing the space-time, which we try to explore, at 
least partially, in the present paper (for the exploration of shadows 
in these space-times, see \cite{CHR,NCS,SD,YNCS}, and for discussions on individual light rays, see \cite{SD}).

The plan of this article is as follows: we start with a brief reminder of the definition of the double Schwarzschild solution in sec.~\ref{sec:double-schw-solut}, and continue with a short description of the main features of our ray-tracing algorithm in sec.~\ref{sec:summ-numer-meth}. Then we go on to discuss in sec.~\ref{sec:light-rays} the properties of individual light rays as they pass through the region between the black holes near the strut. Finally, in sec.~\ref{sec:ray-tracing}, we display the intricate structure of images obtained through the lensing properties of binary system. Sec.~\ref{sec:conclusion} concludes the paper.

\section{Double Schwarzschild solution}
\label{sec:double-schw-solut}
In this section, we collect a few facts on the so-called double 
Schwarzschild solution, a static solution in the family of double Kerr 
solutions. 

It is well known that the stationary axisymmetric Einstein equations 
in vacuum are equivalent to the completely integrable Ernst 
equation, see for instance \cite{ernst,exac,buch} for references. In 
the framework of integrable systems, the Kerr solution describing a 
rotating black hole corresponds to the 2-soliton solution. This led 
to the question whether there is a physical interpretation of higher 
order solitons to the Ernst equation as multi-black holes. In 
\cite{KN} double Kerr solutions were constructed via a B\"acklund 
transformation. For a discussion of this solution, see 
\cite{exac,Manko} and the literature therein. In the present 
paper, we are interested in the static representative in this class of 
solutions. Below we follow the presentation in \cite{Manko}.

In the static axisymmetric case in vacuum, the metric can be chosen 
in Weyl-Lewis-Papapetrou form, see \cite{exac},
\begin{equation} 
	\dd s^{2}=-f 
	\dd t^{2}+\frac{1}{f}\left(e^{2k}(\dd\rho^{2}+\dd z^{2})+\rho^{2} 
	\dd \phi^{2}\right),
	\label{metric}
\end{equation}
where the cylindrical Weyl coordinates are such that $\rho$ measures 
the distance to the symmetry axis parameterized by $z$, and where 
$\partial_t$ and $\partial_\phi$ correspond to the static and the 
axisymmetric Killing vector respectively. The metric functions $f$ 
and $k$ depend on $\rho$ and $z$, but not on $\phi$ and $t$. The 
potential $f$ satisfies the axisymmetric Laplace (or Euler-Darboux) 
equation, $k$ is given for known $f$ in terms of quadratures.

\begin{figure}[h!]
  \includegraphics[width=0.49\textwidth]{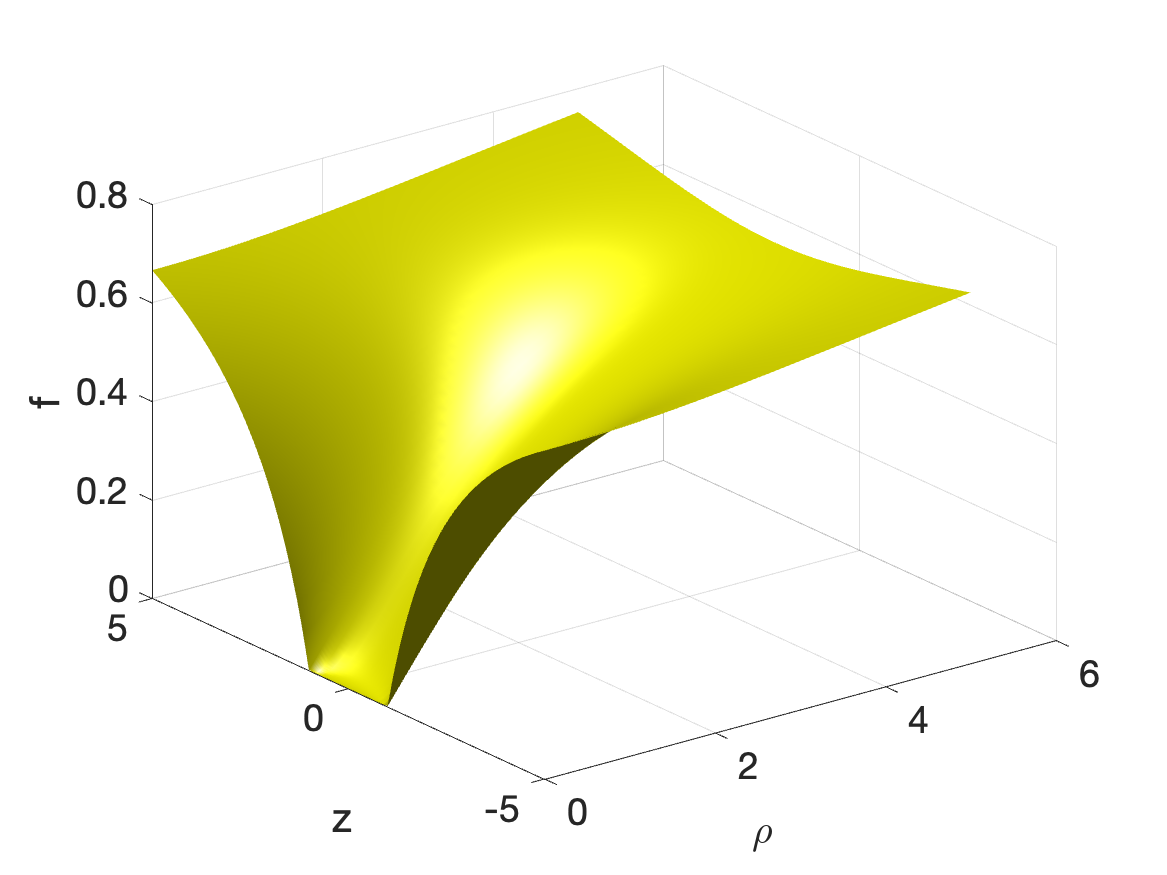}
  \includegraphics[width=0.49\textwidth]{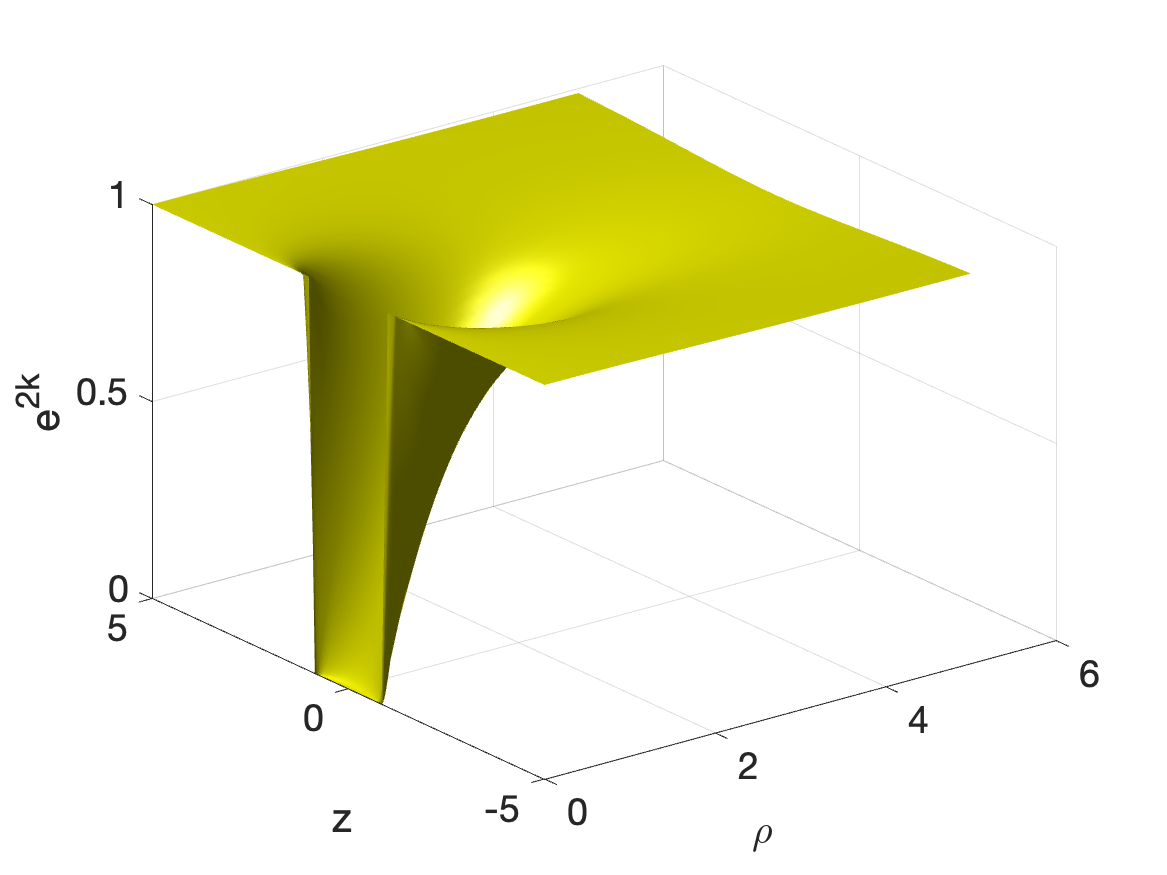}
 \caption{The metric functions (\ref{fk}) for the 
 Schwarzschild solution for $m=1$ , on the left 
 $f$, on the right $e^{2k}$.}
 \label{figschw}
\end{figure}

We recall the well known Schwarzschild solution of mass $m>0$ in this setting,
\begin{equation}
	f= \frac{X  - 1}{X  + 1},\quad e^{2k}=\frac{m^2(X^2 -1)}{r_{+}r_{-}}
	\label{schw},
\end{equation}
where $r_{\pm}=\sqrt{(z\pm m)^{2}+\rho^{2}}$, $X=(r_{+}+r_{-})/(2m)$. 
The metric is equatorially symmetric, i.e., both metric functions in 
(\ref{schw}) are even functions of $z$. The solution is 
asymptotically flat, both $f$ and $e^{2k}$ tend to 1 for 
$z^{2}+\rho^{2}\to\infty$. The horizon is localized on the $z$-axis 
in $[-m,m]$. Both metric functions vanish at the horizon. The 
function $e^{2k}$ can be chosen equal to 1 on the whole axis in the complement of the 
horizon. We show the Schwarzschild metric for $m=1$ in Fig.~\ref{figschw}.

In contrast to the single black hole case, in the 
double Schwarzschild case, the solution is parameterized by three 
parameters $m_{1}$, $m_{2}$ corresponding to the Komar masses of the 
two black holes, and a distance $R_{0}>m_{1}+m_{2}$.
We introduce $M=m_{1}+m_{2}$ and
\begin{equation}
	R_{\pm}=\mp\sqrt{\rho^2+(z+R_{0}/2\pm m_{2})^2},\quad 
	r_{\pm}=\mp\sqrt{\rho^2+(z-R_{0}/2\pm m_{1})^2},
	\label{rR}
\end{equation}
as well as the functions 
\begin{equation}
  \begin{aligned}
	A & 
	=(R_{0}^2-M^2)(R_{+}-R_{-})(r_{+}-r_{-})-4m_{1}m_{2}(R_{+}-r_{-})(R_{-}-r_{+}),
	\\
	B & = 2m_{1}(R_{0}^2-m_{1}^2 + m_{2}^2)(R_{-}-R_{+}) +  
	2m_{2}(R_{0}^2-m_{2}^2 + m_{1}^2)(r_{-}-r_{+}) \\
        & \hspace*{3em}+
    4R_{0}m_{1}m_{2}(R_{+}+R_{-}-r_{+}-r_{-}).
	\label{AB}
      \end{aligned}
\end{equation}
Then the metric functions for the double Schwarzschild solution are 
given in the form 
\begin{align}
	f & 
	=\frac{A-B}{A+B},
	\nonumber\\
	e^{2k} & = 
	\frac{(A^2-B^2)}{16R_{+}R_{-}r_{+}r_{-}(R_{0}^2-(m_{1}-m_{2})^2)^2}.
	\label{fk}
\end{align}

\begin{figure}[h!]
  \includegraphics[width=0.49\textwidth]{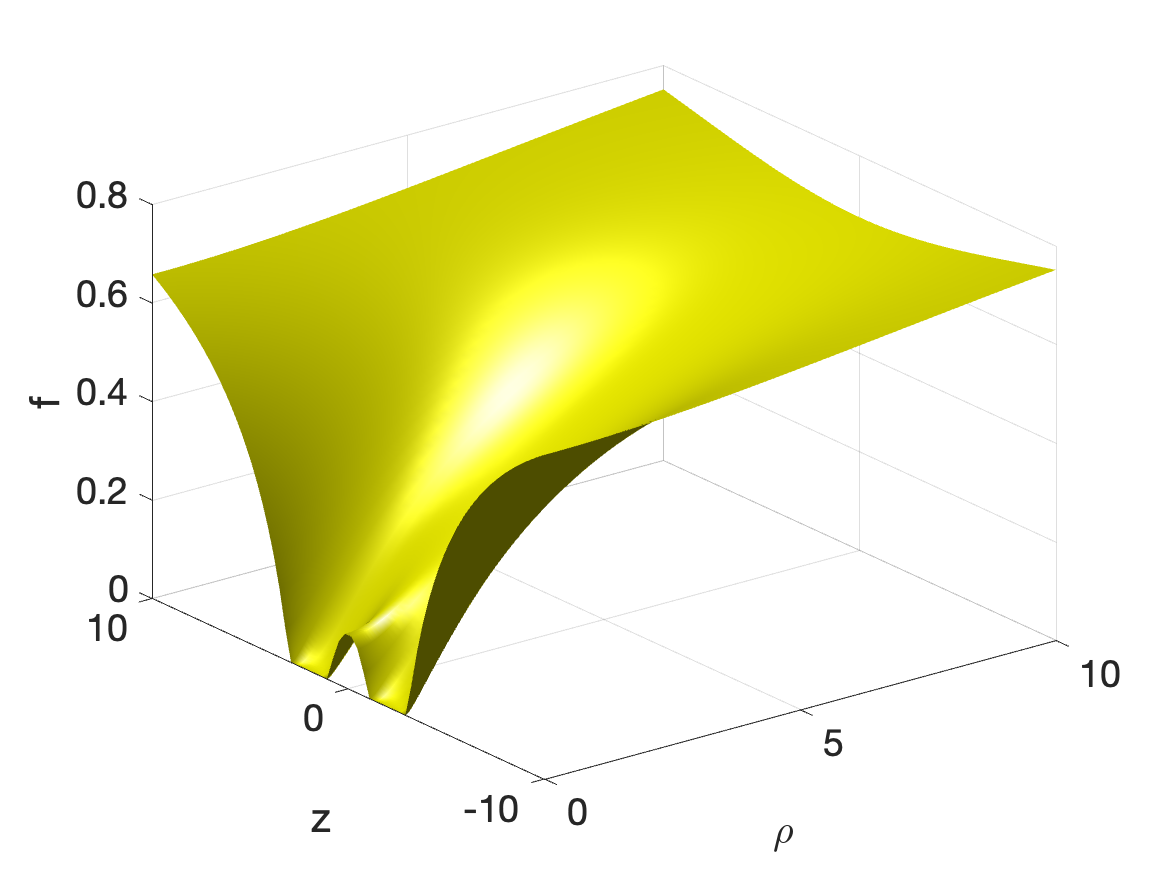}
  \includegraphics[width=0.49\textwidth]{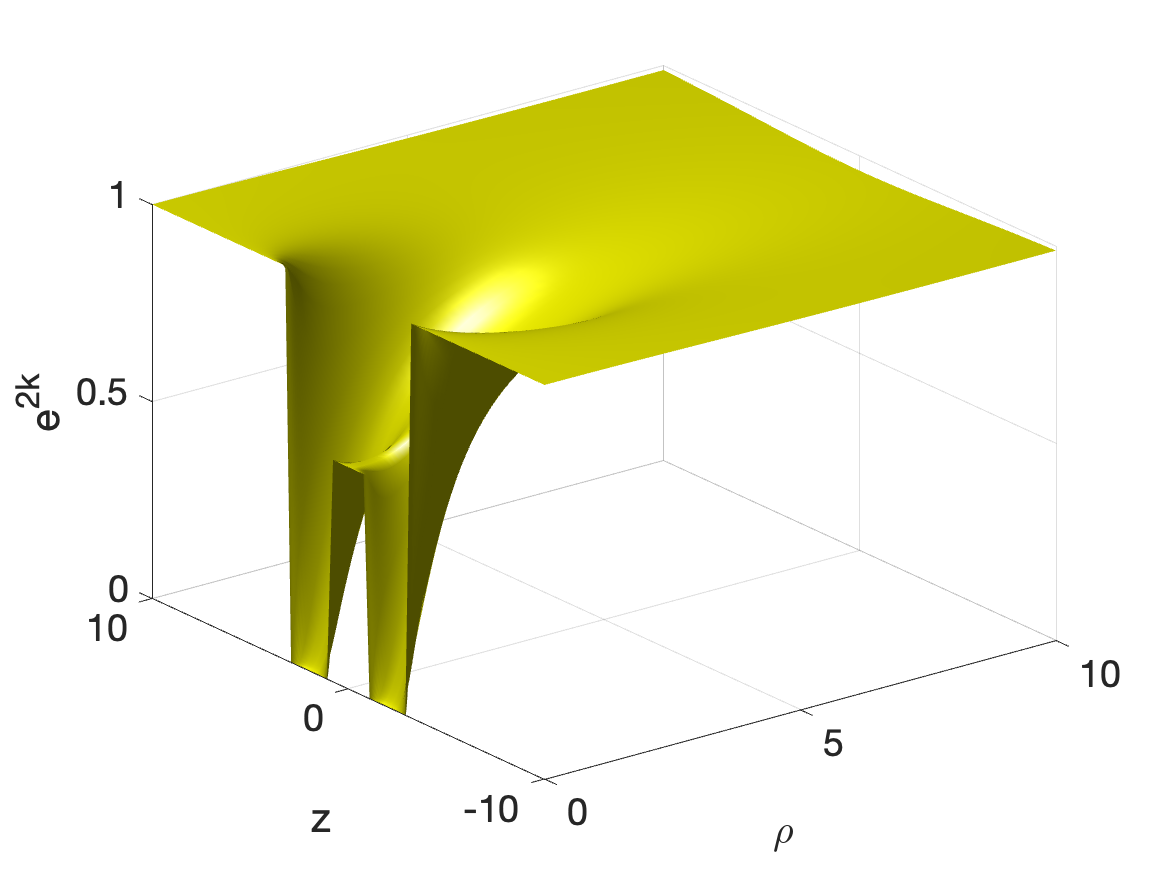}
 \caption{The metric functions (\ref{fk}) for the double 
 Schwarzschild solution for $m_{1}=m_{2}=1$ and $R_{0}=4$, on the left 
 $f$, on the right $e^{2k}$.}
 \label{DSm1}
\end{figure}

In this paper, we are mainly interested in the equal mass case in 
which the solution is equatorially symmetric. We show the metric functions for 
the example $m_{1}=m_{2}=1$ and $R_{0}=4$ in Fig.~\ref{DSm1}. 

It can be seen that there are two horizons, which are once more 
located  on the symmetry axis between $R_{0}/2\pm m_{1}$ and 
$-R_{0}/2\pm m_{2}$, in the present example along the intervals $[1,3]$ 
and $[-1,-3]$.  Both functions, $f$ and $e^{2k}$, vanish there. The 
solution is asymptotically flat. On the regular part of 
the axis, here for $|z|>3$, the metric function $k$ vanishes. However,
this is not the case between the two horizons, for $|z|<1$ in the 
example. There, $k$ is constant, but  not equal to zero, 
which implies that the axis is not 
elementary flat: the circumference of a circle of radius $\rho$ 
around the axis divided by $\rho$ does not tend to $2\pi$ in the 
limit $\rho\to0$ but to a larger value $2\pi e^{-k}$. 
This corresponds to a singularity in the space-time, a \emph{Weyl 
strut}, which keeps the situation static despite two black holes 
attracting each other.

The fact that the circumference of a small circle around the axis is larger than the one of a circle with the same radius in a Euclidean geometry means that there is an ``excess angle'' at the axis. Qualitatively, this seems to indicate that the strut has a repulsive influence on geodesics transverse to the axis. On the other hand, there is the attractive nature of the black holes influencing the geodesics in the direction of the axis. Thus, we expect interesting dynamics of the geodesics and consequently of the imaging properties of the system. It is therefore an important and interesting task to study the 
impact of this Weyl strut on the light propagation between the two 
black holes.

\section{Summary of the numerical method}
\label{sec:summ-numer-meth}

\subsection{Geodesic equations}
The motion of particles (whether mass-less or not) in general space-times is governed by the geodesic equation 
\[
  \frac{\dd ^2 x^\mu}{\dd s^2} + \Gamma^\mu_{\alpha \beta} \frac{\dd  x^\alpha}{\dd s}  \frac{\dd  x^\beta}{\dd s} = 0 ,
\]
where $\{x^0,x^1,x^2,x^3\} = \{t,\rho,\zeta,\phi\}$ in our case, and $s$ is an affine parameter.

It is well known that these equations can be obtained from a variational principle with the action $\mathcal{A} = \int \mathcal{L}\,\dd s$ where the Lagrangian is $\mathcal{L} = g_{\mu\nu}\dot{x}^\mu \dot{x}^\nu$, and the canonical momenta $p_\mu$ are defined by $p^\mu=\dot{x}^\mu=\dd x^\mu /\dd s $. However, these equations take a simpler form in stationary axisymmetric space-times, since the presence of any Killing vector $K^\mu$ leads to a conserved quantity $p_\mu K^\mu$ along the geodesics.

We denote the conserved quantities associated to $\partial_t$ and $\partial_\phi$ by $-E$ (the negative sign is used by convention) and $L$. Since $g_{t \phi}=0$ in double Schwarzschild space-times, they take the simple forms $E= f p^t $ and $L= (\rho^2/f) p^\phi$. In addition, the Lagrangian $\mathcal{L}$ is also a conserved quantity, with $\mathcal{L}=0$ since we are considering null geodesics. Thus, the geodesic equations could be further simplified; however, we will use this property to track the numerical errors and determine the physical relevance of the approximate solution.

Thus, the  geodesic equations in Weyl coordinates can be written as  the following  system of ODEs for the remaining six variables
which we combine into the vector $\vec{y}=(t,\rho,p^\rho,z,p^z,\phi)\in \R^6$:
\begin{equation} \label{ode_geodesics}
  \begin{split}
    \frac{\dd t}{\dd s} &= \frac{E}{f} ,\\
    \frac{\dd \rho}{\dd s} &= p^\rho, \\
    \frac{\dd p^\rho}{\dd s} & = \frac{1}{2h} \left[ -f_\rho (E/f)^2 - h_\rho (p^\rho)^2 + h_\rho (p^z)^2 +  (\rho^2/f)_\rho (Lf/\rho^2)^2 - 2 h_z p^\rho p^z \right], \\
    \frac{\dd z}{\dd s} &= p^z, \\
    \frac{\dd p^z}{\dd s} & = \frac{1}{2h} \left[ -f_z (E/f)^2 + h_z (p^\rho)^2 - h_z (p^z)^2 + (\rho^2/f)_z (Lf/\rho^2)^2 - 2 h_\rho p^\rho p^z \right], \\
    \frac{\dd \phi}{\dd s} &= \frac{f}{\rho^2} L,
  \end{split}
\end{equation}
where $h:=g_{\rho\rho}=g_{zz}=e^{2k}/f$ . 
Thus, the determination of the geodesic passing through the initial point $\vec{y}_0$ amounts to the solution of the initial value problem (IVP)
\begin{equation} \label{eq:ivp}
\left\{ \begin{array}{cc}
       \frac{\dd \vec{y}}{\dd s} = F(\vec{y}), \\
   \vec{y}(0) = \vec{y}_0,
\end{array}  \right.
\end{equation}
where $F:\R^6\to\R^6$ is the function described by the right hand sides of~\eqref{ode_geodesics}. This process is summarized as follows: given initial conditions for all eight variables ($t_0$, $p^t_0$, $\rho_0$, $p^\rho_0$, $z_0$, $p^z_0$, $\phi_0$, $p^\phi_0$), we first obtain the conserved quantities $E$ and $L$. For the remaining six variables encoded by the vector $\vec{y}$ we express the geodesic equations in the form \eqref{eq:ivp} and solve them numerically with the Runge-Kutta method of fourth order within Matlab.

\subsubsection{Numerical errors}
The geodesics are computed iteratively with the Runge-Kutta 4, whose local truncation error is on the order of $O(h^5)$, where $h$ is the step-size at the step $k\in\N$. Matlab's \texttt{ode45} function uses an adaptive method in which $h$ is updated at every step to assure that the estimated local error $e_j$ in every component of the solution $y$ satisfies
$$ |e_j| \leq \max( \texttt{RelTol} *|y_j|, \texttt{AbsTol}), $$
for small tolerances $\texttt{RelTol}, \texttt{AbsTol} >0$. 

However, this control is done over local errors, which means that when trajectories are computed over large time spans the accumulation of errors might lead to a nonphysical approximate solution, even if the local errors are kept under control. This is particularly important for photons that orbit the black holes several times before eventually falling toward the horizons or escaping to infinity.
Therefore we introduce the accuracy $\epsilon>0$ and regard the approximate solution as physically relevant if the condition $| g_{\mu\nu} p^\mu p^\nu |<\epsilon$ is satisfied globally. In practice, this is achieved by choosing the tolerances two orders of magnitude smaller than $\epsilon$. For instance, given $\epsilon=10^{-6}$, the condition described above is met by setting  $\texttt{RelTol}= \texttt{AbsTol} = 10^{-8}$.

\subsection{Initially parallel light rays} 
We are interested in the effects of the Weyl strut on individual light rays, which can be observed with clarity when the light rays are located on the equatorial plane in a system with black holes of equal mass ($m_1=m_2$), since the attraction of the black holes has the same magnitude in this case and therefore the light rays stay on the plane, allowing us to observe the effects caused solely by the Weyl strut.
In order to do this, we shoot a beam of initially parallel light rays passing between the black holes.

Similarly, we shoot a beam of parallel light rays lying in the $xz$-plane in order to observe how they are absorbed by the black holes and especially to observe their behavior when they pass between them. In this case, it is not necessary to impose the condition $m_1=m_2$ in order for the light rays to stay on the $xz$-plane, as one can deduce from \eqref{ode_geodesics} and the fact that the conserved quantity $L$ vanishes for this type of initial conditions (in fact, for any initial condition with $p^\phi_0=0)$. 

Considering the usual relation between Weyl and Cartesian 
coordinates ($x=\rho \cos\phi$, $y=\rho\sin\phi$), the following initial conditions will produce light rays with the properties described above.

\begin{itemize}
\item[(i)] Photons in the equatorial plane directed toward the negative $x$-direction for the same $x_0$ and for several equispaced values $y_0\in [y_{\min},y_{\max}]$. The initial conditions for these light rays have the form
  \begin{equation} \label{eq:type_i}
    \left\{
      \begin{array}{llll}
        t_0 = 0, & \rho_0=\sqrt{x_0^2+y_0^2}, & z_0 = 0, & \phi_0=\arctan{(y_0/x_0)},\\
        p^t_0 , & p^\rho_0 = - \frac{x_0}{\rho_0} , & p^z_0 =0, &  p^\phi_0 = \frac{y_0}{\rho_0^2}.
      \end{array}
    \right.
  \end{equation}
\item[(ii)] Photons on the $xz$-plane directed in the negative $x$-direction for the same $x_0$ and several equispaced values $z_0\in [z_{\min},z_{\max}]$. The initial conditions for these light rays have the form
  \begin{equation} \label{eq:type_ii}
    \left\{
      \begin{array}{llll}
        t_0 = 0, & \rho_0=x_0, & z_0 = z_0, & \phi_0=0,\\
        p^t_0 , & p^\rho_0 = -1, & p^z_0 =0, &  p^\phi_0=0.
      \end{array}
    \right.
  \end{equation}
\end{itemize}
The component $p_0^t$ is obtained as the positive solution of the quadratic equation $g_{\mu\nu}p_0^\mu p_0^\nu=0$, since we are considering photons moving forward in time when studying them individually. We will refer to these conditions as initial conditions of type (i) and (ii), respectively.

\subsection{Ray tracing}
The apparent images seen by a distant observer are simulated by considering a virtual camera at the spatial location $(\rho_c,z_c,\phi_c)$. The setting we consider is that of a \textit{camera obscura}, as described in \cite{DeLeon:2024}, with focal length $f_L$ and a screen of width $d_H$ and height $d_V$ with resolution $I\times J$ pixels, with $I,J>1$. The camera is placed at a distance $R_c:=\sqrt{\rho_c^2+z_c^2} \gg 1$ and its line of sight may be pointing at the origin of the coordinate system (i.e., the barycenter of the binary black hole system) or at one of the black holes, depending on the apparent image we need to simulate. Once these parameters are chosen, we solve the IVP \eqref{eq:ivp} corresponding to each pixel and assign a coloring depending on whether they fall inside one of the black hole horizons or they escape to infinity.

\section{Light Rays}
\label{sec:light-rays}
We start by studying the behavior of light rays moving on the equatorial plane. Since the metric functions are symmetric with respect to the equatorial plane, light rays initially on the plane with $p^{\zeta}=0$ will remain on the plane. 
\subsection{Horizons and photon surfaces}
\begin{figure}[!htb]
    \centering
    \includegraphics[width=0.45\linewidth]{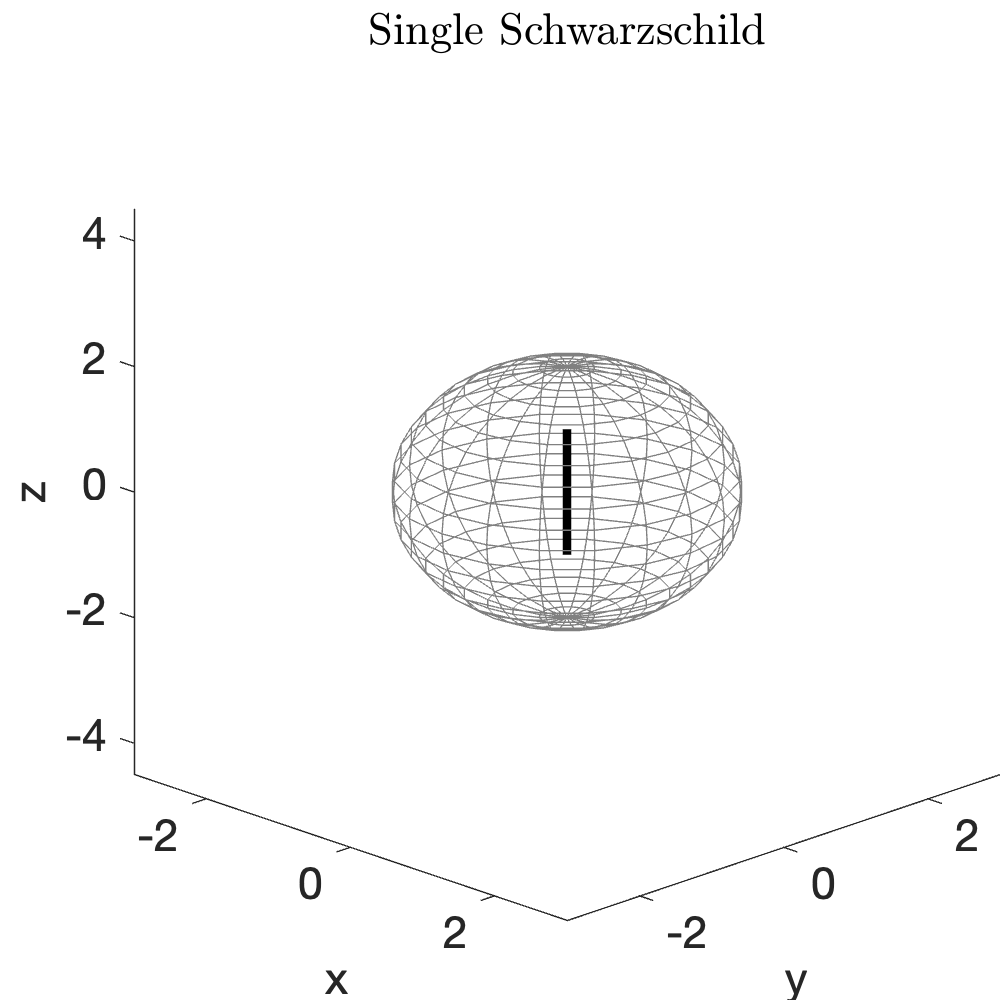}
    \includegraphics[width=0.45\linewidth]{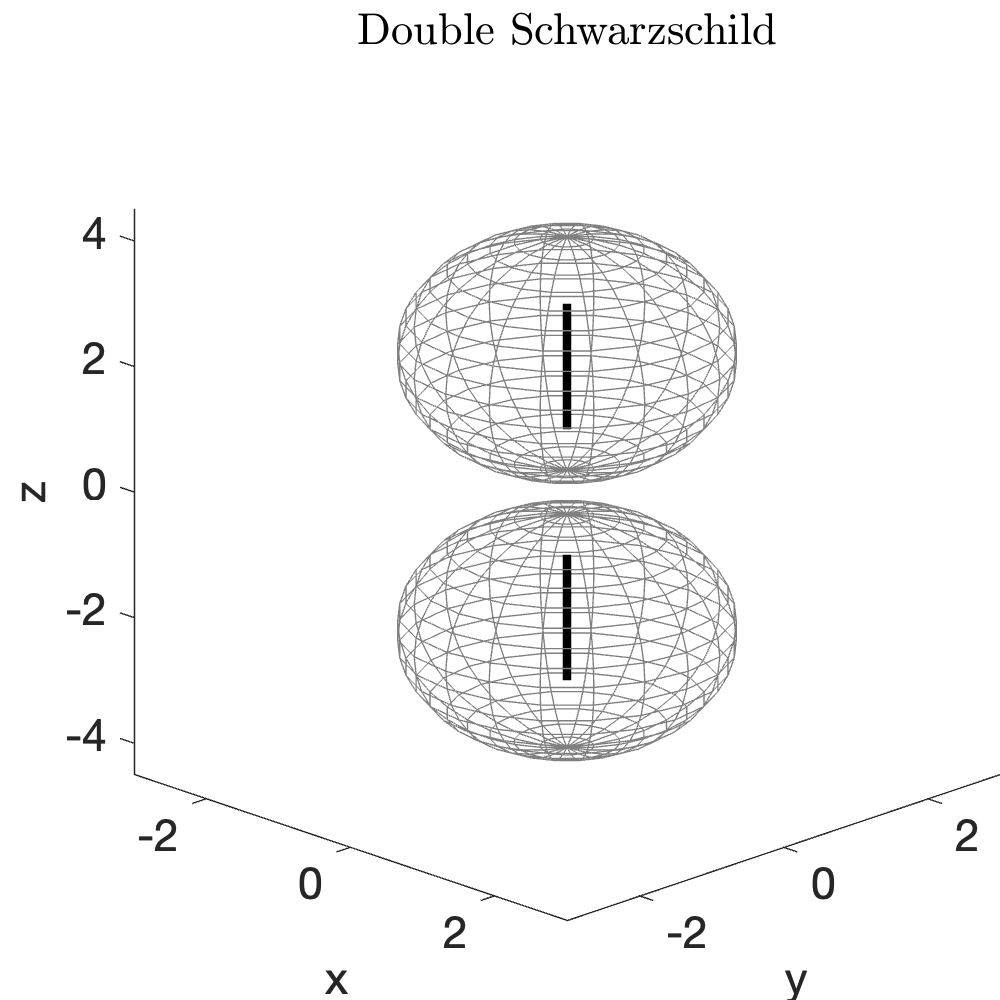}
    \caption{Diagram with the horizon (line on the axis) and photon sphere of each black hole.} \label{fig:diagrams_BHs}
\end{figure}

One key difference with respect to the single Schwarzschild case is 
that none of the light rays on the plane is directed toward any of 
the horizons due to the black holes (BHs being located outside of the plane, as shown in the right hand side of Fig.~\ref{fig:diagrams_BHs}. Therefore, we expect the photons to experience some effects in the neighborhood of the BHs and then escape to infinity. 

We recall that in the Schwarzschild coordinates the spatial 
projection of the horizon  of a single Schwarzschild BH is the sphere 
with radius $r=2m$ centered at the origin, while the photon surface 
is the sphere with radius $r=3m$, also centered at the origin. The Weyl and Schwarzschild coordinates are related via the transformation
\begin{equation}
    \rho = \sqrt{r^2-2mr} \sin\theta, \quad \zeta = (r-m)\cos\theta.
\end{equation}
This means that the horizon in Weyl coordinates is described by the line on the $\zeta$-axis with $-m\leq \zeta\leq m$ and the photon surface is described by the ellipsoid $\rho^2/3m^2+\zeta^2/4 m^2=1$.

The analogous objects in a double Schwarzschild space-time are described as follows: the upper and lower horizons are given by the lines on the $\zeta$-axis with $R_0/2-m_1\leq \zeta \leq R_0/2+m_1$ and $-R_0/2-m_2\leq \zeta \leq -R_0/2+m_2$, respectively. The surfaces are no longer centered at $\pm R_0/2$, as one would expect if one BH did not have any effect on the other. In this case, 
the photon surfaces are ellipsoids (at least numerically)
and their centers are pushed away from the center of the horizon (in the opposite direction of the other BH). 

For instance, in the space-time given by $m_1=m_2=1$ and $R_0=4$ the BHs would 
have their photon surfaces touch at the poles if they were unaffected by the presence of the other BH. However, this is not the case as shown in Fig.~\ref{fig:diagrams_BHs}. The photon sphere was computed numerically, and it seems to be (very close to) an ellipsoid\footnote{As a general remark, we point out that when we refer to geometric objects such as ``ellipsoid'' here, we do not refer to the geometry defined by the space-time under investigation. Instead, these terms simply refer to the coordinates.}.

\subsection{Initial conditions on the equatorial plane}

We study the trajectories of initially-parallel photons on the equatorial plane, i.e., solutions of the IVP \eqref{eq:ivp} with initial conditions of type (i) described by \eqref{eq:type_i} with $x_0=10$ and several equispaced values $y_0\in [y_{\min},y_{\max}]$, where the precise values of $y_{\min},y_{\max}$ and the number of light rays will be indicated within the discussions of each of the following examples. 

\subsubsection{Single Schwarzschild} \label{sec:S_xy}
\begin{figure}[!htb]
    \centering
    \includegraphics[width=0.45\linewidth]{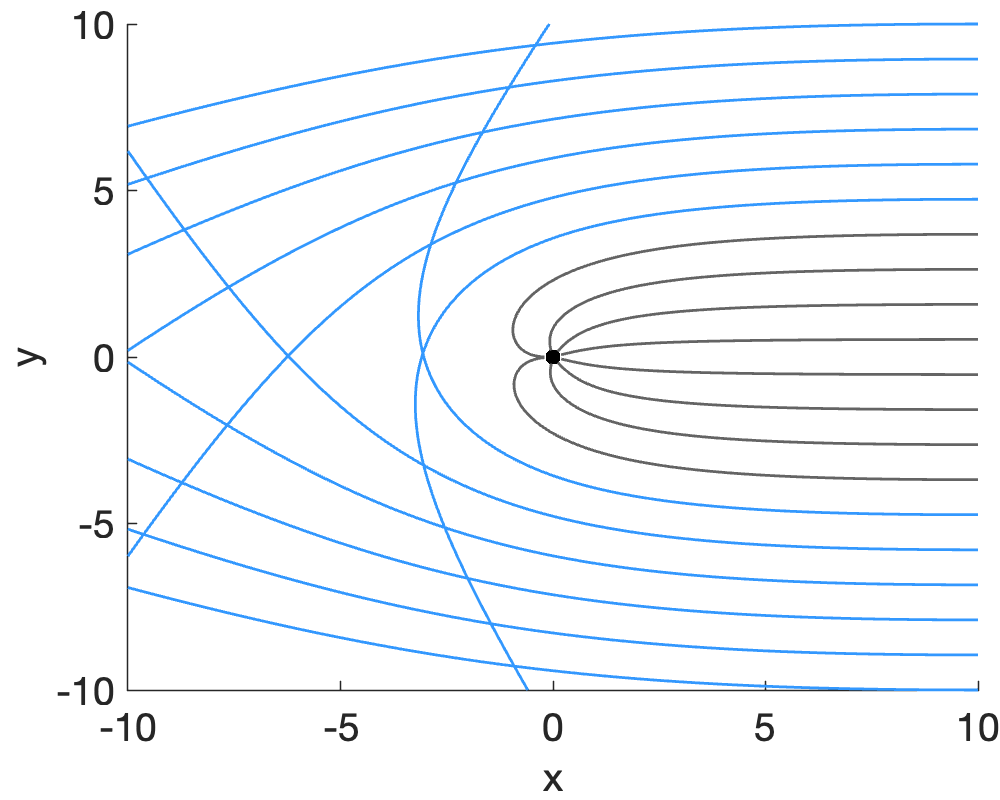}
    \includegraphics[width=0.45\linewidth]{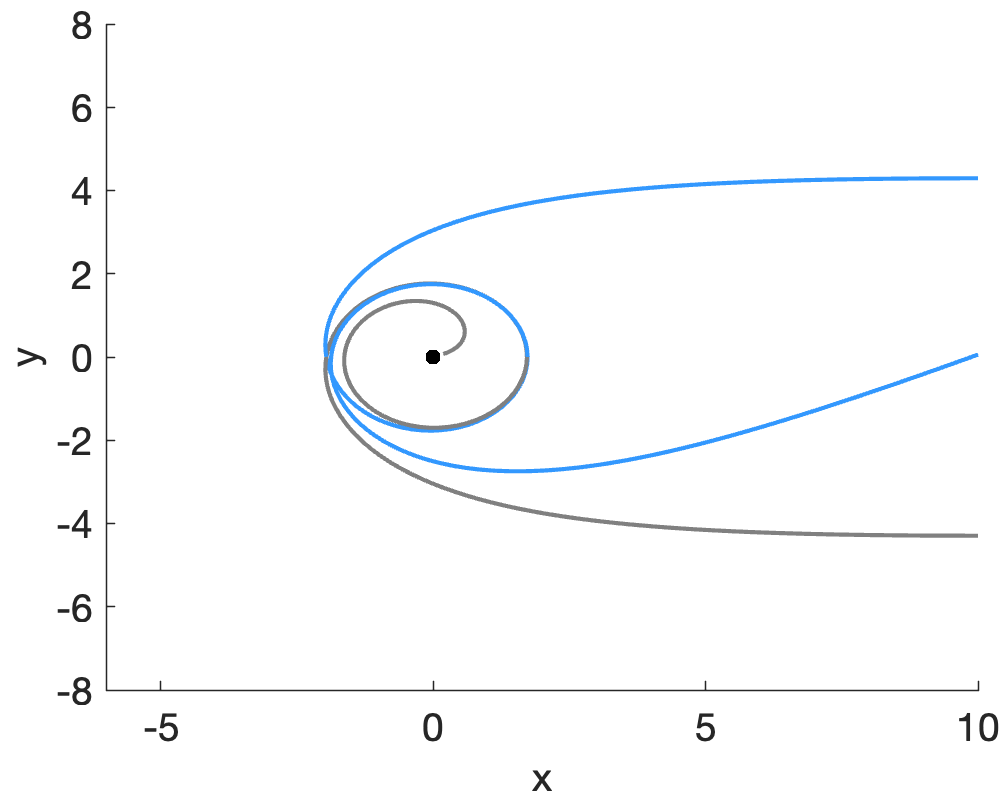}
    \caption{Left hand side: initially parallel light rays on the equatorial plane in a single Schwarzschild space-time. Right hand side: light rays that orbit the BH once and then fall inside the horizon or escape to infinity. The black dot is the 2D projection of the horizon.} \label{fig:schwarzschild_LRs}
\end{figure}

We recall some well known behavior in the single Schwarzschild space-time.
One of the effects on light in this space-time is the bending of light in the direction of the BH. We consider three main situations: (a) the photon is initially pointing sufficiently far away from the BH and thus its trajectory is bent when it passes in a neighborhood of the BH and then it escapes to infinity, (b) the photon is initially pointing sufficiently close to the BH and therefore it crosses the photon sphere and falls inside the horizon, (c) the initial condition is chosen adequately that the photon will orbit the BH indefinitely. In practice, given arbitrary conditions, only cases (a) and (b) are observed since circular photon orbits are unstable, meaning that small perturbations to the adequate initial conditions imply that the photon will eventually fall into the horizon or escape to infinity, possibly after some turns about the BH. Fig.~\ref{fig:schwarzschild_LRs} shows light rays that correspond to case (a) in blue and case (b) in gray. We used initial conditions of type (i) with $x_0=10$ and 20 equispaced values $y_0\in [-2, 2]$ for the left hand side of Fig.~\ref{fig:schwarzschild_LRs}.

\subsubsection{Double Schwarzschild}
\begin{figure}[!htb]
    \centering
    \begin{tabular}{cc}
    \scriptsize $m_1=1,m_2=1,R_0=8$ & \scriptsize $m_1=1,m_2=1,R_0=6$ \\
    \includegraphics[width=0.45\linewidth]{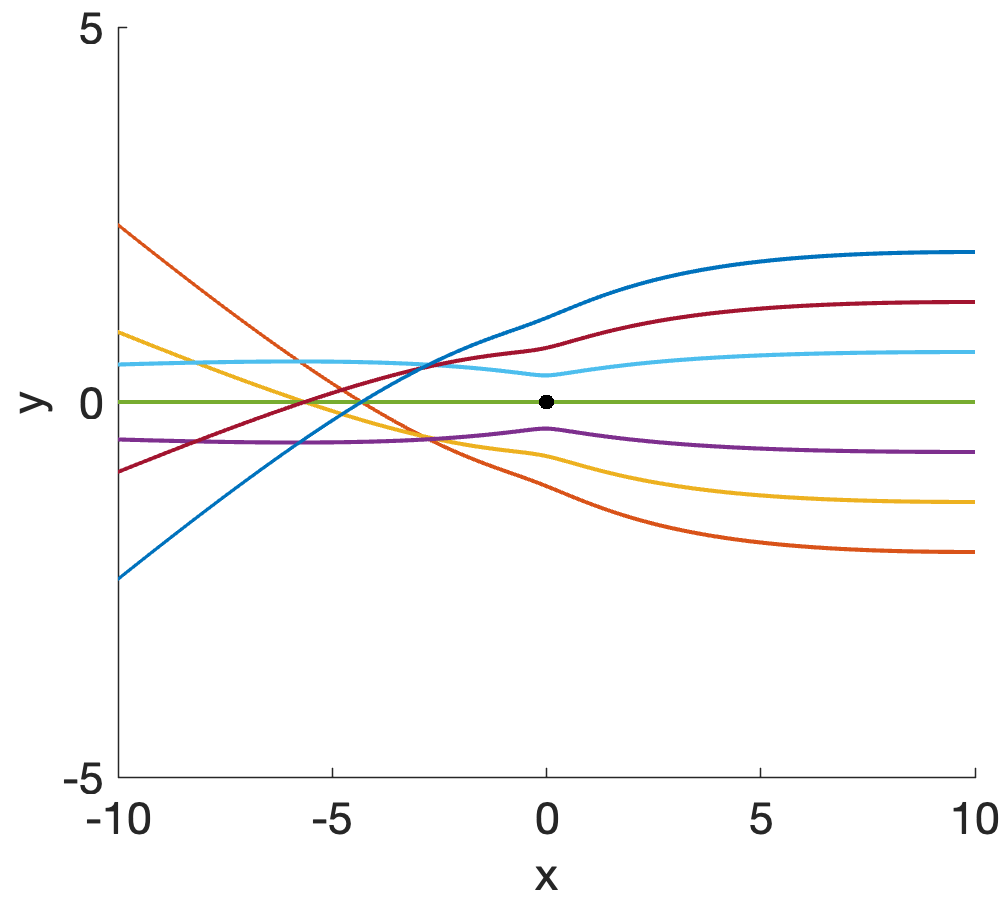} &
    \includegraphics[width=0.45\linewidth]{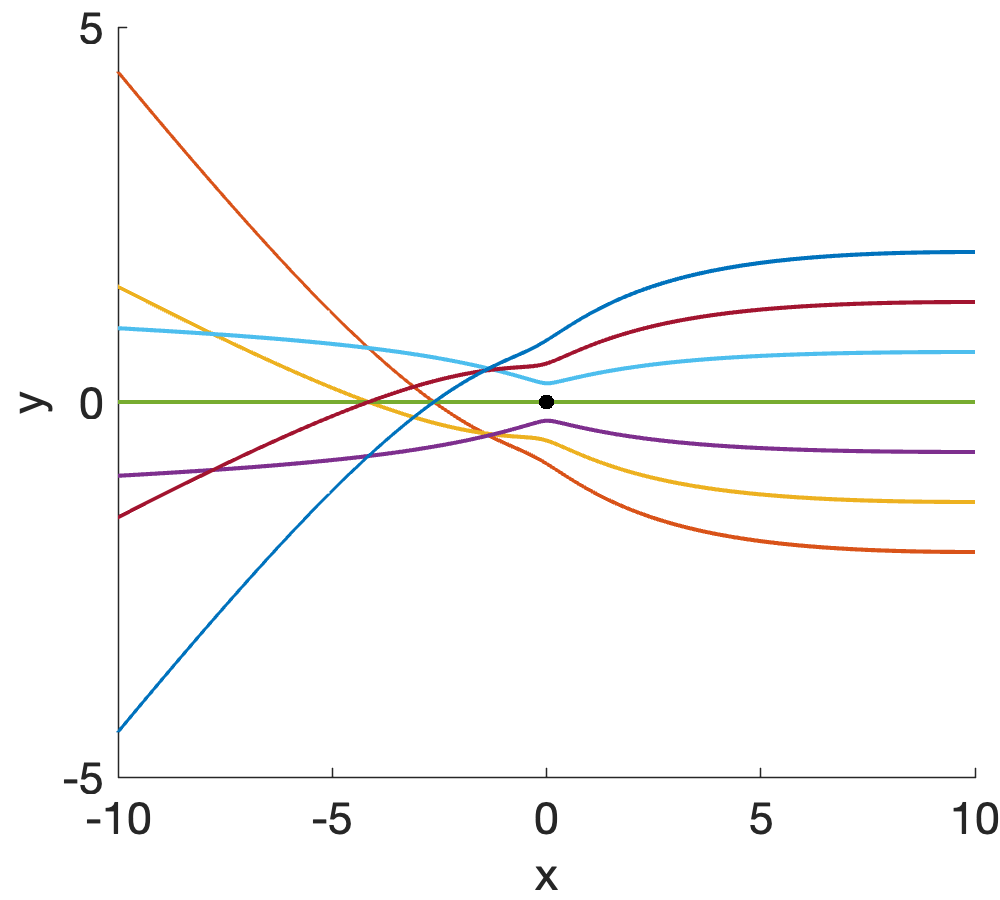} \\
    \scriptsize $m_1=1,m_2=1,R_0=4$ & \scriptsize $m_1=1,m_2=1,R_0=3$ \\
    \includegraphics[width=0.45\linewidth]{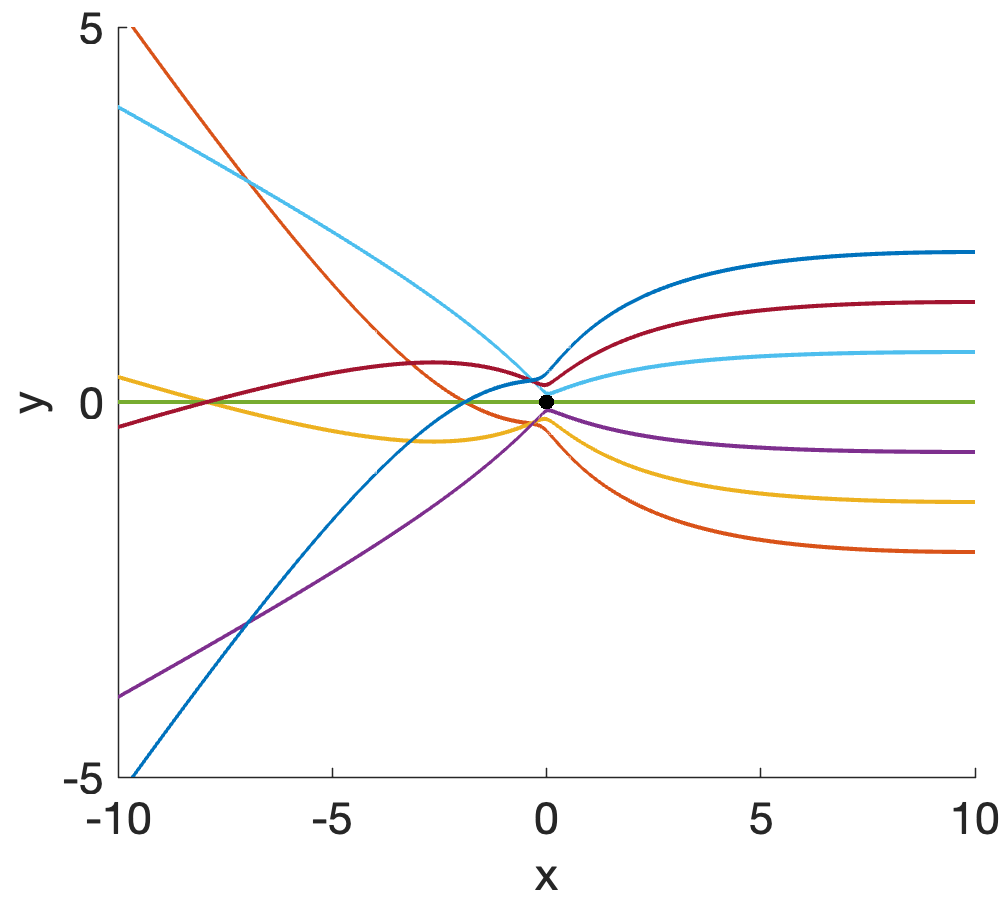} &
    \includegraphics[width=0.45\linewidth]{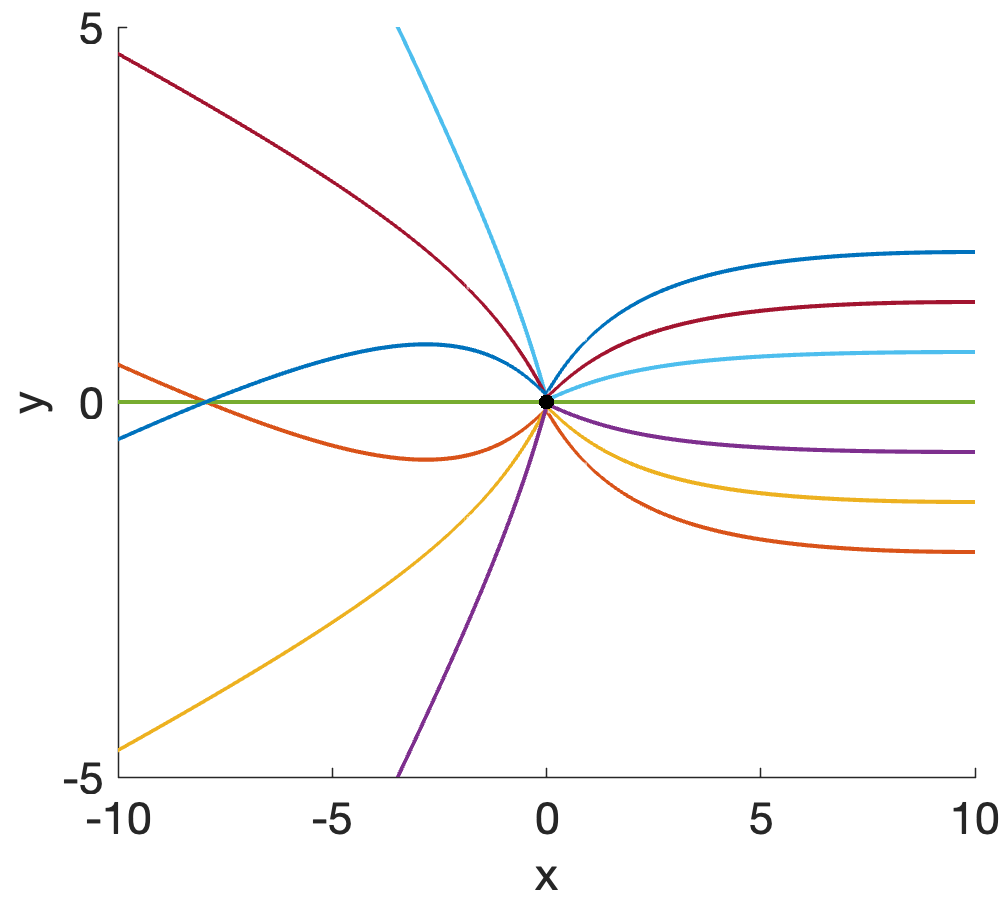}
    \end{tabular}
    \caption{Initially parallel light rays with fixed $m_1=m_2=1$ and decreasing values of $R_0$. The dot is the 2D projection of the horizons.} \label{fig:DS_LRs}
\end{figure}

We reproduce the same experiment as in the previous subsection. We 
expect analogous results if the photon is initially pointing sufficiently far away from the $z$-axis, i.e., if the initial conditions \eqref{eq:type_i} have sufficiently large $|y_0|>0$ and therefore it will produce type (a) trajectories. The other expected behavior is when $y_0=0$, i.e., when the photon is traveling in radial direction and thus it can be deduced from the geodesic equations that the trajectory will be a straight line. Both cases are observed in Fig.~\ref{fig:DS_LRs}, in which we use the same set of initial conditions in all cases. These figures show light rays passing between the BHs (see the diagram in Fig.~\ref{fig:diagrams_BHs}), each system with masses $m_1=m_2=1$ and various values of the separation $R_0$. 
Due to this symmetric configuration, light rays traveling on the equatorial plane will not orbit the BHs nor will they fall inside the horizons, i.e., cases (b) and (c) from the preceding subsection will not be observed in double Schwarzschild space-times if $R_0>2$. However, we shall observe a new effect due to the Weyl strut. 
Fig.~\ref{fig:DS_LRs} shows that how \textit{sufficiently small} $y_0$ has to be in order for the light rays to experience the effects of the Weyl strut will depend on the separation $R_0$. 
If the separation is large enough (e.g., $R_0=8$ as shown in the figure), then most of the light rays will only experience the light bending effect, while those passing close to the axis only experience a minor deflection. Both the attraction and the deflection effects increase as the separation $R_0$ decreases. This is especially visible for the case $R_0=3$ in Fig.~\ref{fig:DS_LRs}. When the photon surfaces merge (e.g., for $R_0= 2$ when the horizons meet), then this light deflection will no longer be visible and instead light rays will fall into the horizon.
We used initial conditions of type (i) with $x_0=10$ and seven equispaced values $y_0\in [-2, 2]$ in all the examples shown in Fig.~\ref{fig:DS_LRs}. 

\subsubsection{Deflected light rays}

\begin{figure}[!htb]
    \centering
    \begin{tabular}{cc}
    \scriptsize $m_1=1,m_2=1,R_0=8$ & \scriptsize $m_1=1,m_2=1,R_0=6$ \\
    \includegraphics[width=0.45\linewidth]{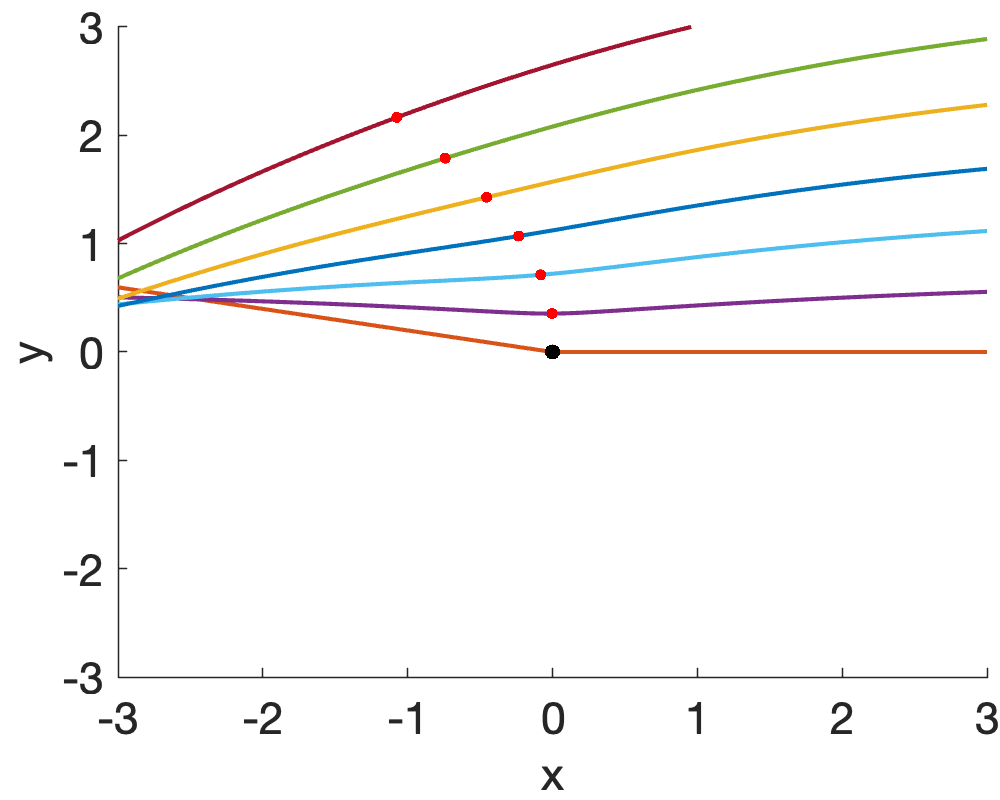} &
    \includegraphics[width=0.45\linewidth]{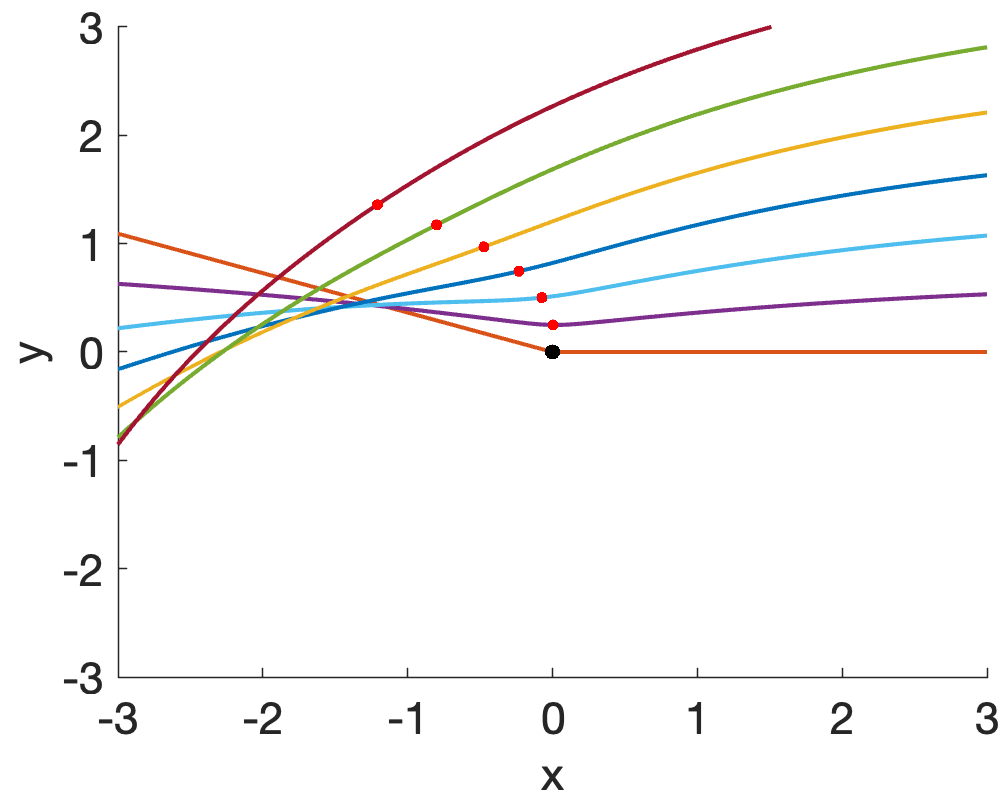} \\
    \scriptsize $m_1=1,m_2=1,R_0=4$ & \scriptsize $m_1=1,m_2=1,R_0=3$ \\
    \includegraphics[width=0.45\linewidth]{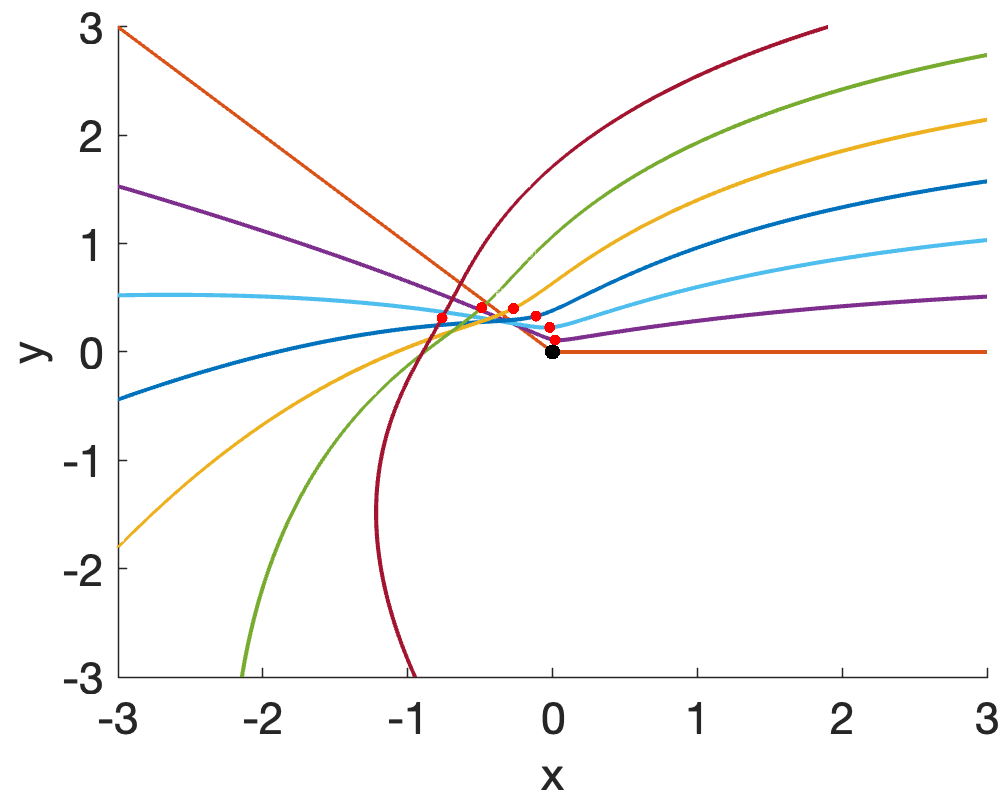} &
    \includegraphics[width=0.45\linewidth]{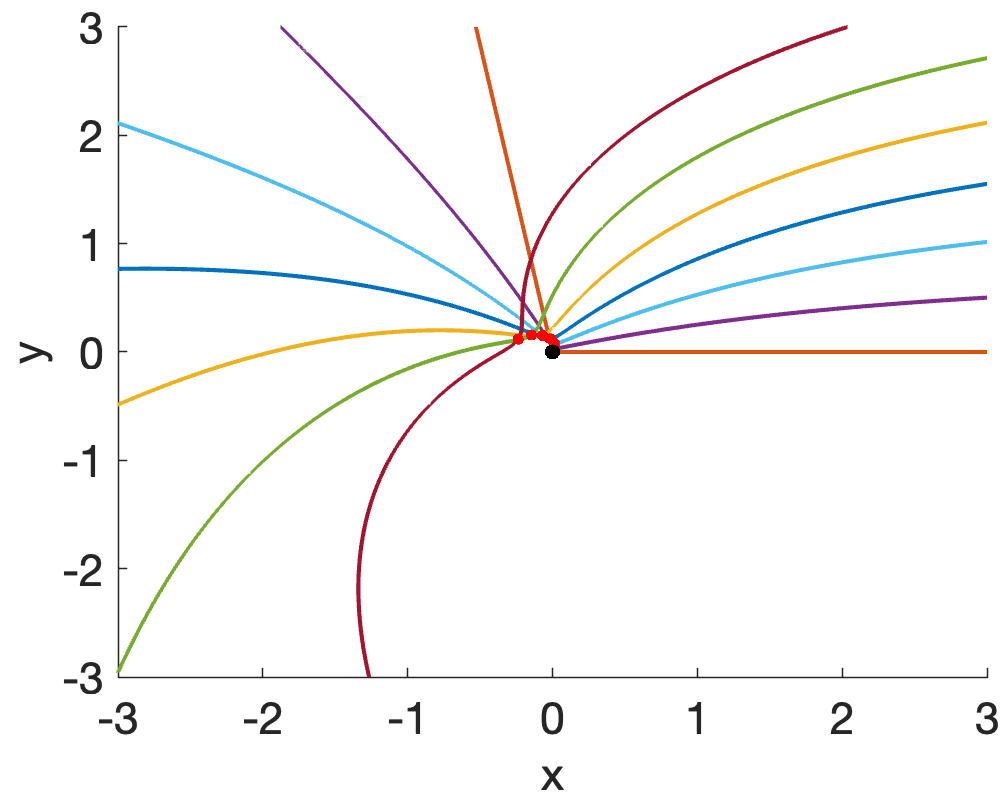}
    \end{tabular}
    \caption{Deflected light rays in the vicinity of the Weyl strut. We consider fixed $m_1=m_2=1$ and decreasing values of $R_0$. } \label{fig:deflected_LRs}
\end{figure}

When light rays are sufficiently far away from the binary system, the behavior is similar to that caused by a single black hole of mass $M=m_1+m_2$, i.e., light rays are bent in the direction of the gravitating source.
However, when light rays are in the vicinity of the Weyl strut a short-range repulsive effect arises. Light rays are deflected when $\rho\to 0$ on the equatorial plane, as observed in Fig.~\ref{fig:deflected_LRs}. We plot light rays corresponding to the initial conditions of type (i) with $x_0=10$ and seven equidistant values $y_0\in [\epsilon,4]$, for some small $\epsilon>0$. The deflection angle increases as $y_0\to 0$, until it reaches a limiting value.
The choice $\epsilon=10^{-6}$ is enough to illustrate the limiting value of the deflection angle in each space-time (further decreasing this value will not show a significant difference; unless a value that is numerically zero is reached and thus the behavior will be that of a photon with a purely radial velocity). 
The effect when $y_0<0$ is analogous since the metric coefficient $g_{t \phi}$ vanishes identically in these space-times.

\begin{figure}[!htb]
    \centering
    \includegraphics[width=0.7\linewidth]{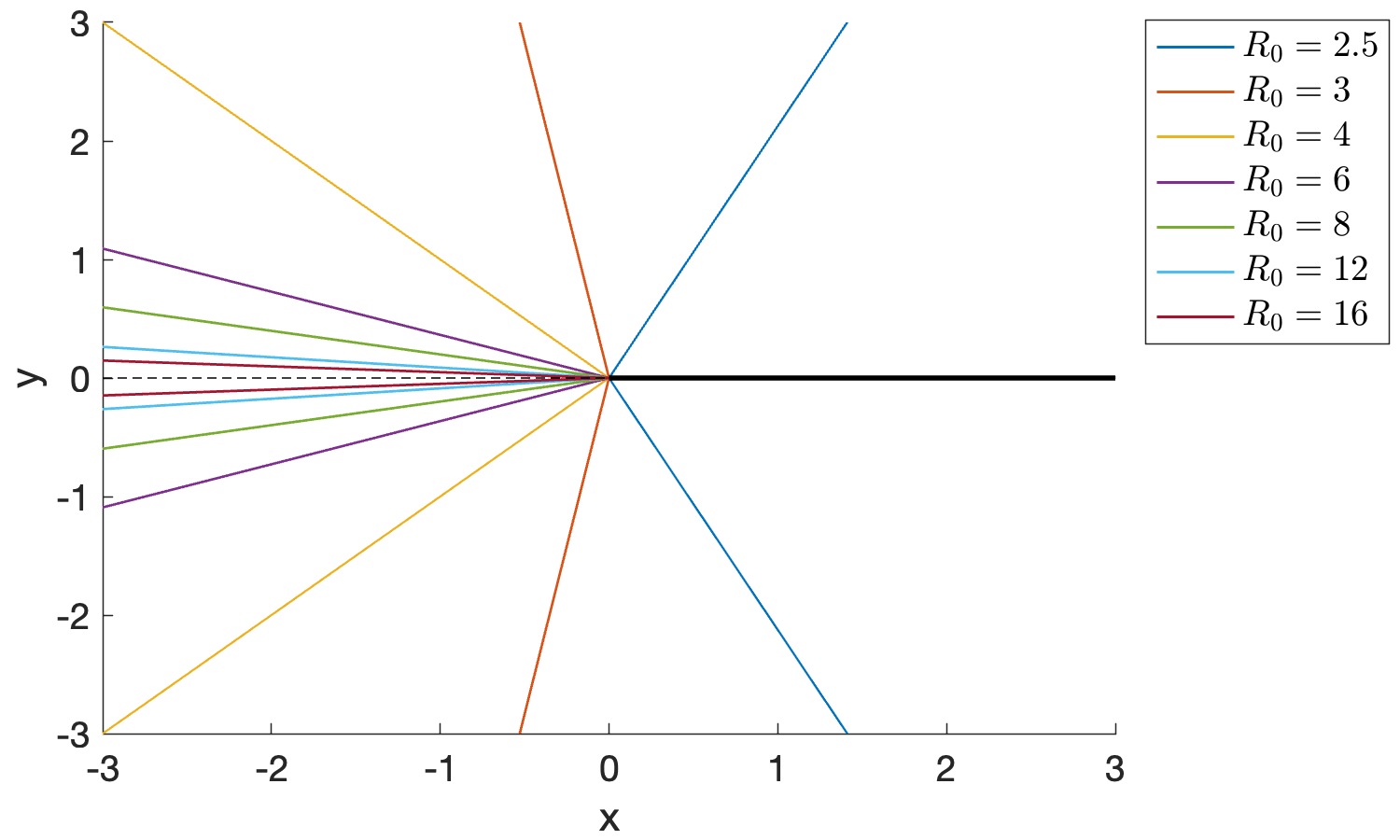}
    \caption{Curves showing the limiting value of the deflection angle for various values of $R_0$. The dashed line indicates a trajectory without deflection.} \label{fig:max_deflection_LR}
\end{figure}

The deflection angle increases as the separation $R_0$ decreases. Fig.~\ref{fig:max_deflection_LR} shows the case $y_0=\pm 10^{-6}$ (enough to illustrate the limiting value of the deflection angle as discussed above) for space-times with $m_1=m_2=1$ and various values of $R_0$. 

\begin{figure}[!htb]
    \centering
    \includegraphics[width=0.55\linewidth]{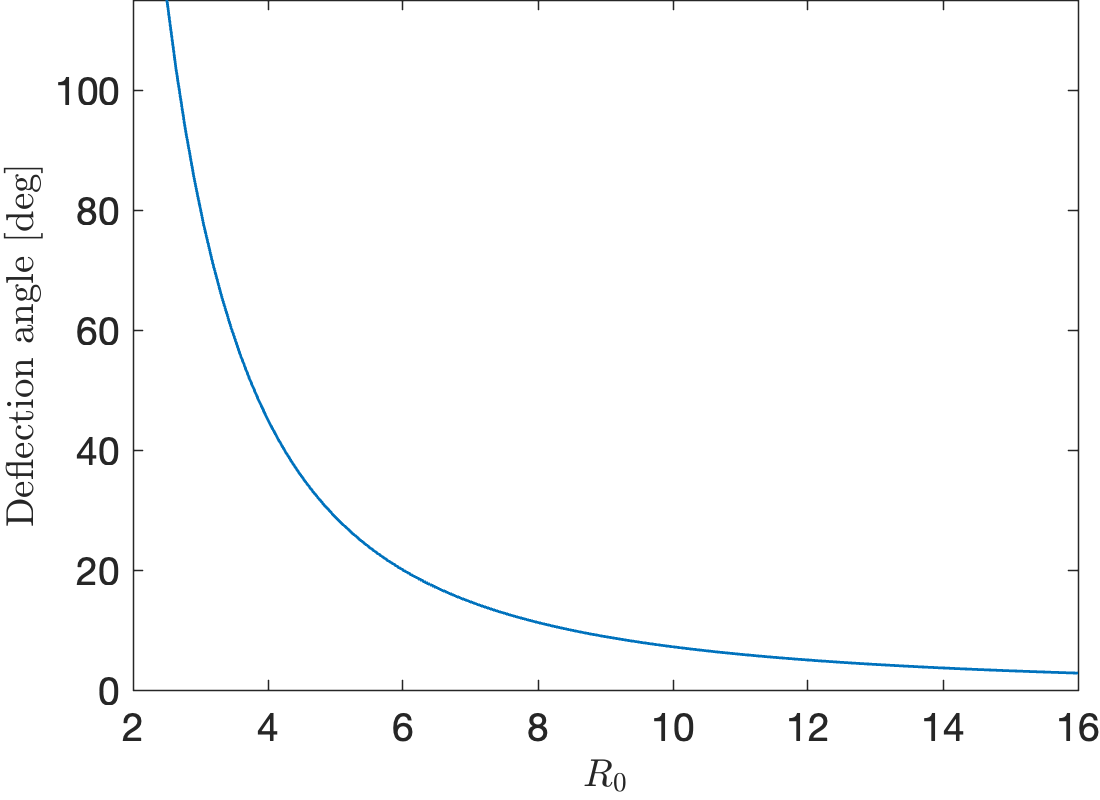}
    \caption{Deflection angle in dependence of $R_0$.} \label{fig:deflection_vs_R0}
\end{figure}

The limiting value of the deflection angle in dependence of $R_0$, which was computed numerically, is shown in Fig.~\ref{fig:deflection_vs_R0} for the separation values $R_0\in [2.5,16]$.

\subsubsection{Caustic}
\begin{figure}[!htb]
    \centering
    \begin{tabular}{cc}
    \scriptsize $R_0=5$ & \scriptsize $R_0=4.5$ \\
    \includegraphics[width=0.45\linewidth]{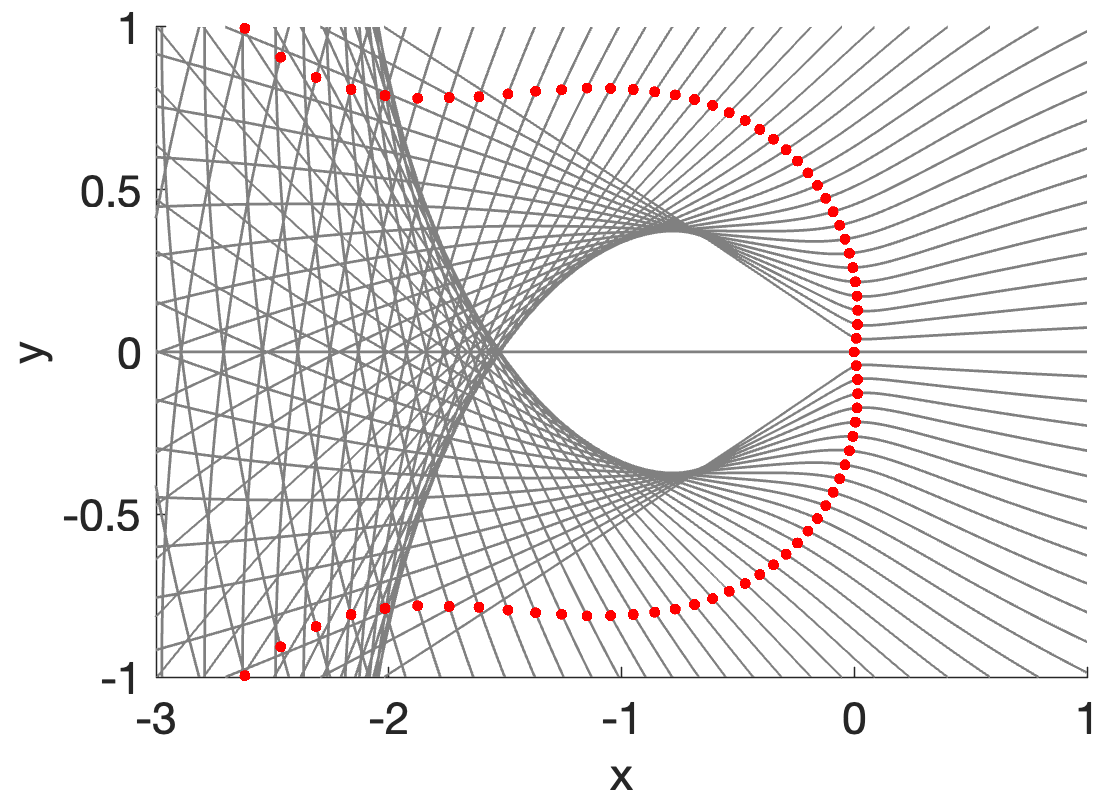} &
    \includegraphics[width=0.45\linewidth]{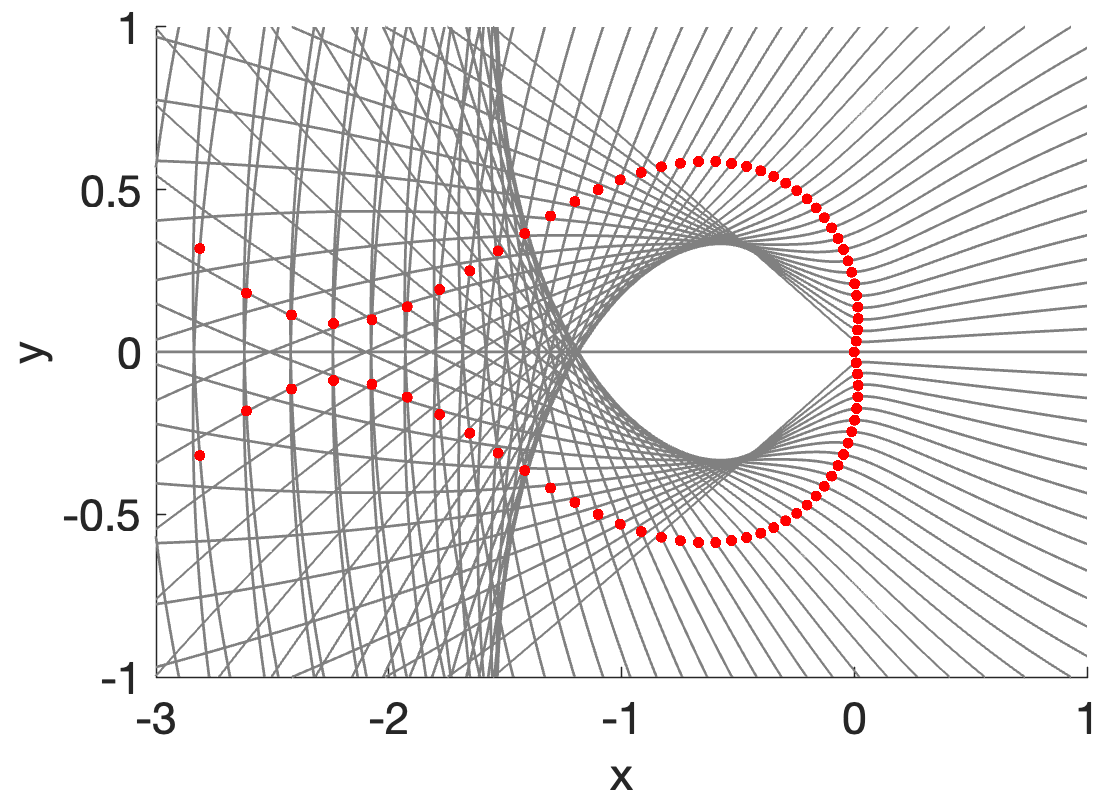} \\
    \scriptsize $R_0=4$ & \scriptsize $R_0=3.5$ \\
    \includegraphics[width=0.45\linewidth]{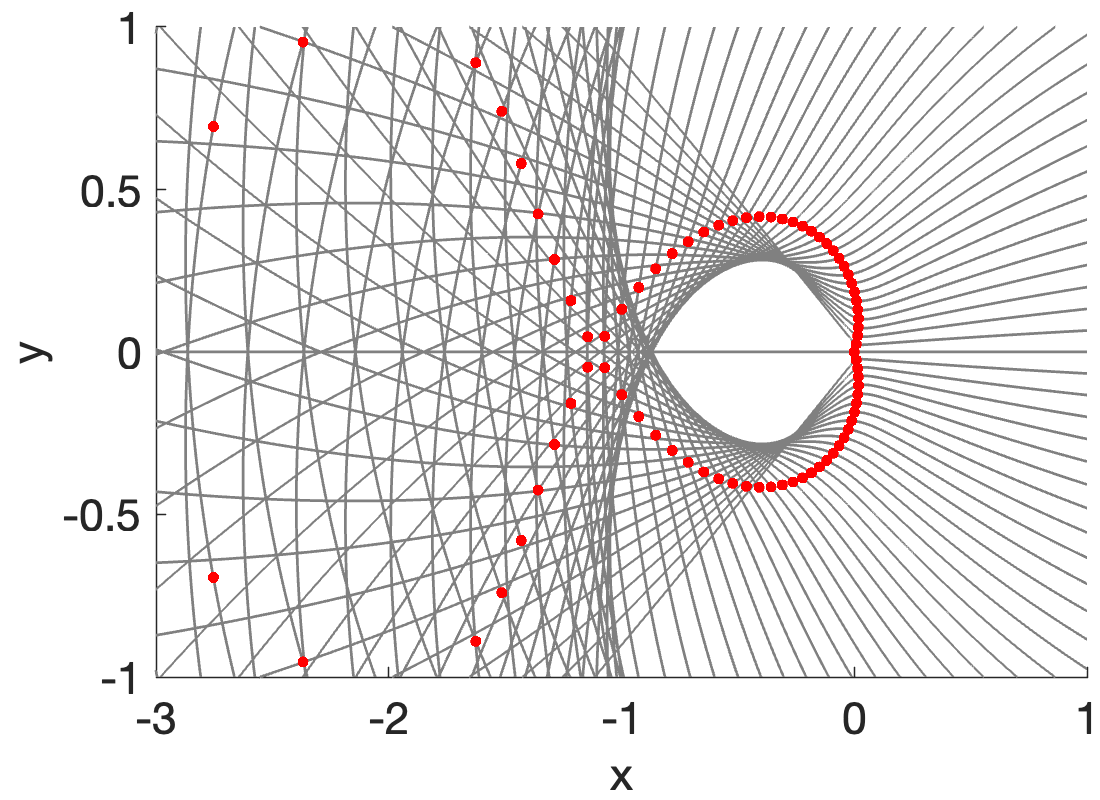} &
    \includegraphics[width=0.45\linewidth]{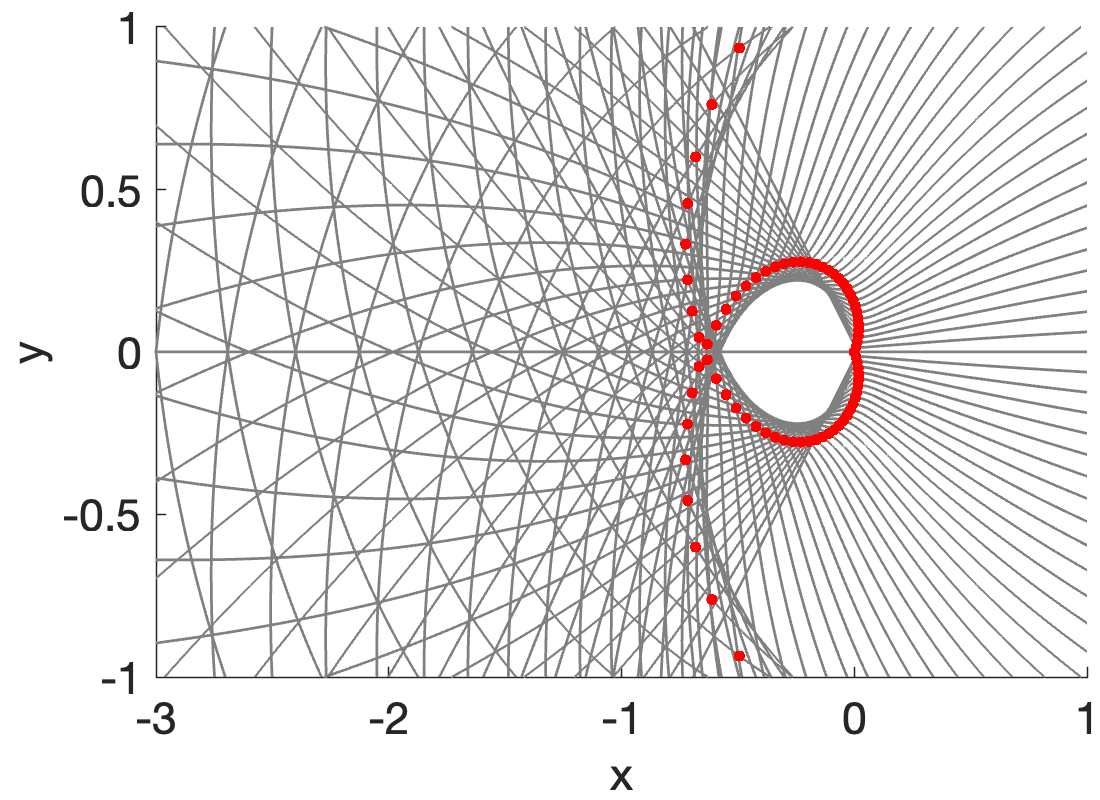} 
    \end{tabular}
    \caption{Light rays with their corresponding \textit{breaking points} (red dots). Apparent caustics are observed in each case. We consider fixed $m_1=m_2=1$ and decreasing values of $R_0$.} \label{fig:caustic}
\end{figure}

Plotting a larger number of light rays will show that an apparent caustic appears about the Weyl strut, as observed in various examples in Fig.~\ref{fig:caustic}.
Moreover, we plot the points in which the light rays \textit{break}. We define the breaking point as the point in the photon's trajectory where $p^{\rho}=0$, i.e., the point of closest approach to the axis (see red dots in the figure).
All the examples shown in Fig.~\ref{fig:caustic} were produced
using initial conditions of type (i) with $x_0=10$ and 91 equispaced values $y_0\in [-7,7]$. 

It can be observed (upon considering a large number of light rays) that the breaking points trace a curve, which we denote as the \textit{breaking point curve}. The shape of the curve is almost a parabola when the separation $R_0$ is large enough, but then
the curve closes about the Weyl strut as the BHs get closer.
\begin{figure}[!htb]
    \centering
    \includegraphics[width=0.6\linewidth]{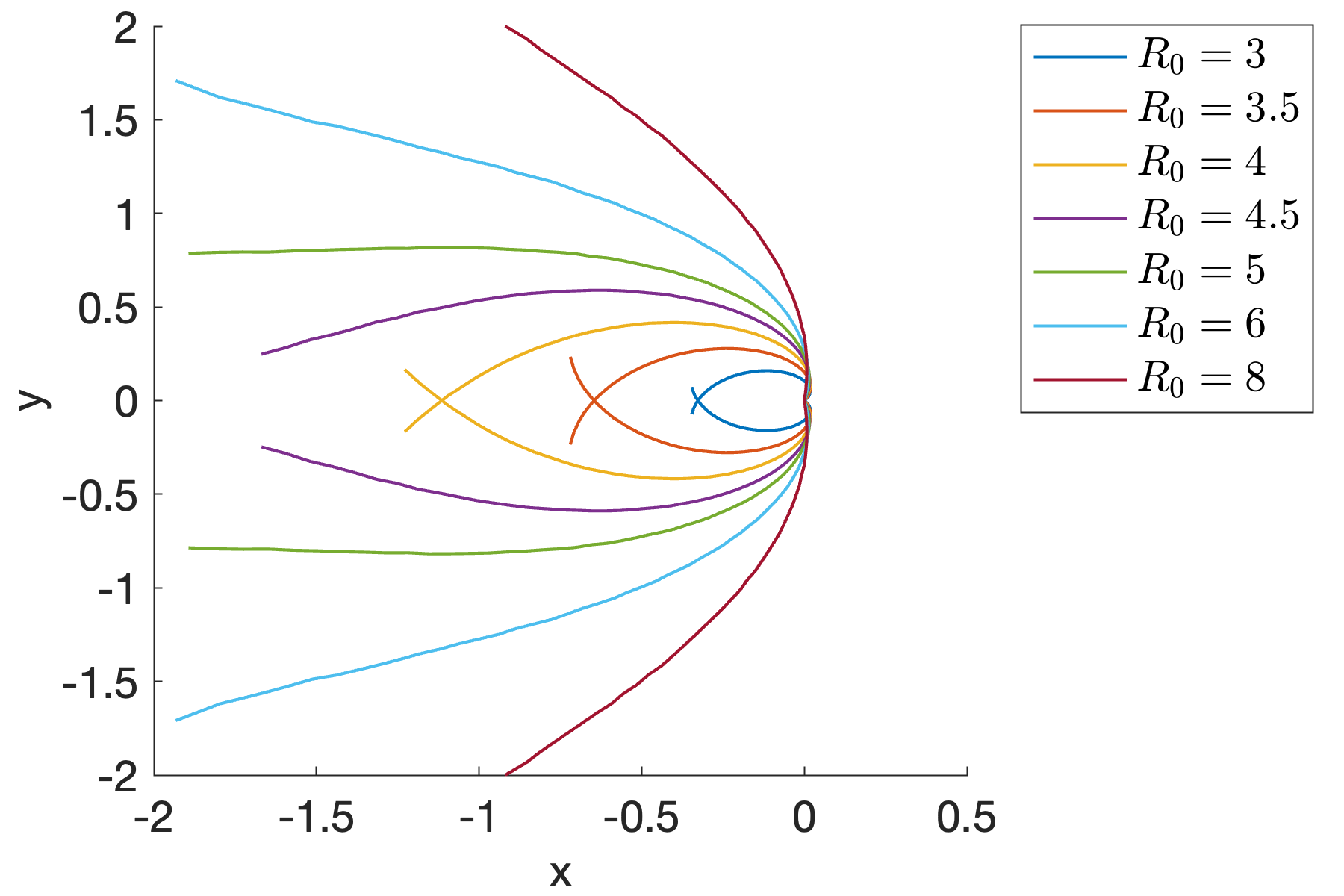}
    \caption{\textit{Breaking point curves} for various values of $R_0$.}
\end{figure}

\subsection{Initial conditions on the $xz$-plane}

We now analyze the trajectory of photons that are initially on the $xz$-plane, i.e., we study the solutions of the IVP \eqref{eq:ivp} with initial conditions of type (ii) described by \eqref{eq:type_ii} with $x_0=10$ and several values $z_0\in [z_{\min},z_{\max}]$. 

\subsubsection{Impact parameters}

\begin{figure}[!htb]
    \centering
    \includegraphics[width=0.45\linewidth]{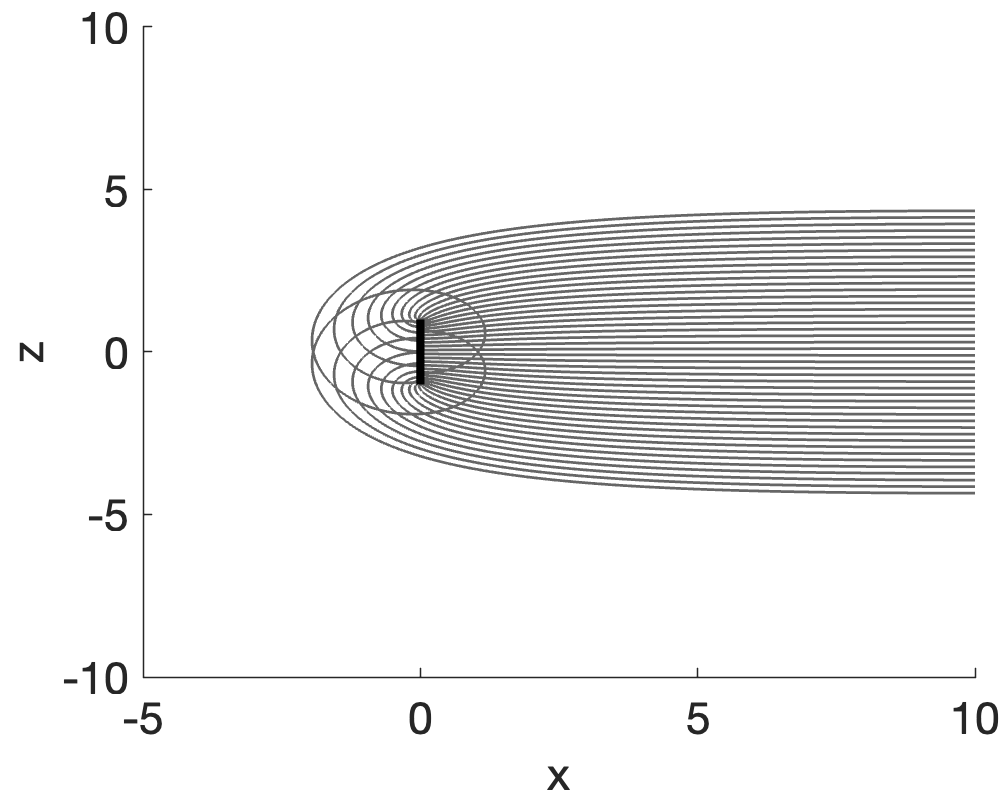}
    \includegraphics[width=0.45\linewidth]{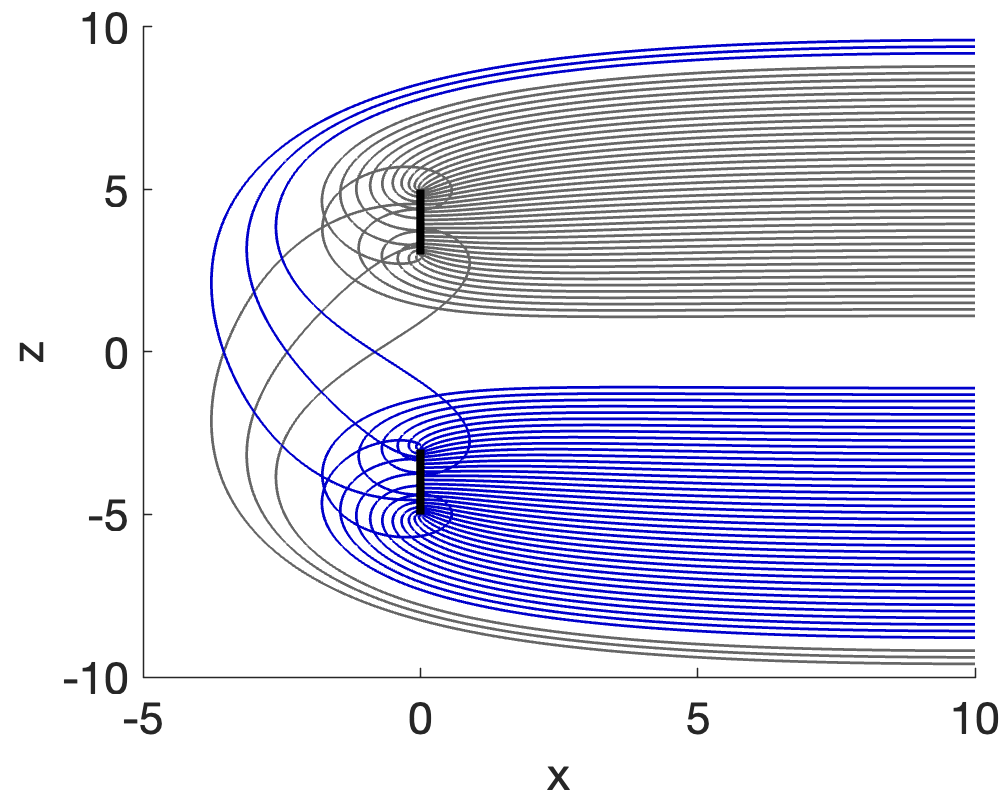}
    \caption{Light rays falling inside the horizons. The left hand side shows light rays in a single Schwarzschild space-time. The right hand side shows light rays in a double Schwarzschild space-time with $R_0=8$, the photons may fall into the upper BH (gray lines) or the lower BH (blue lines).} \label{fig:LRs_xz}
\end{figure}

\begin{figure}[!htb]
    \centering
    \begin{tabular}{cc}
    \scriptsize $R_0=16$ & \scriptsize $R_0=12$ \\
    \includegraphics[width=0.45\linewidth]{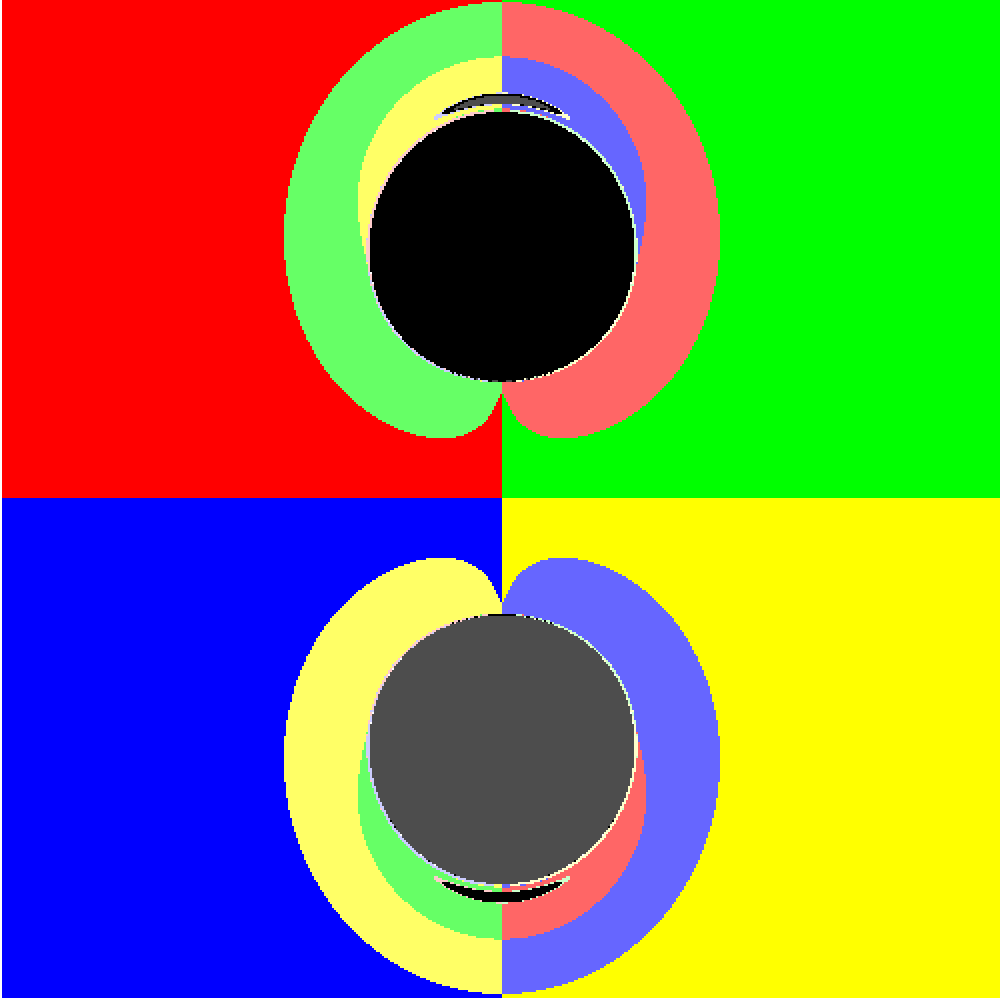} &
    \includegraphics[width=0.45\linewidth]{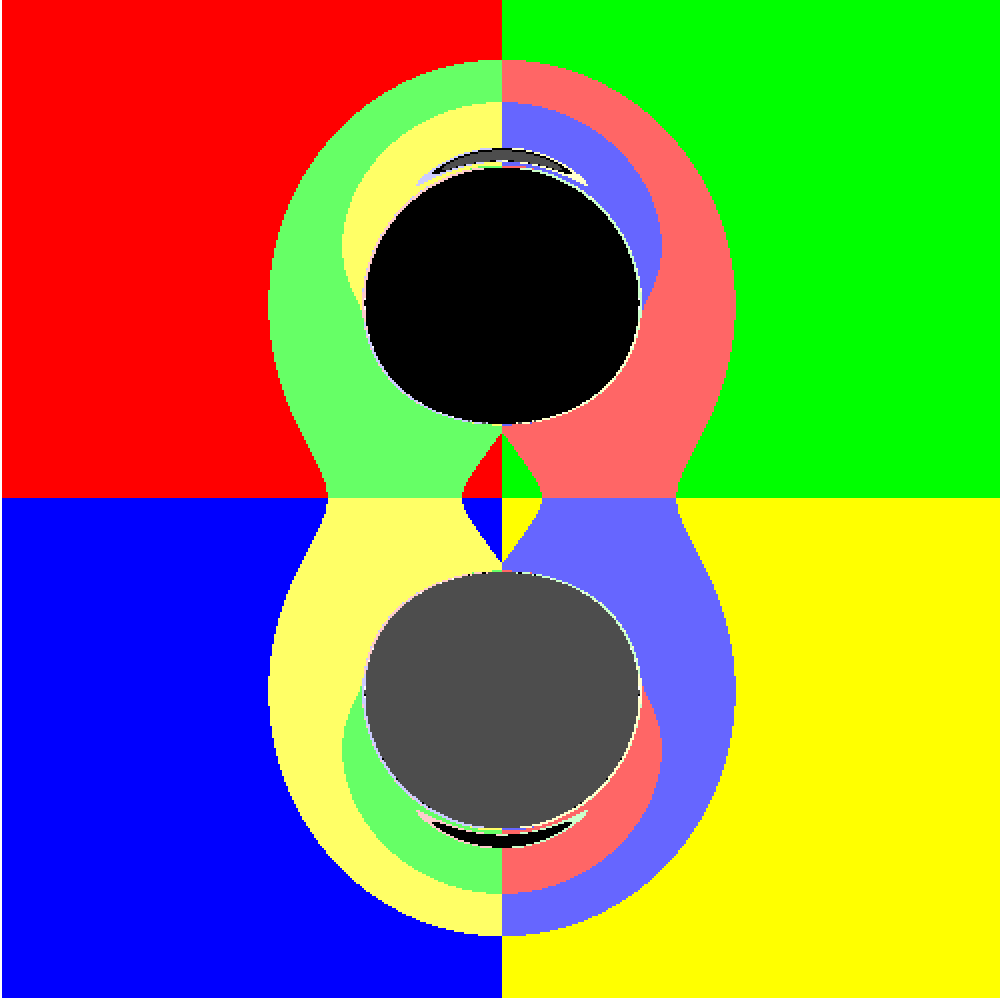} \\
    \scriptsize $R_0=8$ & \scriptsize $R_0=4$ \\
    \includegraphics[width=0.45\linewidth]{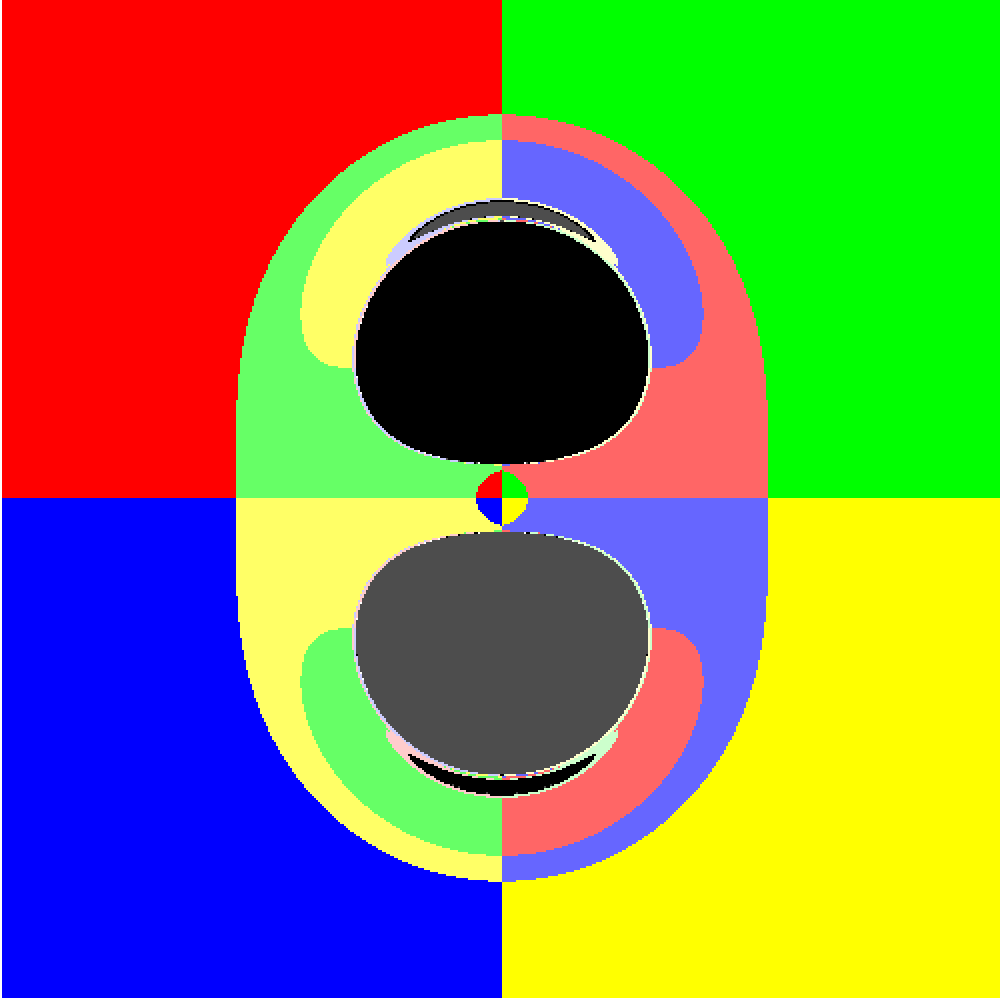} &
    \includegraphics[width=0.45\linewidth]{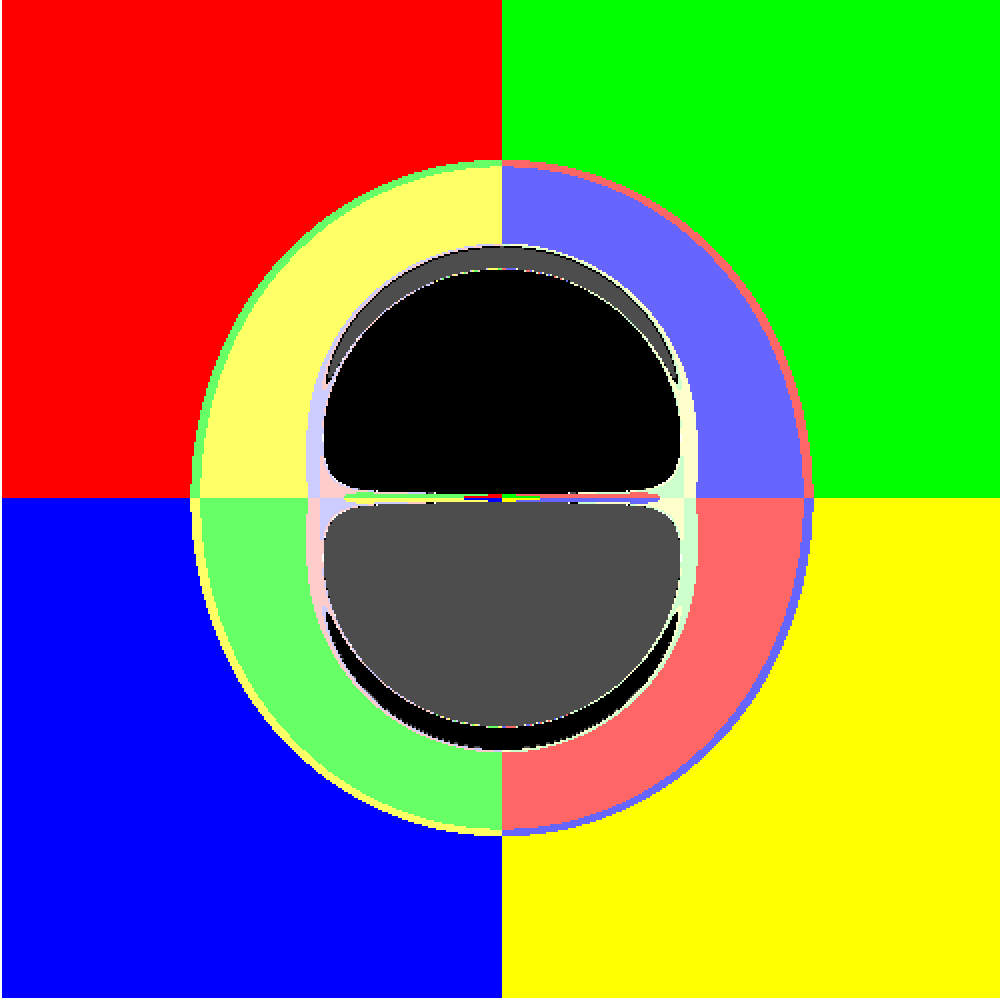} \\
    \scriptsize $R_0=3$ & \scriptsize $R_0=2$ \\
    \includegraphics[width=0.45\linewidth]{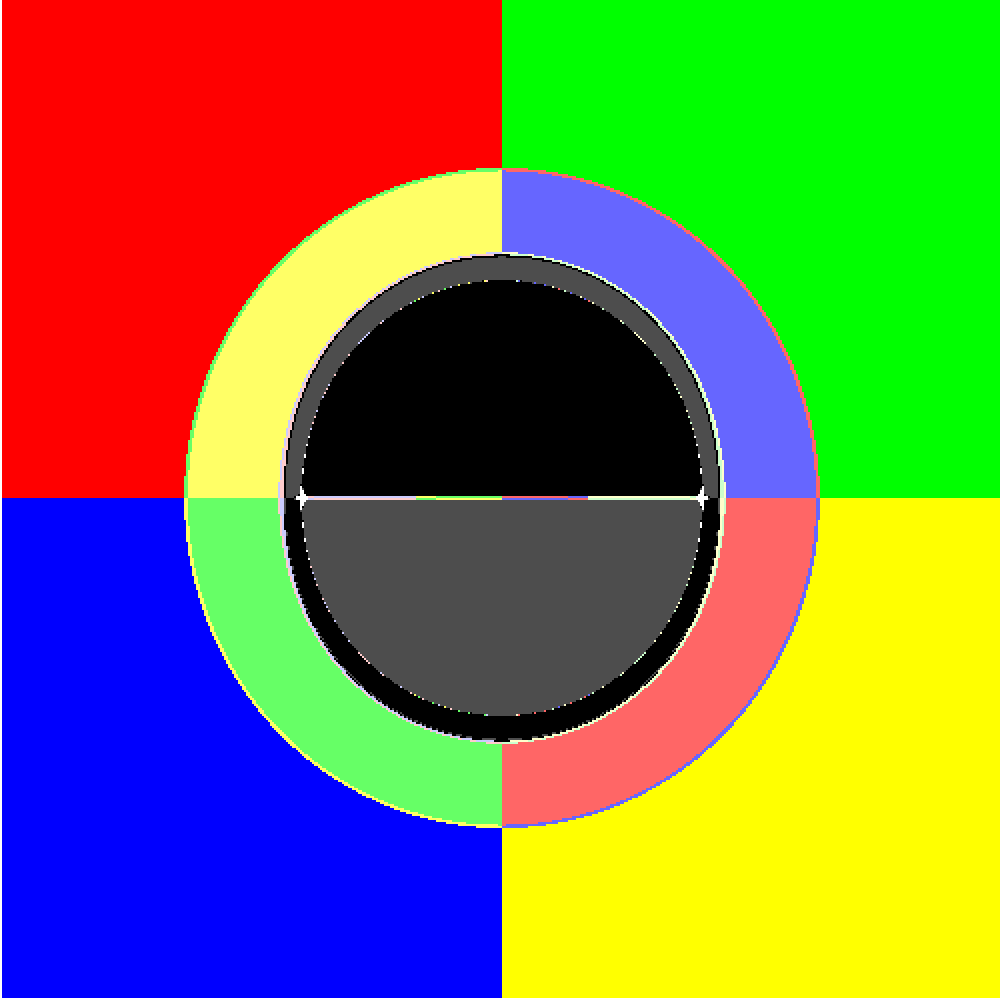} &
    \includegraphics[width=0.45\linewidth]{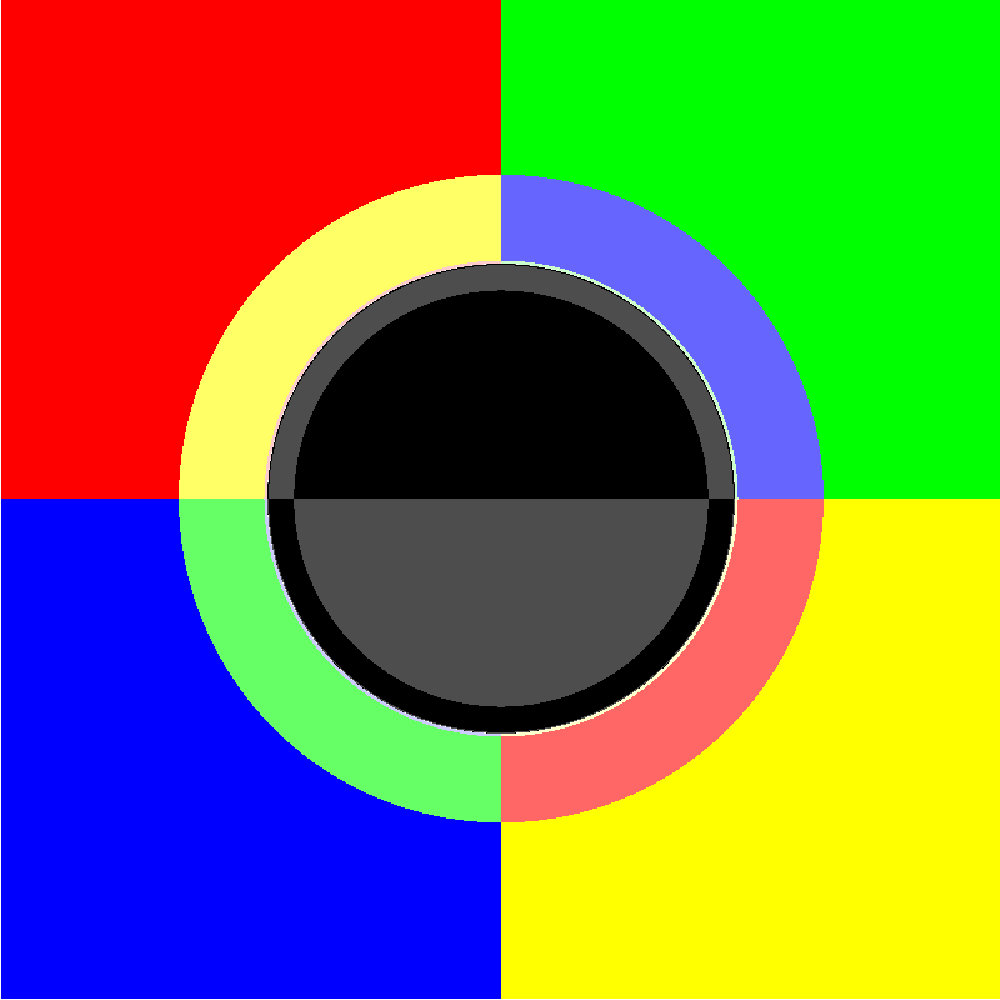}
    \end{tabular}
    \caption{Figures corresponding to space-times with $m_1=m_2=1$ and various values of $R_0$.} \label{fig:DS_shadows}
\end{figure}

Fig.~\ref{fig:LRs_xz} shows light rays with initial conditions of type (ii) with $x_0=10$ and 100 equispaced values $z_0\in [-10,10]$, for both single and double Schwarzschild space-times. Only the light rays falling into the horizons are plotted. 
In the case of a single Schwarzschild BH one can talk of the impact 
parameter $b$, which is the threshold separating light rays that will 
be absorbed by the black hole from those that will escape to infinity 
(given a constant $x_0>0$ for all the light rays), i.e., initial 
conditions with $|z_0|<b$ and $x_0>0$ will fall into the horizon. This impact parameter will be an indication of the BH shadow's size seen by a distant observer. 
On the other hand, double Schwarzschild space-times display various \textit{impact parameters} on the $xz$-plane, as seen in Fig.~\ref{fig:LRs_xz}. These additional impact parameters would correspond to a vertical section of higher order copies of the BH shadows, see Figs.~\ref{fig:DS_shadows} and \ref{fig:DS_nth_images}. For instance, the right hand side of Fig.~\ref{fig:LRs_xz} shows light rays corresponding to a section of the primary and secondary copies of the BH shadows in a space-time with $R_0=8$. There are infinitely many such parameters, as one can deduce from the discussion of trajectory classes in Subsection \ref{sec:LR_classes} (see \cite{SD} for a discussion in terms of Cantor-like sets).

\subsubsection{Light rays orbiting the photon spheres}
We are interested in the different types of trajectories that photons may follow before falling into the horizons or escaping to infinity, which can take more complicated forms than those shown in Fig.~\ref{fig:LRs_xz}.

Before describing the different types of trajectories of photons directed toward binary black holes, we briefly recall the following classification in single black hole space-times: (a) photons escaping to infinity, (b) photons falling inside the horizon, (c) photons traveling on the surface of the photon sphere perpetually. Types (a) and (b) could occur after the photons are initially traveling close to the photon sphere, but eventually collapsing into one of the two types. The higher order images perceived by a distant observer correspond to photons of type (a) that orbit the BH before reaching the observer.

We consider the following classification for the different types of trajectories of photons directed toward the binary black holes:
(a) photons escaping to infinity, (b.1), (b.2) photons falling inside the horizon of the upper (respectively lower) BH, (c.1), (c.2) photons traveling on the surface of the upper (respectively lower) photon sphere perpetually, (c.3) photons following an eight-shaped trajectory perpetually, (c.4) photons traveling on the surface of a larger photon sphere perpetually. See \cite{SD} for further discussions of trajectories of type (c). 

Trajectories of types (a) and (b) can present more convoluted patterns in these space-times, since the photons could initially travel following one of the four trajectories of type (c.1)-(c.4) as well as any of the possible combinations. Thus, there are infinitely many patterns that the photon could follow before eventually collapsing into types (a) or (b).

\begin{figure}[!htb]
    \centering
    \includegraphics[width=0.45\linewidth]{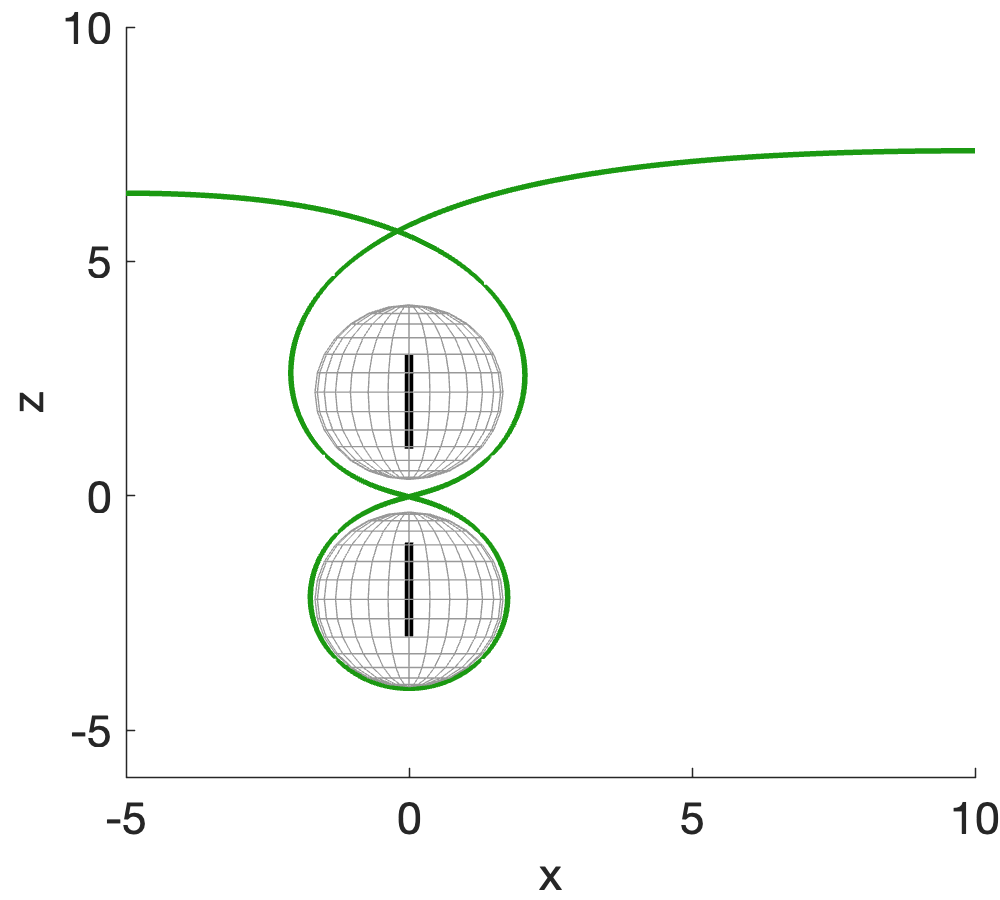}
    \includegraphics[width=0.45\linewidth]{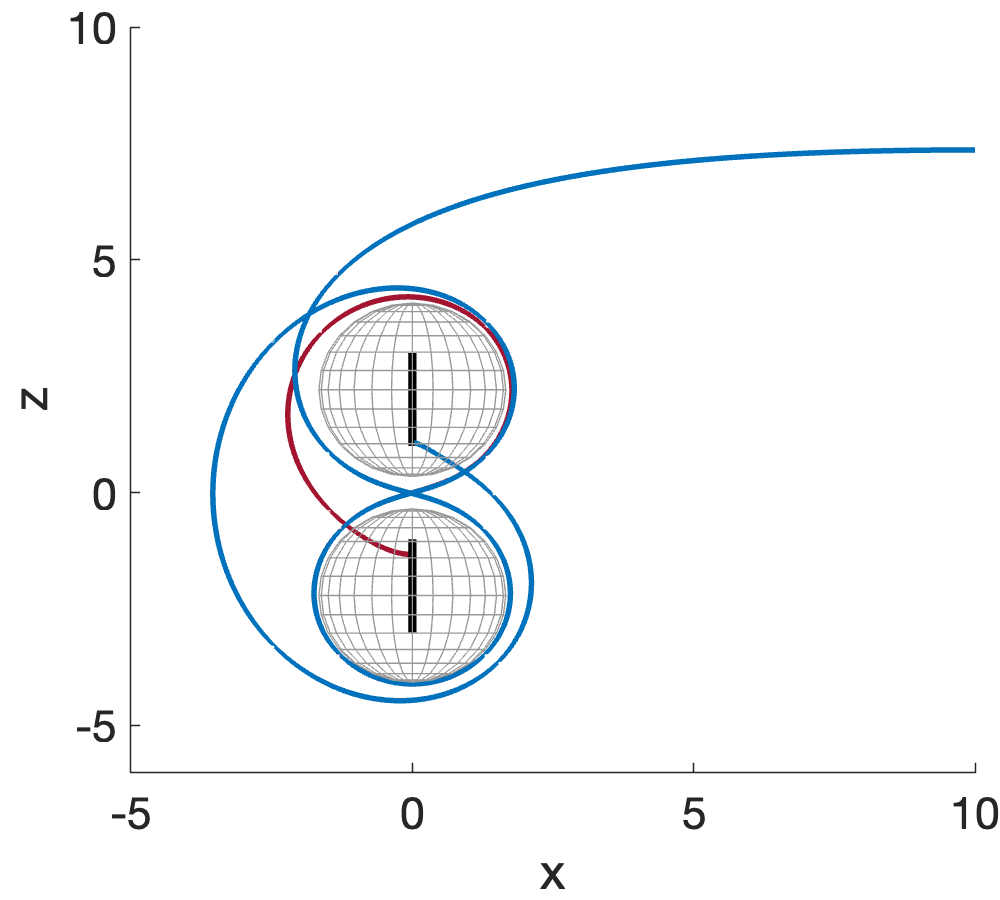}
    \caption{Some light rays and two photon spheres in a space-time with $m_1=m_2=1$ and $R_0=4$. The left hand side shows a light ray that eventually escapes to infinity. The right hand side shows examples of light rays orbiting the two BHs until they eventually fall inside one of the horizons.} \label{fig:8_shape}
\end{figure}

\begin{figure}[!htb]
    \centering
    \includegraphics[width=0.45\linewidth]{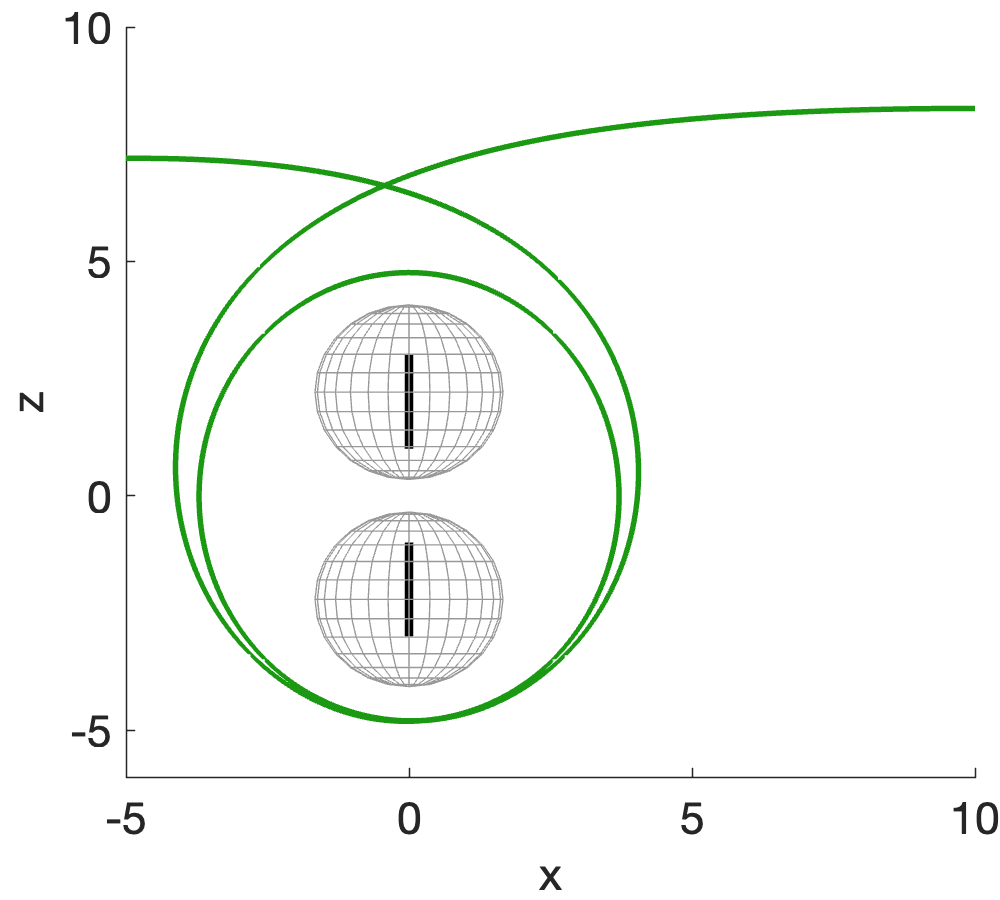}
    \includegraphics[width=0.45\linewidth]{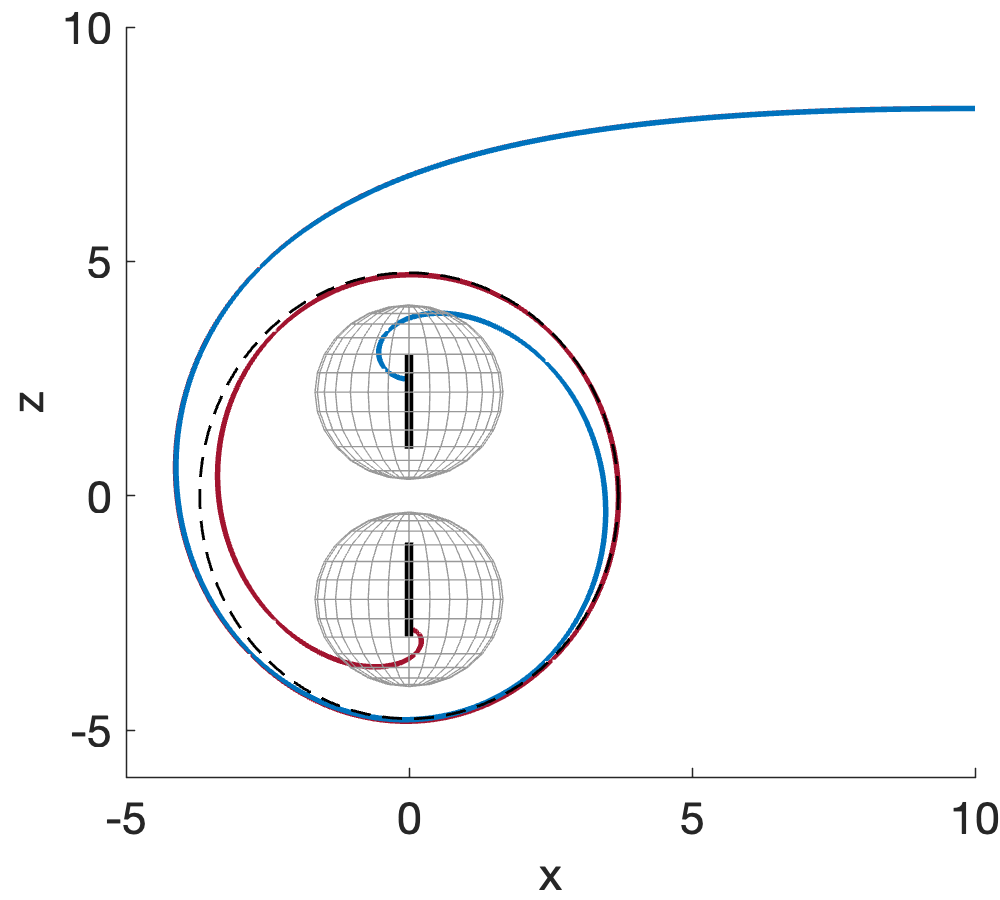}
    \caption{Light rays orbiting the larger photon sphere (dashed elliptic section) generated by the binary system with $m_1=m_2=1$ and $R_0=4$. The left hand side shows a light ray that eventually escapes to infinity. The right hand side shows light rays that eventually fall inside one of the horizons.} \label{fig:large_PS}
\end{figure}

Fig.~\ref{fig:8_shape} shows light rays that initially follow an eight-shaped trajectory before escaping to infinity (left hand side) or falling inside either the upper or the lower horizons (right hand side). Similarly, Fig.~\ref{fig:large_PS} shows light rays that initially orbit the large photon sphere before they escape to infinity or fall inside one of the horizons. In all cases, we chose initial conditions of type (ii) with $x_0=10$ and values $z_0$ that were determined numerically.

\subsubsection{Trajectory classes} \label{sec:LR_classes}
The type of trajectories that reach either horizon can be described by a free group with two generators, but before discussing this description let us analyze a simpler toy model first.
Let us consider a binary system but replacing the lower BH in Fig.~\ref{fig:diagrams_BHs} by a non-gravitating spherical object. The set of possible trajectories reaching this object are described by the free group $G=\langle u \rangle$, in which the neutral element $e$ represents the photons hitting the non-gravitating object directly (1st order image), $u$ the photons that make a half-loop around the BH (2nd order image), $uu$ the photons that make two half-loops around the BH (3rd order image) and in general, $u^m$ will represent the photons making $m$ half-loops around the BH before reaching the non-gravitating object.

A similar analysis is done for the type of trajectories falling inside either horizon of the binary system of BHs.
Here we talk about $m$-th order shadows (for both BHs) instead of $m$-th order images.
Photons that initially follow trajectories of type (c.1)--(c.4) or combinations thereof (before falling inside one of the horizons) can be described by the free group
\begin{equation*}
    G:= \langle u,d \rangle 
\end{equation*}
generated by the elements $u,d$.
Here, $u$ represents half-loops around the upper (\textit{up}) BH and $d$ represents half-loops around the lower (\textit{down}) BH.
The primary shadows (corresponding to photons falling directly into the horizons without orbiting any of the BHs) are represented by the neutral element $e$ of $G$; the secondary shadows correspond to photons that make a half-loop around the upper (lower) BH, which we represent by the element $u$ ($d$), before falling inside a horizon.
In general, the $m$-th order shadows will be described by the elements of $G$ of length $m-1$. For instance, the red (blue) trajectory in Fig.~\ref{fig:8_shape} corresponds to a 4th (5th) order shadow and it is represented by the element $uddu$ ($uddud$).
From this identification, it can be observed that the subset of $m$-th order shadows will contain $2^{m-1}$ different copies.

To be more precise about the number of half-loops around each BH, let us define $ \Delta \varphi\in\R$ as the angular difference between the vector tangent to the light ray in the direction of the photon’s motion at the initial step and the tangent vector at the final step, which is the step when the photon crosses to the other plane (the upper plane is the half-plane with $z>0$ and the lower plane is the half-plane with $z<0$). With this definition, we define the number of half-loops around the BH within a certain plane by $\lceil | 2 \Delta\varphi | \rceil$. When the photon crosses to the other plane, this counting is restarted for the computation of the number of half-loops around the other BH. This process is repeated until the photon falls inside one of the horizons. 
There is no plane crossing in the last stage (the photon falls inside the horizon instead), therefore the counting of half-loops is not performed there.

\begin{figure}[!htb]
    \centering
    \includegraphics[width=0.95\linewidth]{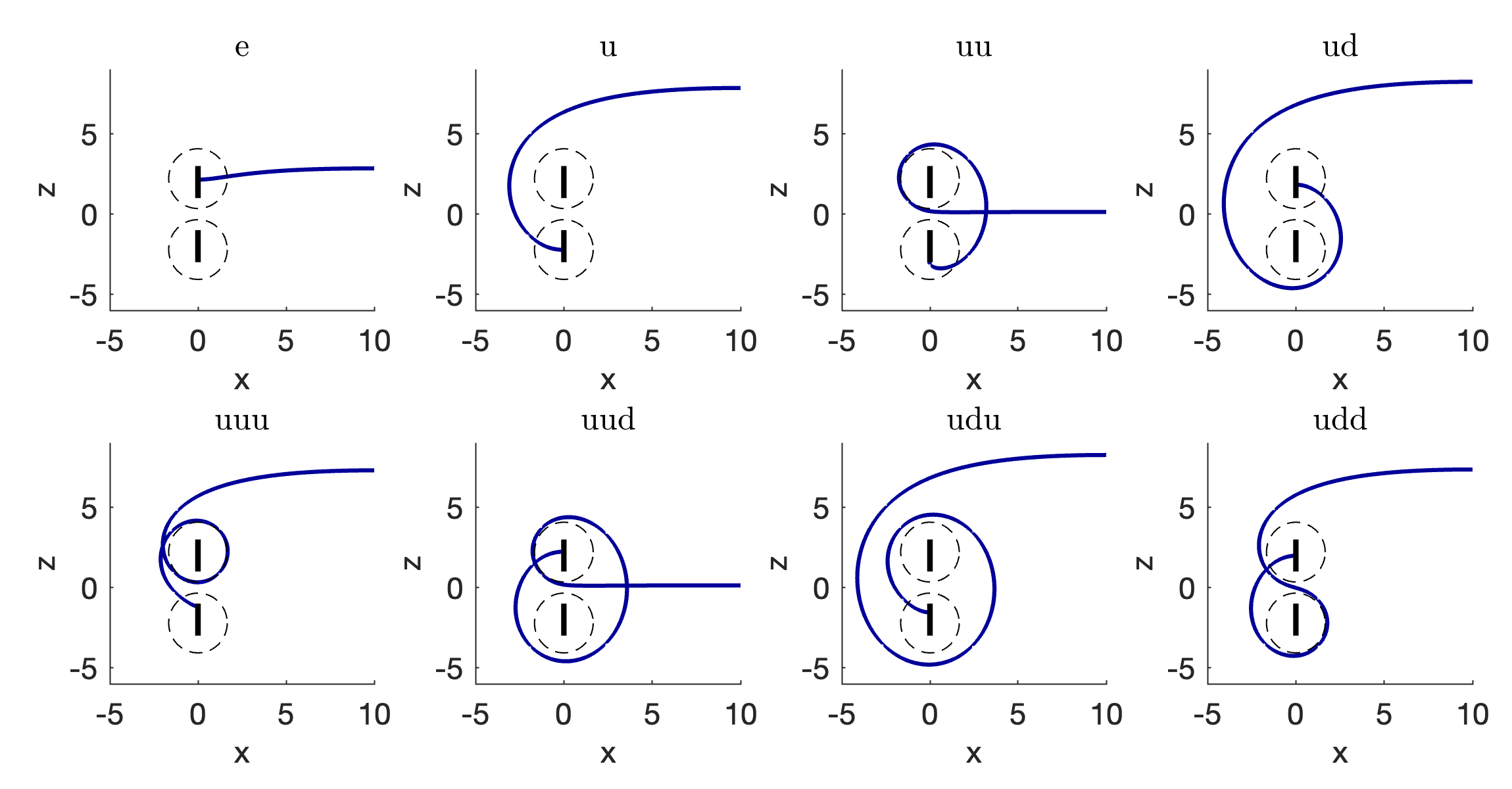}
    \caption{Representatives of trajectory classes whose photons start at $z_0>0$ (from equatorial symmetry of the metric, analogous behaviors are expected when $z_0<0$). The solid lines are the horizons and the dashed lines are the sections of the photon spheres.} \label{fig:LR_classes}
\end{figure}

\section{Ray tracing}
\label{sec:ray-tracing}
The following figures show the region from which the light rays reaching the observer originated. We assume that the source is a sphere of radius $R_\infty\gg 1$ centered at the origin emitting red, green, blue and yellow light from the regions indicated by Table \ref{tab:colors}. We shall refer to this light source as the \textit{celestial sphere}, in analogy to the term used in astronomy.
\begin{table}[!htb]
\begin{tabular}{|c|c|c|c|}
\hline
$\zeta>0,\sin\phi>0$ & $\zeta>0,\sin\phi<0$ & $\zeta<0,\sin\phi>0$ & $\zeta<0,\sin\phi<0$ \\ \hline
red & green & blue & yellow \\ \hline
\end{tabular}
\caption{Coloring of the celestial sphere by region.} \label{tab:colors}
\end{table}

The distant observer is a virtual camera located on the equatorial plane, pointing directly at the origin and located at a distance $R_c\gg 1$.
For simplicity we choose $R_\infty=R_c=20$, see the diagrams in Figs. \ref{fig:CS_quadrant} and \ref{fig:CS_center}, but we could impose $R_\infty\gg R_c$ if we were interested in a more realistic coloring for celestial sphere.

\subsection{Single Schwarzschild}
\begin{figure}[!htb]
    \centering
    \begin{tabular}{cc} 
    \includegraphics[width=0.45\linewidth]{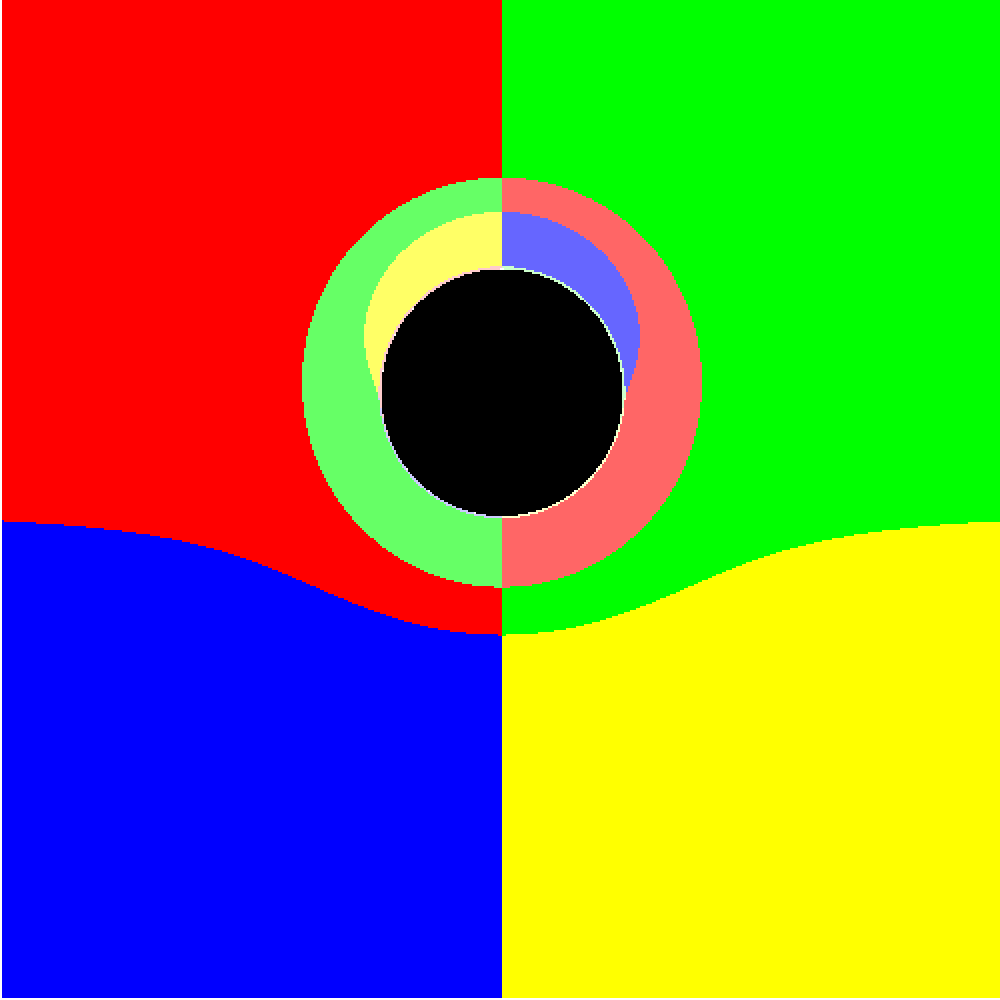} &
    \includegraphics[width=0.5\linewidth]{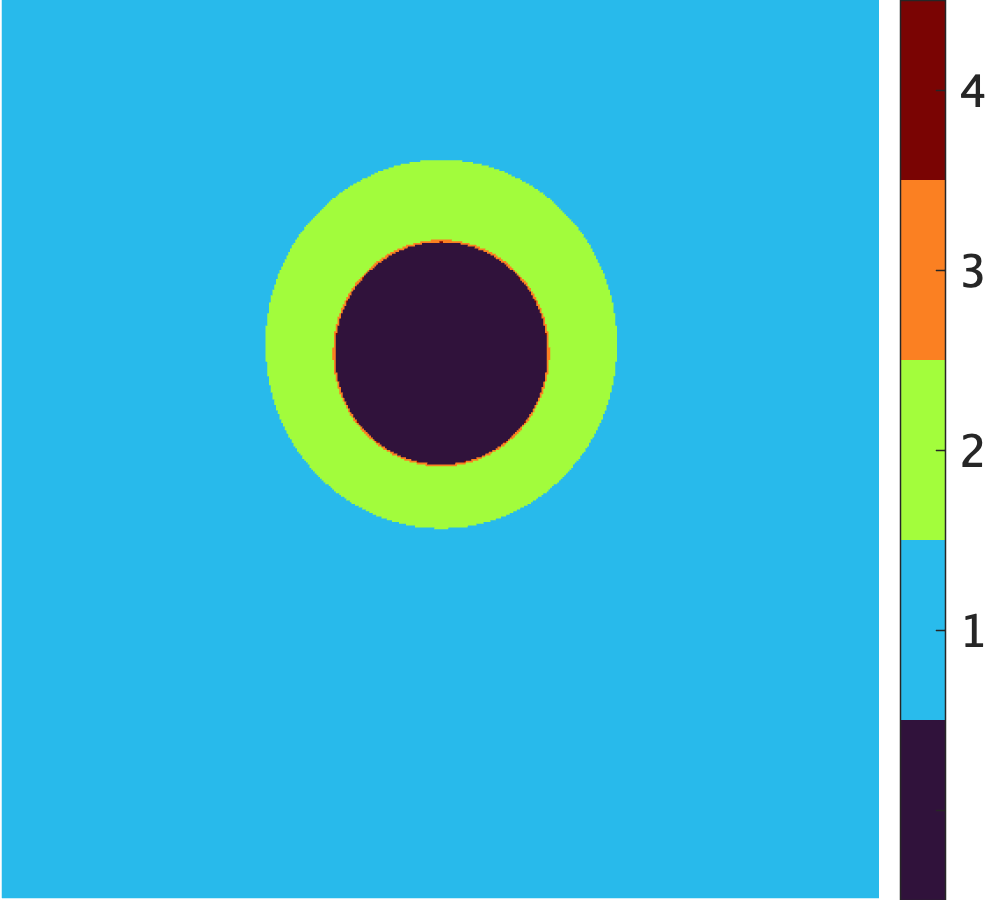}
    \end{tabular}
    \caption{Image simulations with fixed $m_1=1$, $m_2=0$ and $R_0=8$. Left side: celestial sphere with an artificial coloring by quadrants. Right side: monochromatic coloring of the primary, secondary and tertiary copies of the celestial sphere. } \label{fig:RT_single_BHs}
\end{figure}

We recall the gravitational lensing caused by the presence of a single Schwarzschild black hole. Unlike the usual examples in which the black hole is placed at the origin of the coordinate system (see left hand side of Fig.~\ref{fig:diagrams_BHs}), Fig.~\ref{fig:RT_single_BHs} corresponds to single black holes located on the $z$-axis above the equatorial plane. This is equivalent to the double Schwarzschild metric with $m_2=0$. 
We choose this camera setting in order to isolate the optical effects caused by the camera not pointing toward the BH (left hand side of Fig.~\ref{fig:RT_single_BHs}) and those caused by the presence of a second BH. However, if we assign a monochromatic coloring the parts corresponding to the primary and higher order copies of the celestial sphere images, this camera setting will produce similar results to those taken by a camera pointing directly toward the BH. Thus, this coloring will provide additional information on the optical effects due to the presence of a second BH and a Weyl strut.

\subsection{Double Schwarzschild}

We now study the optical effects of a second black hole on the apparent image of the background space using the same settings described above, i.e., we simulate the image perceived by a virtual camera on the equatorial plane pointing toward the origin (the midpoint between the two BHs).
In order to recognize the primary and secondary copies of the celestial sphere with more clarity, we show the primary copies in bright colors and each subsequent copy is shown in paler colors. Moreover, we color the shadows of the lower black hole in gray in order to differentiate the various shadow copies appearing in these images.

\subsection{Higher order images}

\begin{figure}[!htb]
    \centering
    \begin{tabular}{cc} 
    \scriptsize $R_0=16$ & \scriptsize $R_0=12$ \\
    \includegraphics[width=0.45\linewidth]{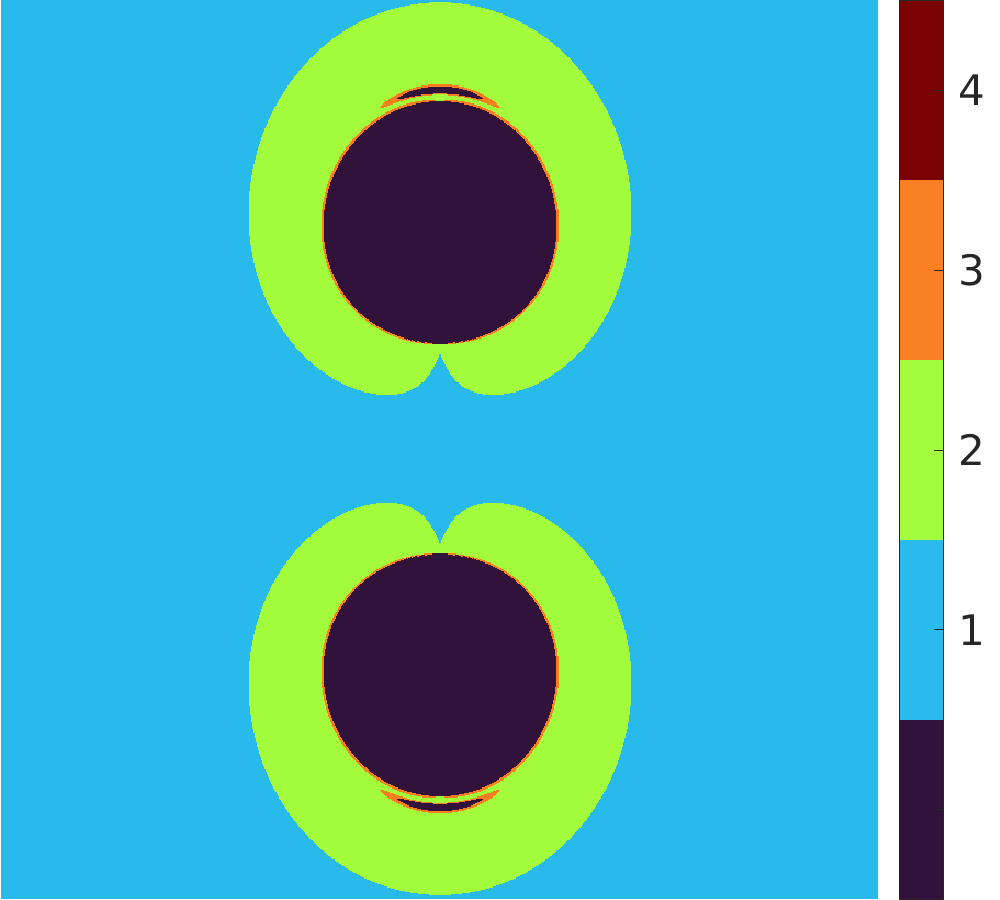} &
    \includegraphics[width=0.45\linewidth]{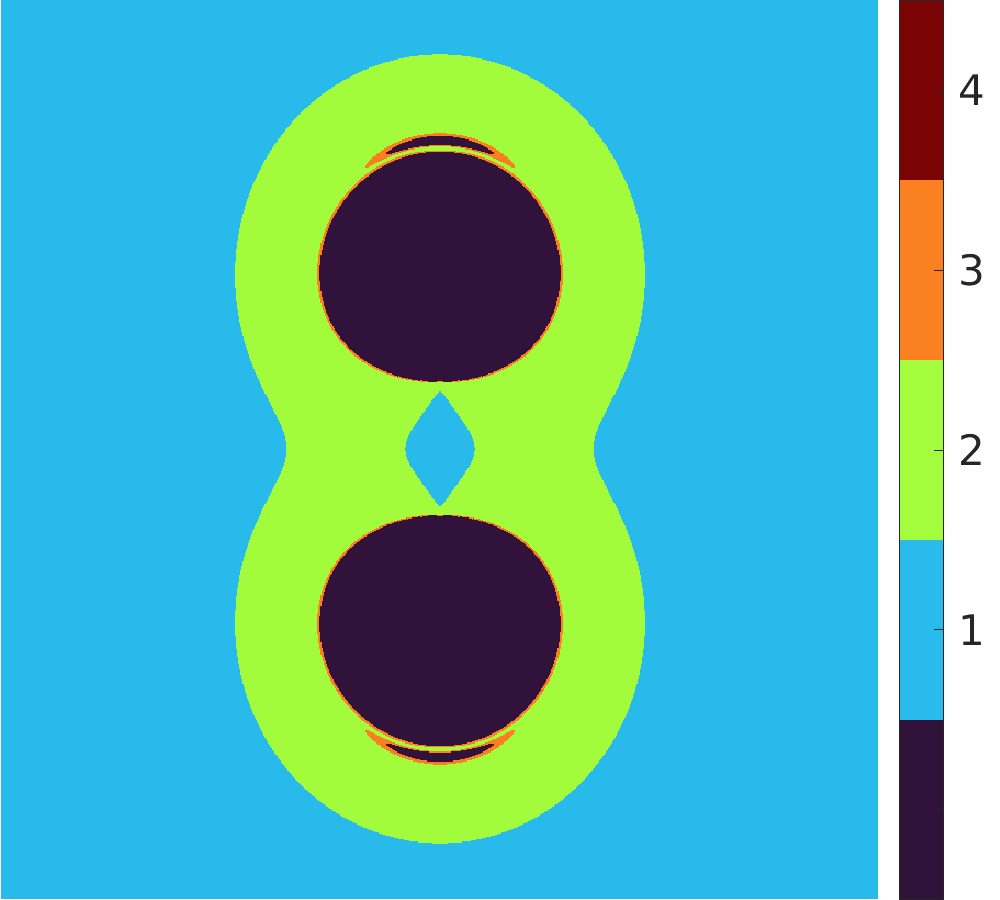} \\
    \scriptsize $R_0=8$ & \scriptsize $R_0=4$ \\
    \includegraphics[width=0.45\linewidth]{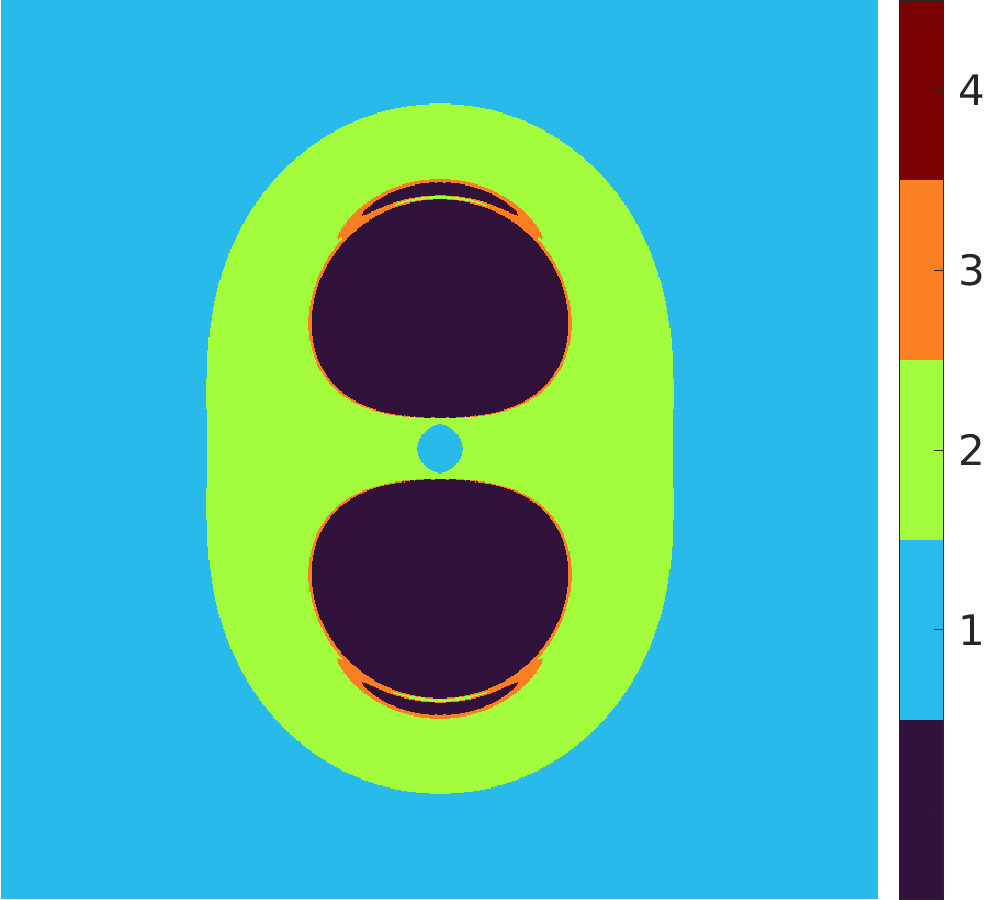} &
    \includegraphics[width=0.45\linewidth]{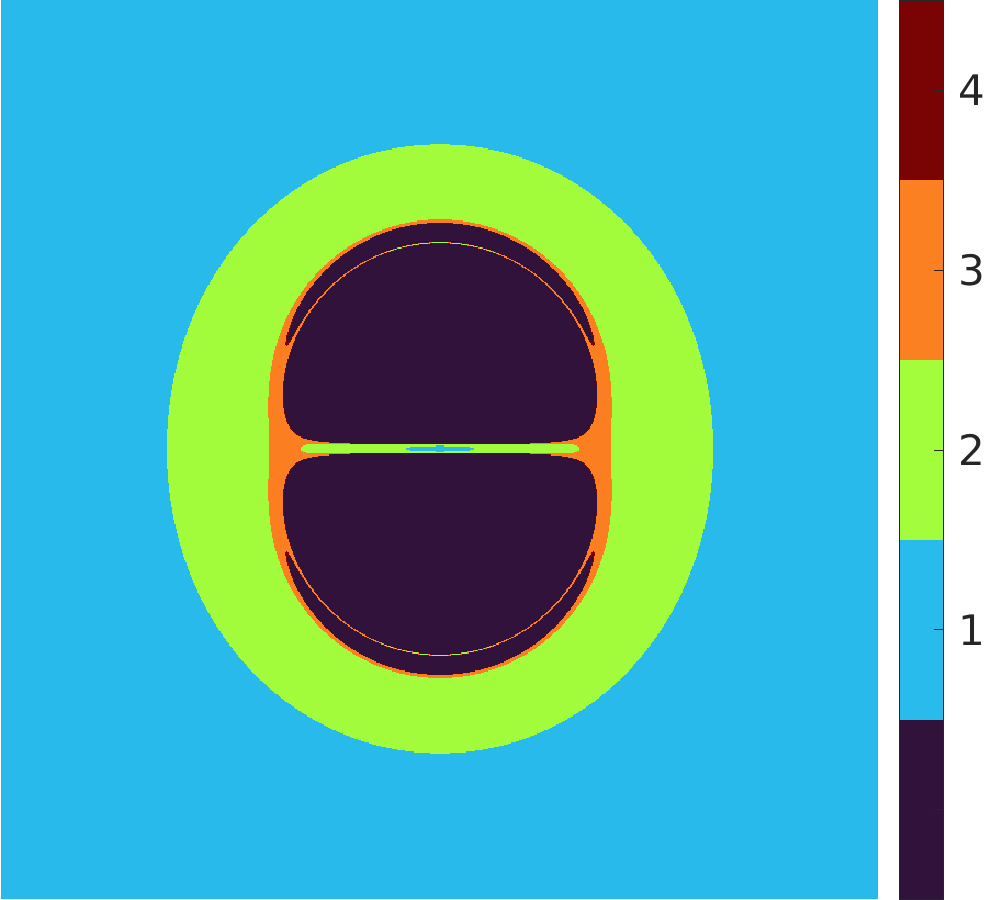} \\
    \scriptsize $R_0=3$ & \scriptsize $R_0=2$ \\
    \includegraphics[width=0.45\linewidth]{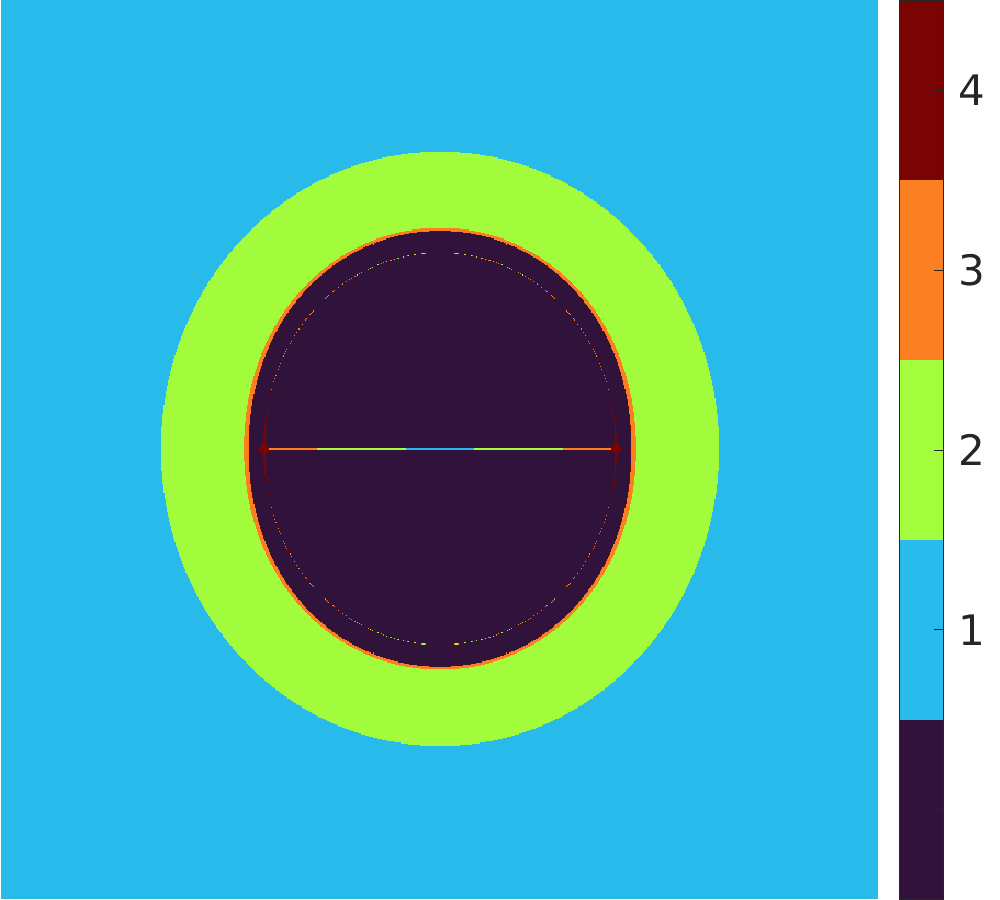} &
    \includegraphics[width=0.45\linewidth]{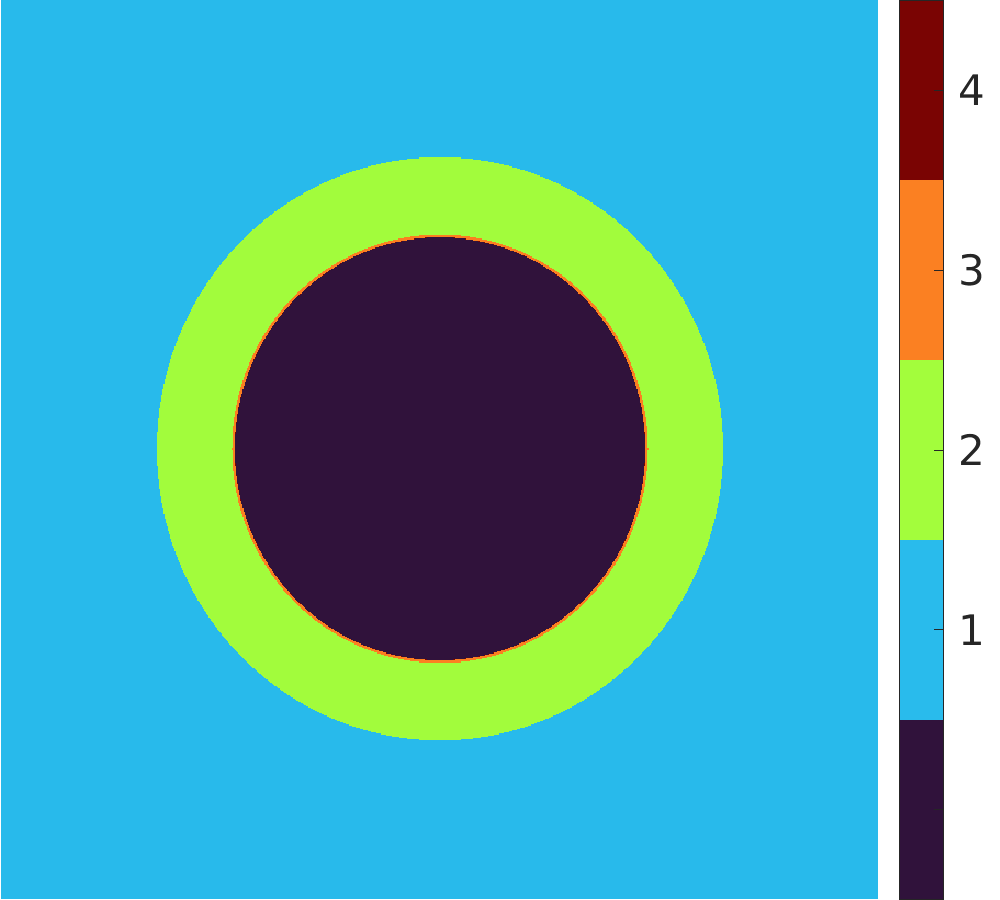} \\
    \end{tabular}
    \caption{Shadows of the black holes and the multiple copies of the celestial sphere. The blue part shows the primary copies, the green part shows the secondary ones and so forth.} \label{fig:DS_nth_images}
\end{figure}

The tertiary copies are difficult to visualize in the previous figure. Therefore, it is convenient to show each copy in a single color in order to appreciate the higher order ones. The $n$-th order copy is described by the expression 
$$n=\lceil |\phi_* | /\pi \rceil, $$ 
where $\phi_*$ is the value of the $\phi$-coordinate of the light rays corresponding to each pixel when they reach the celestial sphere.
Fig.~\ref{fig:DS_nth_images} shows various copies of the celestial sphere.
It is interesting to note that the area of the secondary and tertiary copies are larger compared to that with a single black hole. The reason is that some of the light rays that would otherwise escape to infinity after being deflected by one black hole manage to orbit the other one, as shown in the left hand side of Figs. \ref{fig:8_shape} and \ref{fig:large_PS}.
The rest of the observed effects can be divided in cases with large separations $R_0$, short separations and intermediate values.
(i) Cases with large $R_0$: the photons passing near the strut are 
either absorbed by one of the black holes or they are repelled by the strut and sent back to infinity without orbiting any of the black holes; only a small portion manages to orbit the nearest black hole before escaping to infinity. This is the reason one observes cusps in the examples with $R_0=12,16$. 
(ii) Cases with small $R_0$:
as the distance $R_0$ decreases, the effects of the strut are only visible on the photons traveling near the equatorial plane (examples $R_0=3,4$), until the horizons eventually merge (at $R_0=2$) and then the apparent image is equivalent to that produced by a single black hole with a larger mass.
(iii) Cases with intermediate values: these cases are interesting since part of the primary image is located within the secondary image (examples $R_0=8,12$). This occurs in the region where
the repulsive effect of the strut overcomes the attractive effect of the black holes, i.e., the photons are traveling sufficiently close to the strut. However, this is a short-ranged effect and the attractive effect of the black holes becomes dominant as the distance to the strut increases, as shown in Fig.~\ref{fig:deflected_LRs} for light rays traveling on the equatorial plane.

\subsection{Comparison between single and double Schwarzschild}
\begin{figure}[!htb]
    \centering
    \begin{tabular}{cc}
    \scriptsize $m_1=1,m_2=0,R_0=16$ & \scriptsize $m_1=1,m_2=1,R_0=16$ \\
    \includegraphics[width=0.45\linewidth]{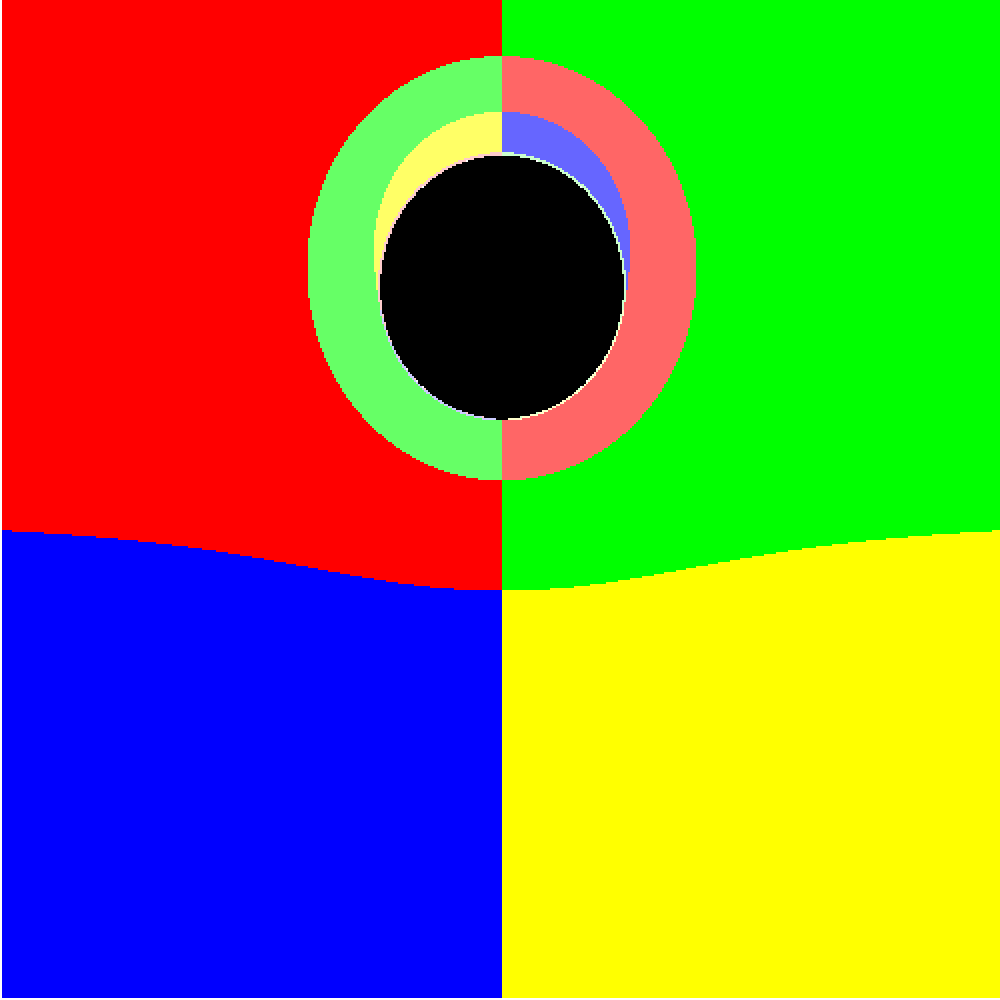} &
    \includegraphics[width=0.45\linewidth]{figures/DS_fig_R16_fl05.png} \\
    \scriptsize $m_1=1,m_2=0,R_0=8$ & \scriptsize $m_1=1,m_2=1,R_0=8$ \\
    \includegraphics[width=0.45\linewidth]{figures/Single_Schwarzschild_fig_R8.png} &
    \includegraphics[width=0.45\linewidth]{figures/DS_fig_R8_fl05.png} \\
    \scriptsize $m_1=1,m_2=0,R_0=4$ & \scriptsize $m_1=1,m_2=1,R_0=4$ \\
    \includegraphics[width=0.45\linewidth]{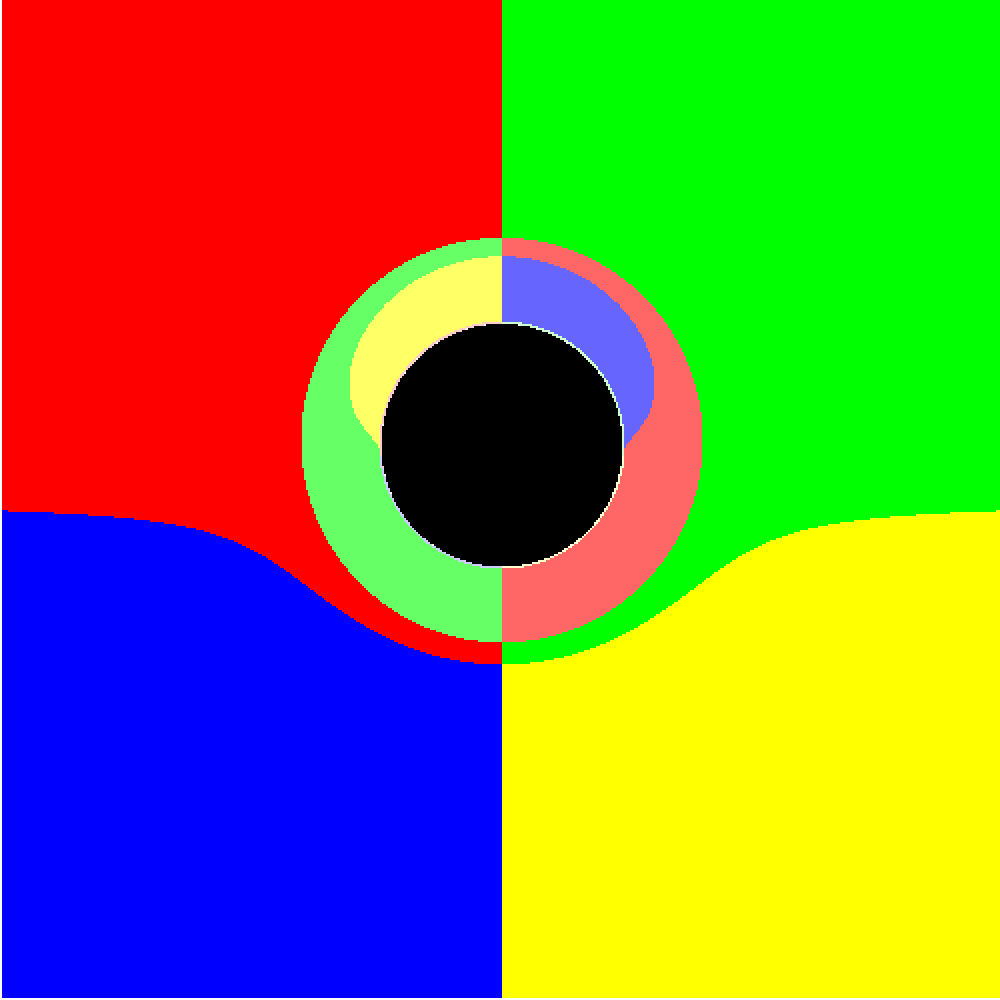} &
    \includegraphics[width=0.45\linewidth]{figures/DS_fig_R4_fl05.png} 
    \end{tabular}
    \caption{Gravitational lensing caused in single and binary black hole space-times.} \label{fig:1S_DS}
\end{figure}

We compare the gravitational lensing caused by a single black hole and a system of binary black holes. 
Since we consider symmetric cases here ($m_1=m_2$), we focus the discussion on the upper half of the apparent images in Fig.~\ref{fig:1S_DS} near the center.
More precisely, we focus on the horizontal strip from the center of the image to the primary shadow, since the rest of the image will not present major differences. 
The following are the main observed differences: (i) The upward deflection is reduced due to the presence of the other black hole attracting the photons and therefore partially straightening the light rays. More precisely, the photons that depart near the equatorial plane stay close to the plane. (ii) The sidewards deflection of the photons passing near the equatorial plane increases when the photons are sufficiently far away from the strut. This is well observed in Fig.~\eqref{fig:1S_DS} when the size of the secondary images in the left hand side and those in the right hand side are compared. (iii) The photons passing sufficiently near the strut (but distant enough from either black hole to avoid falling inside the horizons) are defocused as clearly shown by the individual light rays in Fig.~\ref{fig:deflected_LRs}; thus, photons that should have been attracted by the upper black hole are repelled by the strut. This is particularly visible for the case $R_0=8$ in Fig.~\ref{fig:1S_DS} in which the photons are violently repelled by the strut, overcoming the attractive effect of either black hole, which can be seen from the fact that they reach the celestial sphere without orbiting the black holes.

\subsection{Zoom near the shadow}
The following images show the behavior of light near the secondary 
shadows of the black holes. 
A third copy of the upper black hole's shadow is observed. This copy corresponds to light rays that are first deflected by the upper black hole, then they are deflected by the lower one and then they finally fall toward the upper one, i.e., trajectories of class $ud$ in Fig.~\ref{fig:LR_classes}.

\begin{figure}[!htb]
    \centering
    \includegraphics[width=0.45\linewidth]{figures/DS_fig_R8_fl05.png}
    \includegraphics[width=0.45\linewidth]{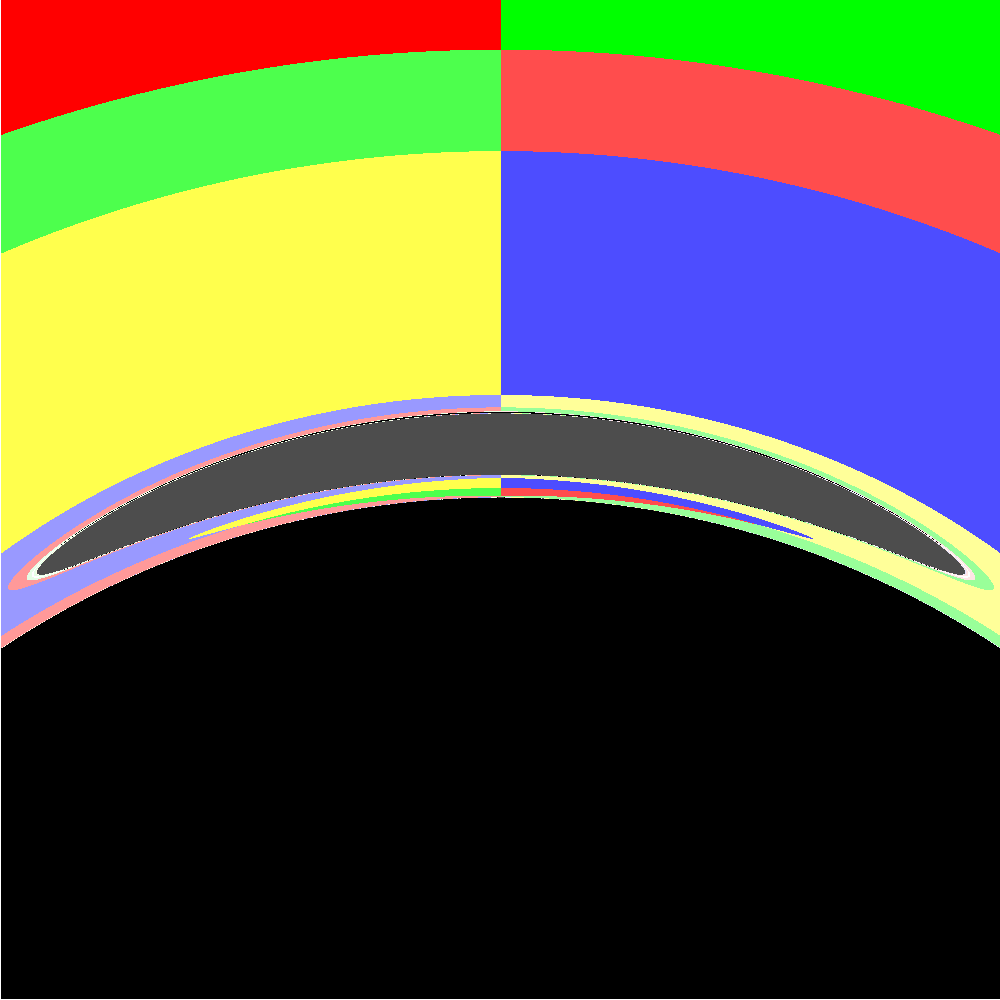}
    \includegraphics[width=0.45\linewidth]{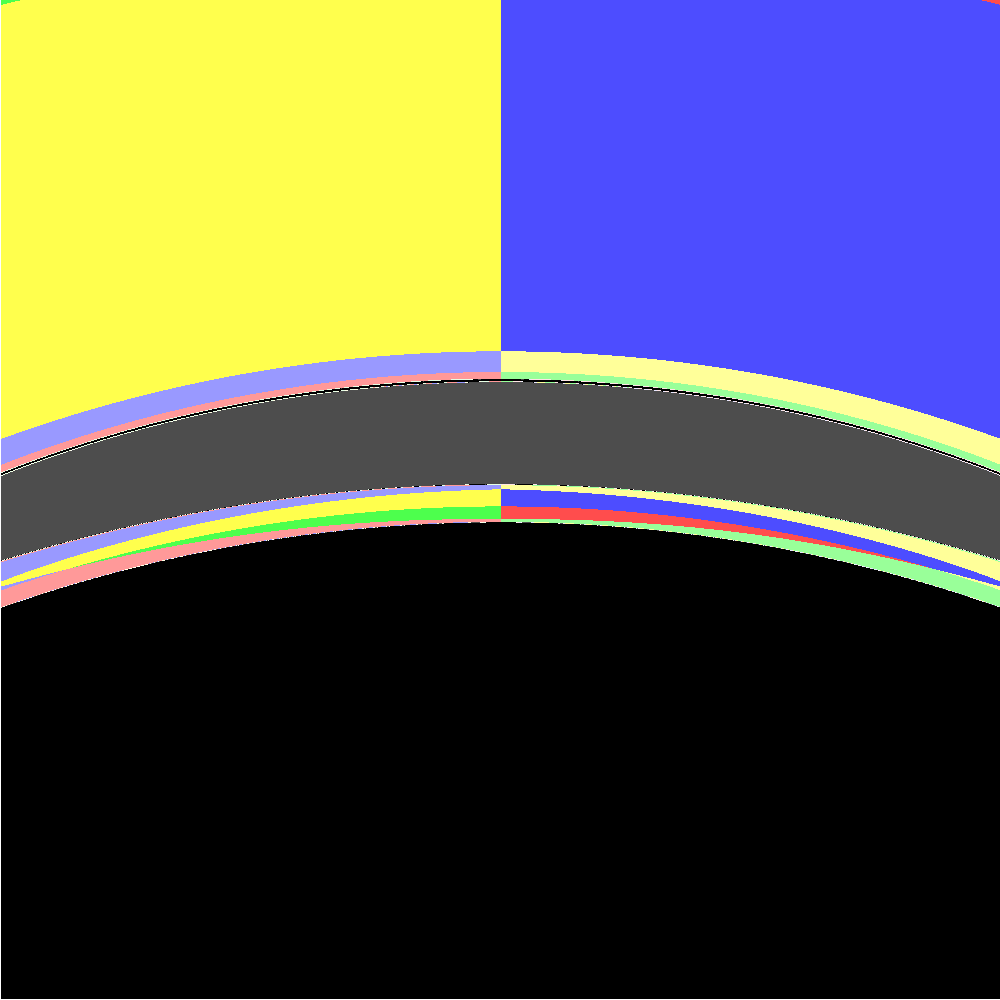}
    \includegraphics[width=0.45\linewidth]{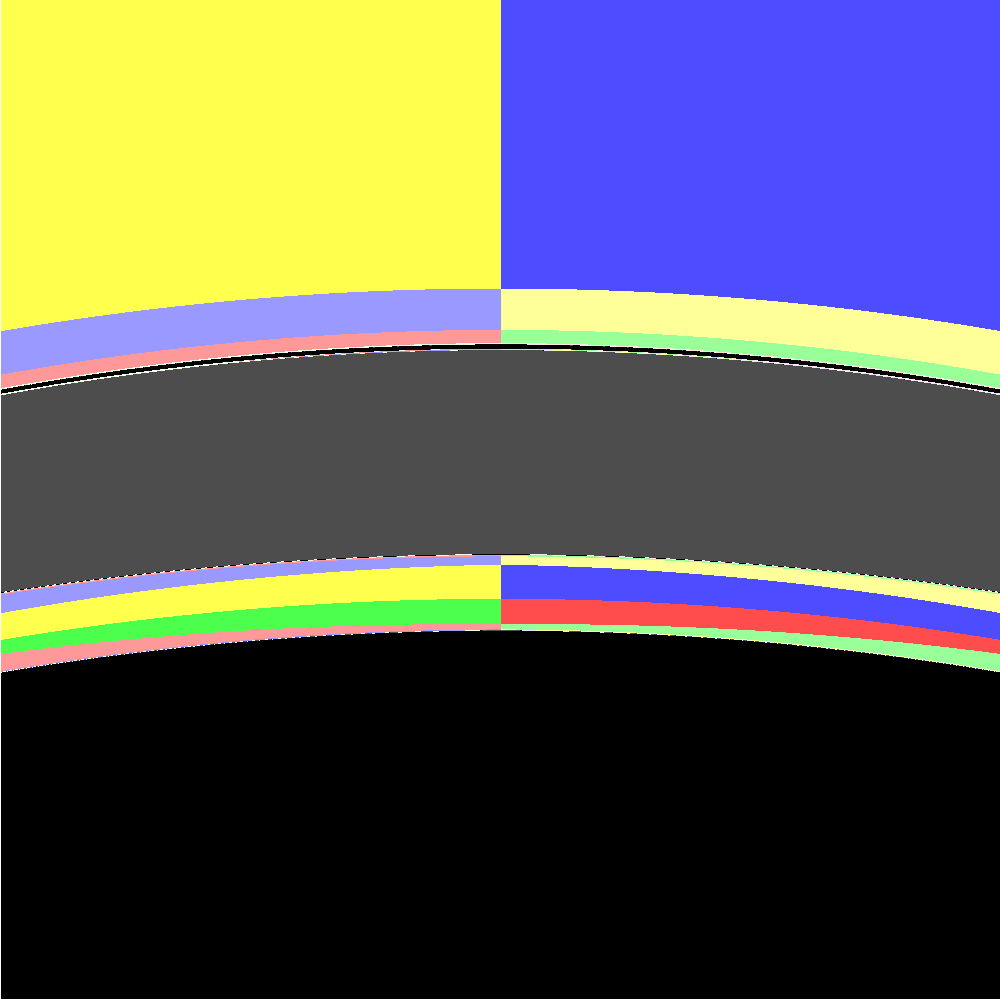}
    \includegraphics[width=0.45\linewidth]{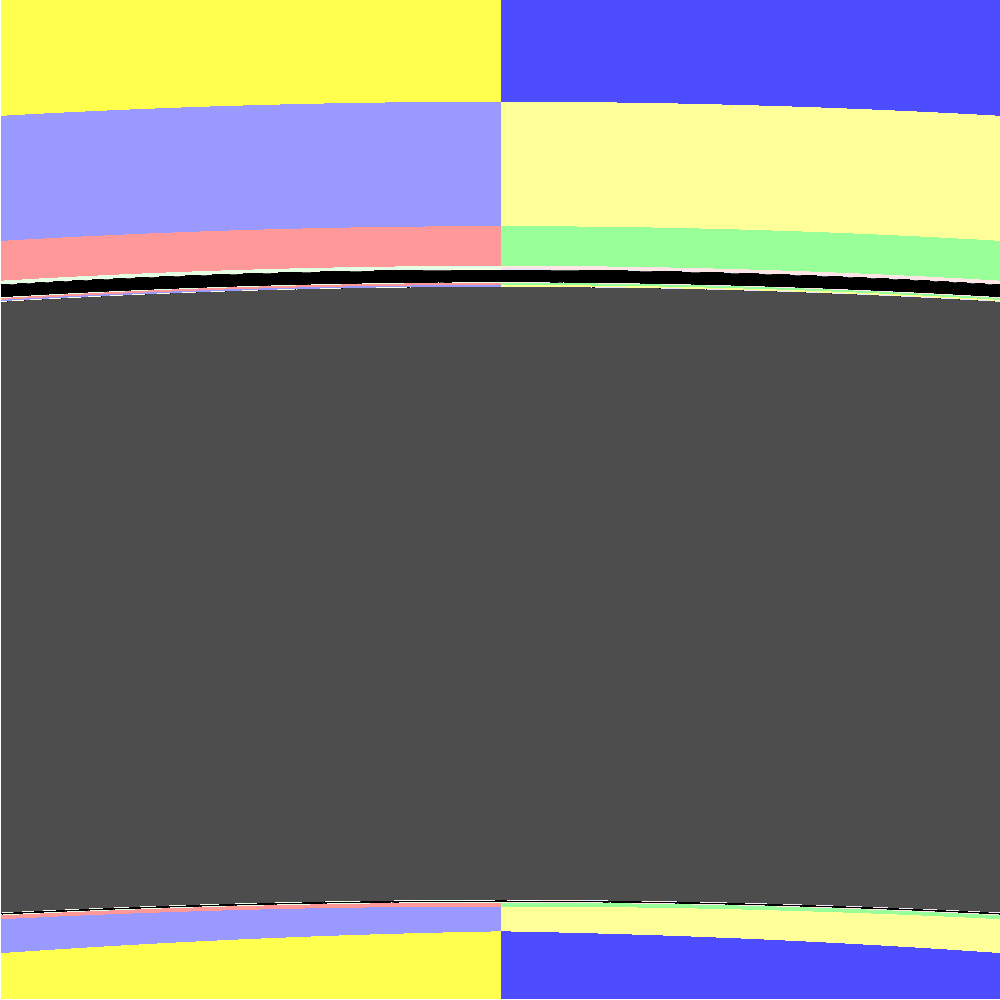}
    \includegraphics[width=0.45\linewidth]{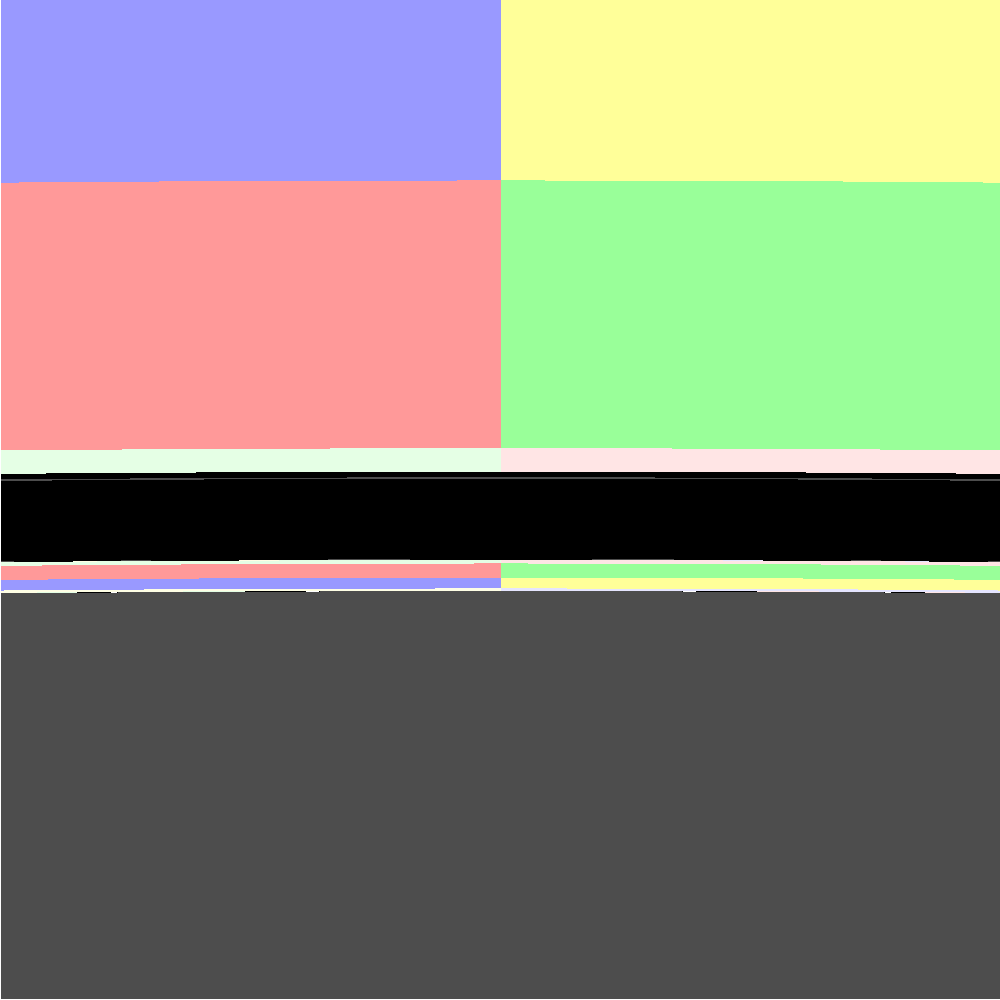}
    \caption{Zoom near the secondary copy of the shadow of the lower black hole (colored in gray). The chosen angular apertures of the virtual camera (from left to right and from top to bottom) are: $90.0^\circ$, $18.92^\circ$, $11.42^\circ$, $5.72^\circ$, $1.91^\circ$ and $0.29^\circ$.}
\end{figure}

\subsection{Trajectories from the virtual camera}

\begin{figure}[!htb]
    \centering
    \includegraphics[width=0.4\linewidth]{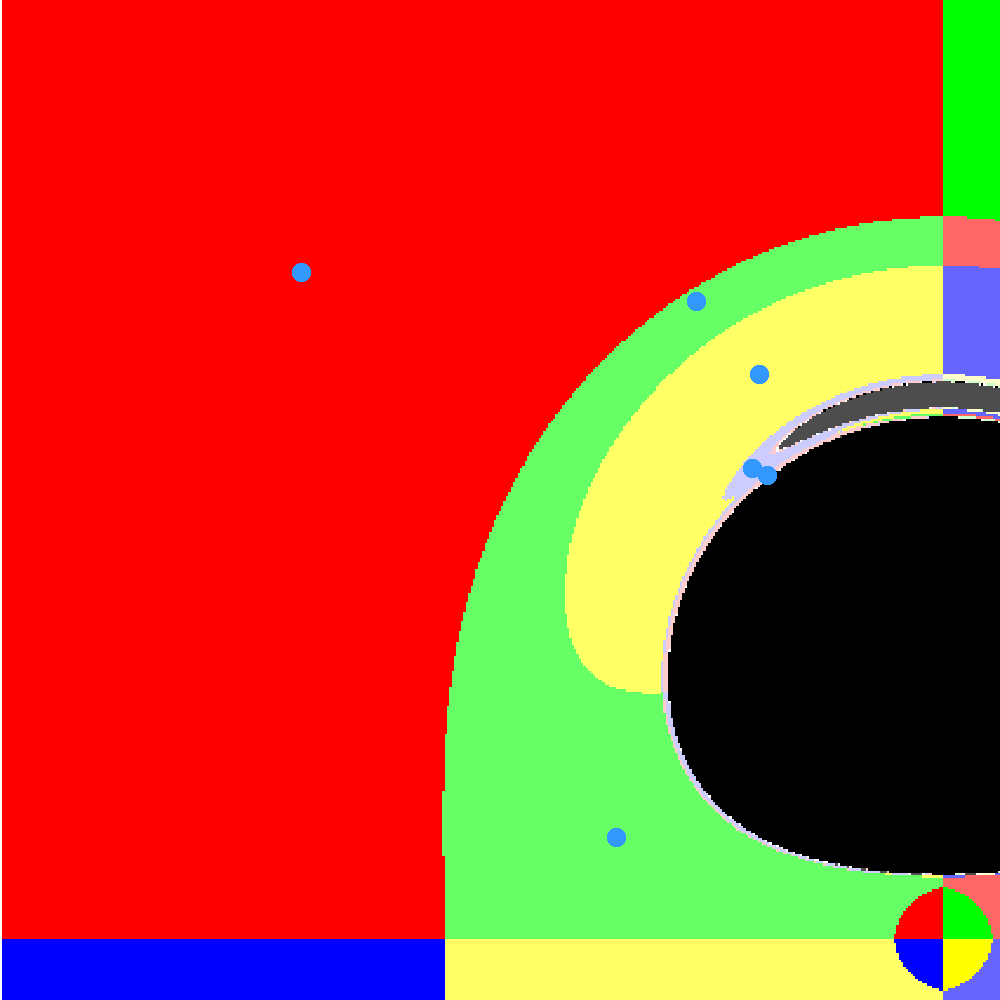}
    \includegraphics[width=0.4\linewidth]{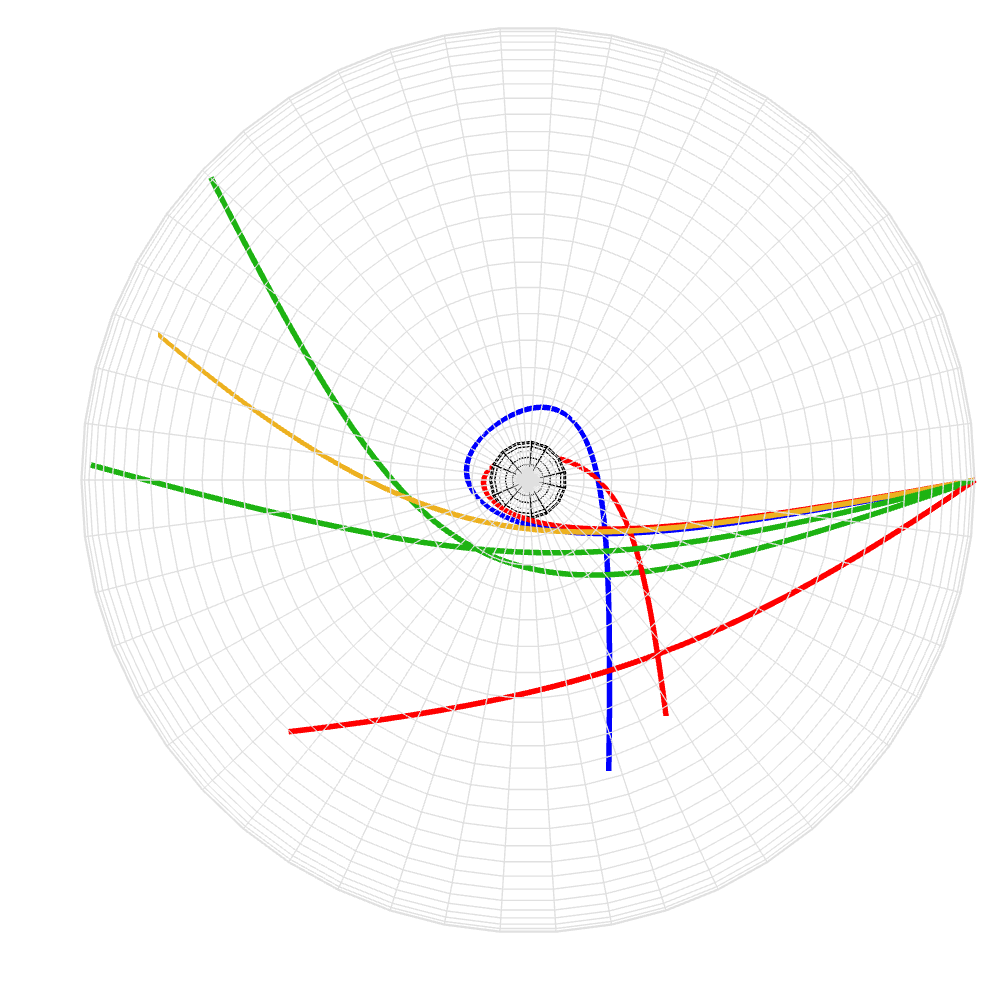}
    \includegraphics[width=0.4\linewidth]{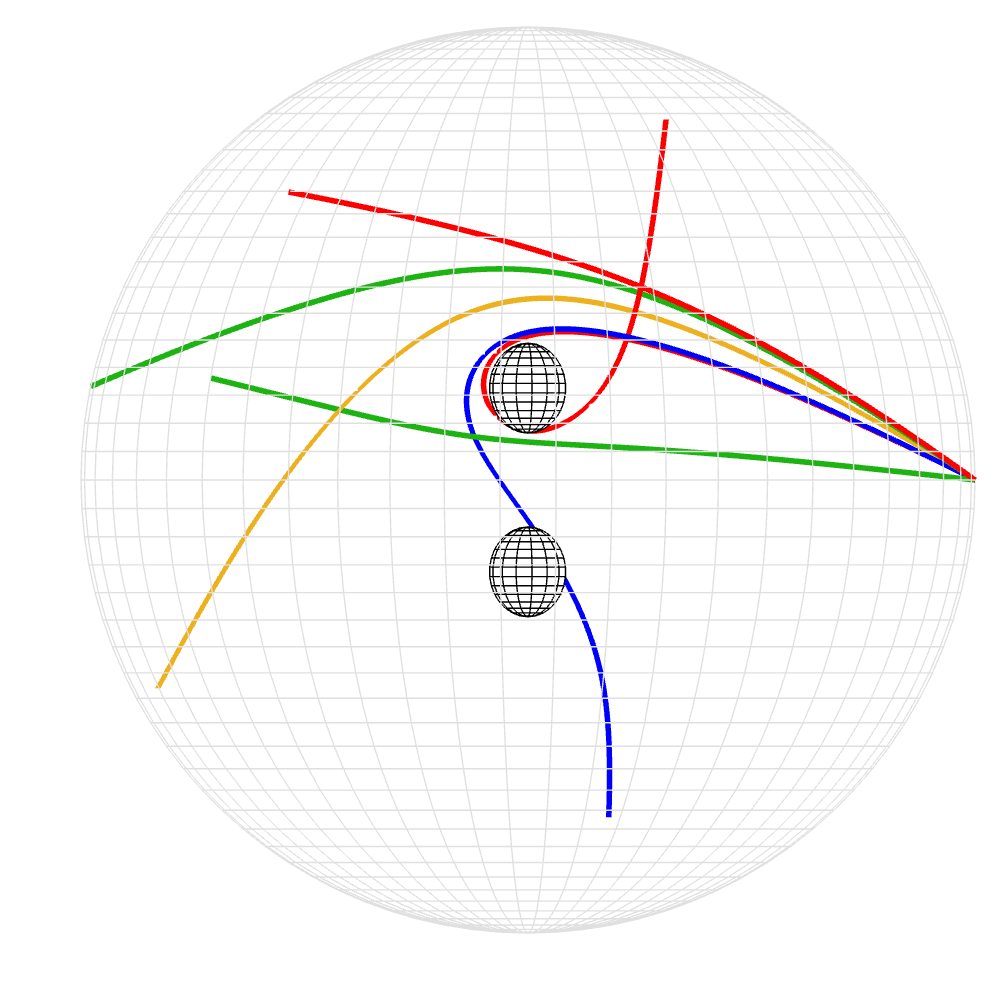}
    \includegraphics[width=0.4\linewidth]{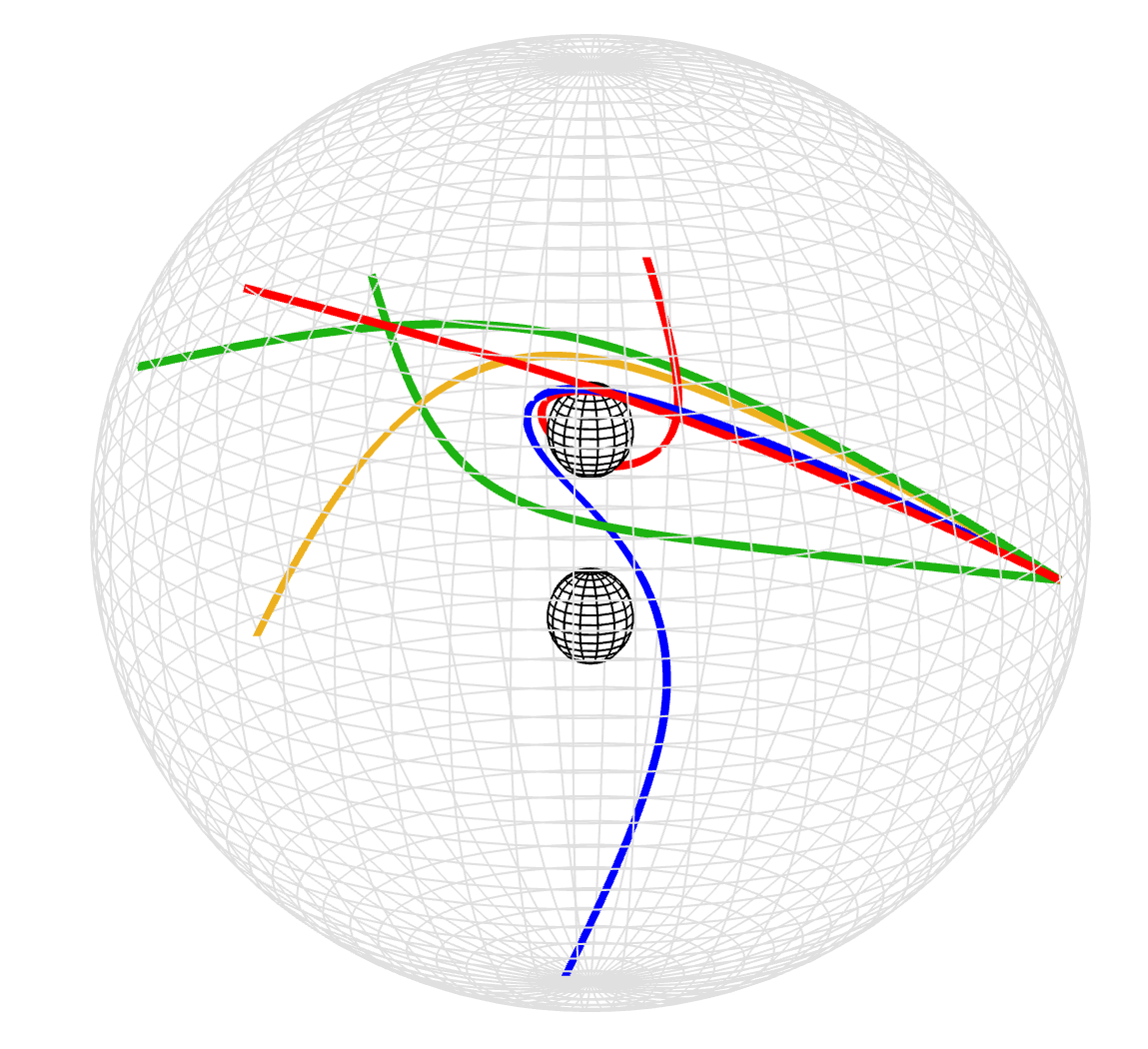}
    \caption{Trajectories of some light rays reaching different areas of the celestial sphere in a space-time with $m_1=m_2=1$ and $R_0=8$. The $xy$ projection, $xz$ projection and a 3D view are shown.} \label{fig:CS_quadrant}
\end{figure}

Fig.~\ref{fig:CS_quadrant} shows light rays corresponding to pixels in the primary, secondary and tertiary copies of the celestial sphere. The noticeable differences appear from the tertiary copy onward, since most of these trajectories correspond to photons traveling under the gravitational influence of both BHs and thus they will present more convoluted trajectories, see Fig. \ref{fig:LR_classes} and discussions in Subsection \ref{sec:LR_classes}.

\begin{figure}[!htb]
  \centering
  \includegraphics[width=0.3\linewidth]{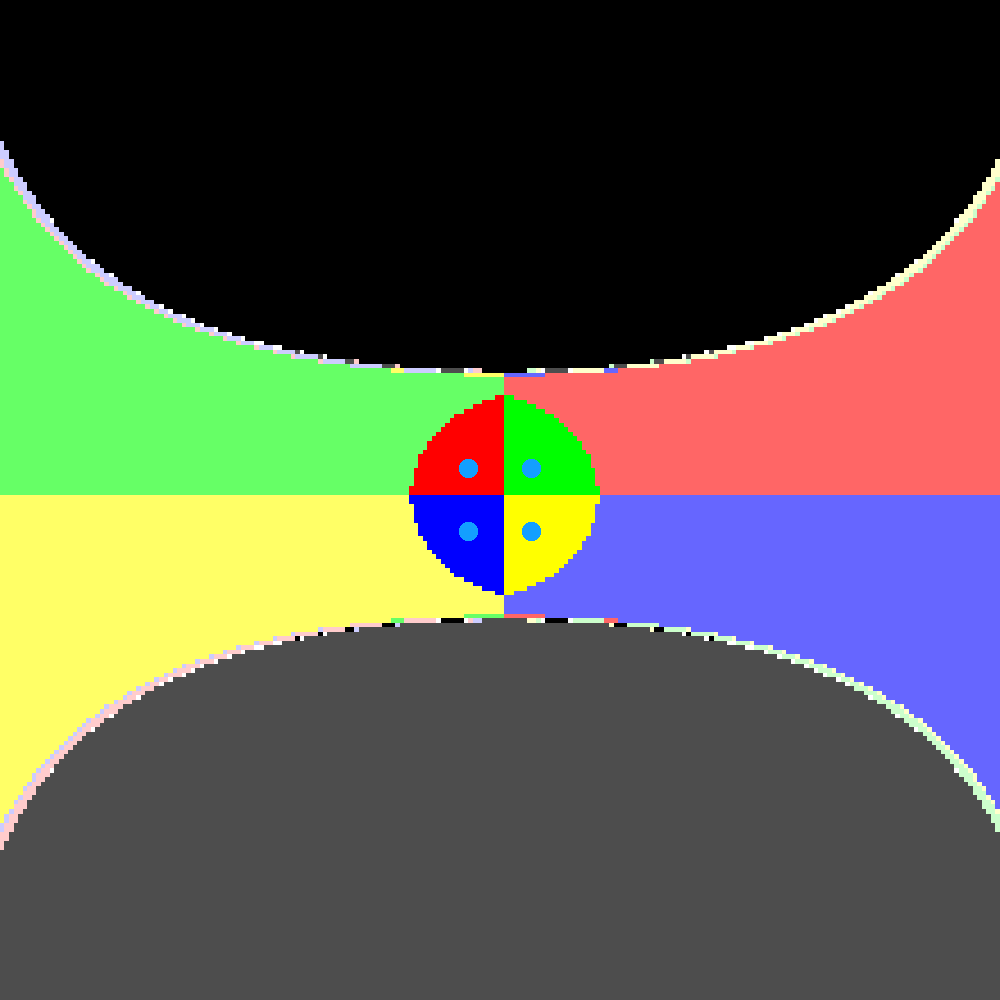}
  \includegraphics[width=0.3\linewidth]{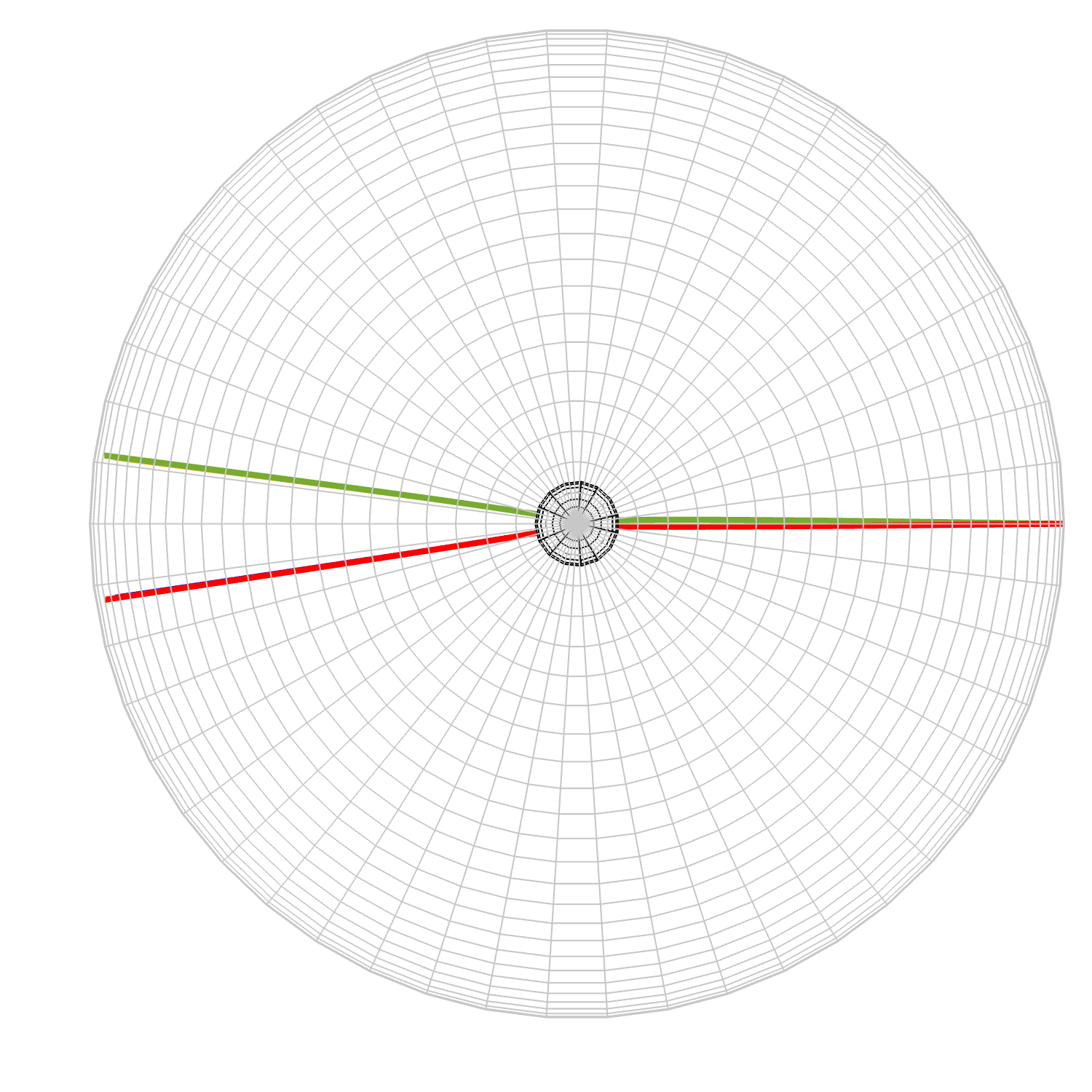}
  \includegraphics[width=0.3\linewidth]{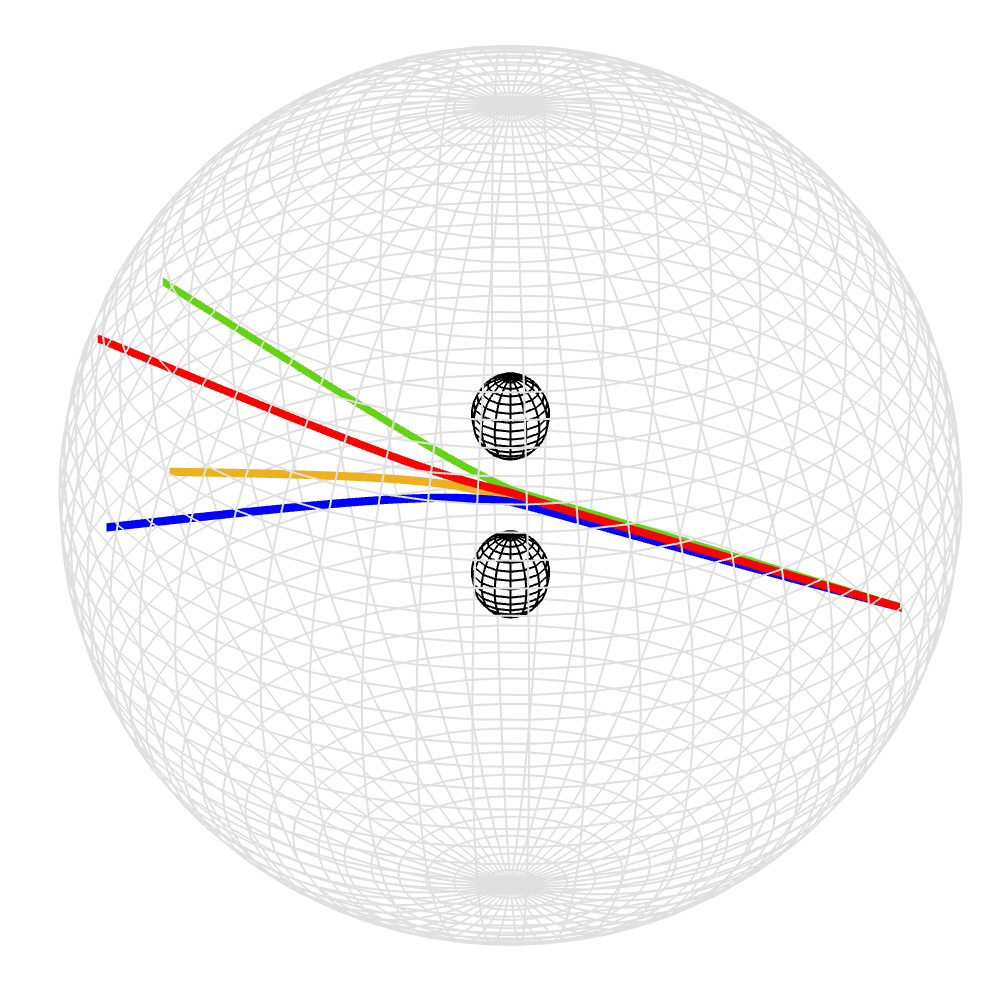}
  \caption{Trajectories of some light rays reaching different areas of the celestial sphere passing through the middle. The $xy$ projection, $xz$ projection and a 3D view are shown.} \label{fig:CS_center}
\end{figure}

Fig.~\ref{fig:CS_center} shows light rays corresponding to pixels in the primary image that suffer a defocusing effect due to the Weyl strut. These photons travel sufficiently close to the strut, but sufficiently distant from the photon spheres. This means that the effect of the strut overcomes the attractive force of either BH and repels the photons to infinity.

\section{Conclusion}
\label{sec:conclusion}
In this paper, we have presented a detailed study of ray tracing in 
the double Schwarzschild solution. The focus was on the comparison of 
null geodesics in the well known single black hole case with the 
double Schwarzschild solution. It was shown that black holes with a 
large spatial distance between them essentially are very similar to 
single black holes, the Weyl strut separating them shows only weak 
effects. However, if the black holes get closer, its defocusing 
effect becomes more and more noticeable. The Weyl strut acts as a 
concave lens with stronger effects the closer the holes are. This 
has also pronounced effects on the shadow of the black holes.

The presence of the Weyl strut between the two black holes has a 
strong effect on the dynamics of the null geodesics. The combination of the repulsion of the strut and the attraction of the individual black holes leads to interesting behavior of the trajectories. The location of higher order images inside lower order ones is reminiscent of chaotic maps, and so it is a natural question as to whether the behavior of the light rays in the double Schwarzschild solution is chaotic. The answer to this question is not easy to find, and we leave this to further investigation.

In this paper, we have concentrated on equal mass configurations where 
the equatorial plane is such that the attractive forces of the two 
black holes just compensate. This allows a clear identification of 
the effects of the Weyl strut in this plane. In future work, we will 
also study non-symmetric configurations.

The double Schwarzschild solution is a particular case of the double 
Kerr solution \cite{KN,Manko} and the only static member of this 
class of solutions. In the presence of rotating black holes, we 
expect frame dragging effects as known from the Kerr solution to 
appear in addition to the effects of the Weyl strut discussed in the 
present paper. It is also expected that the Weyl strut becomes weaker 
the faster the holes rotate, the optimum in this sense being two 
counter-rotating extreme holes. To explore such effects will be the 
subject of future work.

\end{document}